\documentclass[aps, prd, twocolumn, lengthcheck, superscriptaddress, showpacs, letterpaper, nofootinbib]{revtex4-1}

\usepackage{amsfonts}
\usepackage{amsmath}
\usepackage{amssymb}
\usepackage{xcolor}
\usepackage{graphicx}
\usepackage{subfigure}
\usepackage{longtable}
\usepackage{slashed}
\usepackage{hyperref}
\usepackage{amsthm}
\usepackage{mathrsfs}
\usepackage{color}
\usepackage{epsfig}
\usepackage{epstopdf}

\def\be{\begin{eqnarray}}\def\ee{\end{eqnarray}}

\def\la{\langle}\def\ra{\rangle}
\def\be{\begin{eqnarray}}\def\ee{\end{eqnarray}}
\def\lsim{\mathrel{\rlap{\lower3pt\hbox{\hskip1pt$\sim$}}
     \raise1pt\hbox{$<$}}} %less than or approx. symbol
\def\gsim{\mathrel{\rlap{\lower3pt\hbox{\hskip1pt$\sim$}}
     \raise1pt\hbox{$>$}}} %greater than or approx. symbol

\allowdisplaybreaks

\begin{document}

\title{Topology and emergent symmetries in dense compact star matter}

\author{Yong-Liang Ma}
\email{ylma@ucas.ac.cn}
%\affiliation{College of Physics, Jilin University, Changchun, 130012, China}
\affiliation{School of Fundamental Physics and Mathematical Sciences,
Hangzhou Institute for Advanced Study, UCAS, Hangzhou, 310024, China}

\author{Wen-Cong Yang}
\affiliation{School of Fundamental Physics and Mathematical Sciences,
Hangzhou Institute for Advanced Study, UCAS, Hangzhou, 310024, China}
\affiliation{College of Physics, Jilin University, Changchun, 130012, China}

\begin{abstract}
It has been found that the topology effect and the possible emergent scale and hidden local flavor symmetries at high density reveal a novel structure of the compact star matter. The $N_f\geq2$ baryons can be described by the skyrmion in the large $N_c$ limit and there is a robust topology change in the skyrmion matter approach to dense nuclear matter. The hidden scale and local flavor symmetries which are sources introducing the lightest scalar meson---dilaton---and lowest lying vector mesons into to nonlinear chiral effective theory are seen to play important roles in understanding the nuclear force. We review in this paper the generalized nuclear effective theory (G$n$EFT), which applicable to nuclear matter from low density to the compact star density, constructed with the robust conclusion from the topology approach to dense matter and emergent scale and hidden local flavor symmetries. The topology change at density larger than two times saturation density $n_0$ encoded in the parameters of the effective field theory is interpreted as the hadron-quark continuity in the sense of Cheshire Cat Principle. A novel feature predicted in this theory that has not been found before is the precocious appearance of the conformal sound velocity in the cores of massive stars, although the trace of the energy-momentum tensor of the system is not zero. That is, in contrast to the usual picture, the cores of massive stars are composed of quasiparticles of fractional baryon charges, neither baryons nor deconfined quarks. Hidden scale and local flavor symmetries emerge and give rise a resolution of the longstanding $g_A$ quench problem in nuclei transition. To illustrate the rationality of the GnEFT, we finally confront the generalized effective field theory to the global properties of neutron star and the data from gravitational wave detections.
\end{abstract}

\maketitle

\section{Introduction}

Although it has been investigated for several decades, there is no consensus on how to describe the equation of state (EoS) of dense nuclear matter relevant to compact stars~\cite{Brown:2001nh,Holt:2014hma,Drews:2016wpi,Baym:2017whm,Ma:2019ery,Li:2019xxz,Ma:2020nih,Lovato:2022vgq}. We do not certainly know what are the constituents involved and how do the symmetries of quantum chromodynamics (QCD) evolve in medium. The resolution of these questions has strong impacts on the most fundamental issues of particle and nuclear physics that have defied theorists, for example, the mechanism of chiral symmetry breaking and emergence of nucleon mass.

In the past decade, we studied the dense nuclear matter using a generalized nuclear effective field theory (G$n$EFT) including the lowest lying vector mesons $\rho$ and $\omega$ and the lightest iso-scalar scalar meson $\sigma$, in addition to the nucleon and pion considered in the standard chiral effect field theory (S$\chi$EFT)~\cite{Paeng:2015noa,Paeng:2017qvp,Ma:2018jze,Ma:2018xjw}. Implemented with the robust conclusions of the medium modified hadron properties obtained from the topology structure of QCD at large $N_c$ limit, we found that at density relevant to the cores of massive stars, the sound velocity saturates the approximate conformal limit although the scale symmetry is not restored and the system is still in the confined phase~\cite{Ma:2018qkg,Ma:2020hno}. That is, there could be a pseudoconformal structure in the cores of massive stars (see Refs.~\cite{Ma:2019ery,Ma:2020nih,Rho:2021zwm,Ma:2021nuf,Lee:2021hrw,Rho:2022uns} for reviews). Although the existing of the conformal sound velocity in massive neutron stars is in stark contrast to the previous beliefs~\cite{Bedaque:2014sqa,Tews:2018kmu,Moustakidis:2016sab,Alsing:2017bbc}, it is now observed in more and more models~\cite{McLerran:2018hbz,Jeong:2019lhv,Kapusta:2021ney,Zhao:2020dvu,Margueron:2021dtx} and allowed by the constraints from the neutron star data~\cite{Fujimoto:2022ohj,Marczenko:2022jhl}.

To obtain the pseudoconformal structure of dense nuclear matter, two significant ingredients are considered in the G$n$EFT, the hidden symmetries which are invisible in the matter-free space---hidden scale symmetry and hidden local flavor symmetry---and topology of QCD at large $N_c$.

The local flavor symmetry---hidden local symmetry (HLS)~\cite{Bando:1984ej,Bando:1987br,Harada:2003jx}---provides an effective field theory (EFT) approach of vector mesons $\rho$ and $\omega$ in the framework of the nonlinear realization of chiral symmetry. And, the breaking scale symmetry offers a source for introducing the scalar meson---regarded as dilaton---\'a la Crewther and Tunstall~\cite{Crewther:2013vea,Cata:2018wzl,Crewther:2020tgd}. In the compact star matter, due to the strong correlations among its constituents, these hidden symmetries may emerge and these emergent symmetries affect the compact star properties in either direct or indirect way.

When considered in the large $N_c$ limit, baryon can be regarded as the topology soliton---skyrmion---in a nonlinear field theory~\cite{Skyrme:1961vq}. This is an alternative approach to nuclear physics other than the widely used models including nucleon as an explicit fermionic field nowdays. Using the skyrmion approach, people found that when the nuclear matter is squeezed to a certain high density, the constituent of matter is changed from the baryon number-1 object to the baryon number-1/2 object, i.e., there is a topology change~\cite{Kugler:1988mu,Kugler:1989uc}. The existing of this topology change is robust although the density where it happens---denoted as $n_{1/2}$---is model dependent~\cite{Lee:2003eg,Park:2003sd,Park:2008zg,Ma:2013ooa,Ma:2013ela,Ma:2016gdd}. An interesting conclusion which has not been observed in other approaches is that, after $n_{1/2}$, some hadron properties such as the pion decay constant in medium $f_\pi^\ast$ and effective nucleon mass $m_N^\ast$ become density independent~\cite{Ma:2013ooa,Ma:2013ela}. Moreover, it is found that at $n\gsim n_{1/2}$, the sound velocity saturates the conformal limit $v_s^2/c^2 \simeq 1/3$~\cite{Shao:2022njr} although $f_\pi^\ast$ and $m_N^\ast$ are not zero.

Since the skyrmion approach to nuclear matter takes tremendous numerical simulations and includes obscure mathematics, the approach to nuclear matter using the EFTs including baryon fields as explicit degrees of freedom are widely used now. We implement the model independent observations from the skyrmion approach and the effects of the emergent symmetries to the medium modified parameters in G$n$EFT through the (extended) Brown-Rho scaling~\cite{Brown:1991kk}. By using the $V_{lowk}$ RG approach implementing the strategy of Wilsonian renormalization group flow~\cite{Bogner:2003wn}, with respect to the constraints from the nuclear matter around saturation density $n_0 \approx 0.16~$fm$^{-3}$, we construct the equation of state (EoS) of compact star matter which has pseudoconformal symmetry, i.e., the sound velocity at the density relevant to the compact stars satisfies the conformal limit $v_s^2/c^2 \simeq 1/3$ but the conformal symmetry is not restored. The predictions of such constructed pseudoconformal model (PCM) satisfy all the constraints from terrestrial experiments and astrophysical observations.

In this contribution, complementary to Refs.~\cite{Ma:2019ery,Ma:2020nih,Rho:2021zwm,Ma:2021nuf,Lee:2021hrw,Rho:2022uns}, we will review the key points of the pseudoconformal structure of dense nuclear matter with special interests on the emergent scale and local symmetries and topology constituents of the compact star matter.
%effect of the topology change, the constituents of the pseudoconformal matter and the patterns of the symmetries involved.

\section{Hidden symmetries and hadron resonances}

In the S$\chi$EFT of nuclear physics, the hadron degrees of freedoms are nucleon and pions. However, it has been recognized long time ago that the hadron resonances are crucial for reproducing the empirical date of nuclear matter around saturation density, such as the iso-scalar scalar meson $\sigma$ and vector mesons $\rho$ and $\omega$ in the Walecka model~\cite{Serot:1984ey}.

Another reason to include the hadron resonances in G$n$EFT is that although the finite nuclei as well as infinite nuclear matter can be fairly accurately accessed by nuclear EFTs---``pionless or pionful"---anchored on relevant symmetries they are expected to break down at some high density relevant to, say, the interior of massive stars. For example, when applying the S$\chi$EFT to nuclear matter where the power counting in density is $O(k_F^q)$ people found that even for the normal nuclear matter, the expansion requires going to $\sim q=5$~\cite{Holt:2014hma} therefore the more loops should be considered and more parameters are involved. This makes the calculation involved and ambiguities hard to control.

With the above considerations, we include the iso-scalar scalar meson $\sigma$ and vector mesons $\rho$ and $\omega$ in G$n$EFT. Note that, it is recently found that the iso-vector scalar mesons $\delta$ (denoted as $a_0(980)$ in particle physics) also affects the EoS through the symmetry energy in a sizable way~\cite{Li:2022okx,Miyatsu:2022wuy}. However, since we do not have any idea on how to construct an EFT for them and their structures are still under debate, we will not consider them in the present work.

In the literature, the hidden symmetries which are not visible in the matter-free space provide sources for including hadron resonances in the effective theories. Explicitly, the observed hadron resonances list in particle data group booklet~\cite{Workman:2022ynf} indicates that the approximate chiral symmetry in QCD breaks to the vector channel and pions can be regarded as the Nambu-Goldstone bosons of the broken axial symmetry. The local flavor symmetry---hidden local symmetry (HLS)~\cite{Bando:1984ej,Bando:1987br,Harada:2003jx}---suggests an EFT approach of vector mesons in the framework of the nonlinear realization of chiral symmetry. And, the hidden scale symmetry offers a source for introducing the iso-scalar scalar meson to chiral EFTs \'a la Crewther and Tunstall~\cite{Crewther:2013vea,Cata:2018wzl}. In the compact star matter, due to the strong correlations among its constituents, these hidden symmetries may emerge and these emergent symmetries affect the compact star properties in either a direct or an indirect way.

\subsection{Hidden scale symmetry}

The QCD Lagrangian in chiral limit is invariant under scale transformation. Therefore, the trace of the momentum-energy tensor $\theta_\mu^\mu$ vanishes at classical level, i.e, $\langle\theta_\mu^\mu\rangle=0$. This invariance is broken by trace anomaly at quantum level
\be
\theta_\mu^\mu & = & \frac{\beta(\alpha_s)}{4\alpha_s}G_{\mu\nu}^aG^{a\mu\nu} + (1+\gamma_5)\sum_{q=u,d,s} m_q \bar{q}{q},
\ee
where $m_q$ is the quark mass. Since the trace anomaly has the quantum number of vacuum, it has long been regarded as the source of the iso-scalar scalar meson in effective models~\cite{Isham:1970xz,Ellis:1970yd,Schechter:1980ak,Crewther:2013vea,Cata:2018wzl}.

In the construction of the effective model of the iso-scalar scalar meson using the trace anomaly as its source, the only constraint on the effective Lagrangian comes from the anomaly match. However, to build an EFT of scalar meson \'a la Weinberg, one should set up the power counting mechanism. This was finalized by the pioneer work of Crewther and Tunstall~\cite{Crewther:2013vea,Cata:2018wzl}.

An alternative to the CT scheme is the framework proposed by Golterman and Shamir (GS) in the large $N_c$ and large $N_f$ Veneziano limit~\cite{Golterman:2016lsd}. Although the IR structure is presumably different in both approaches, the GS scheme and CT scheme are found of the same form to NLO once $\beta^\prime$ in CT and $\Delta n_f = |n_f^c-n_f|$ in GS are related. In this work, we follow the CT scheme.

The basic idea of Crewther and Tunstall (CT for short) is that, the iso-scalar scalar meson can be regarded as the pseudo-Nambu-Goldstone boson of the spontaneous breaking of the scale symmetry driven by the explicit scale symmetry breaking. Provided that there is a nonperturbative infrared fixed point (IRFP) $\alpha_{IR}$ in QCD, the mass of the scalar meson, here dilaton, is generated by the explicit scale symmetry breaking which is encoded in the departure from the IRFP and the current quark mass. The magnitude of the mass is proportional to the deviation from the IRFP $(\Delta \alpha = \alpha_{IR}-\alpha_s)$ and the current quark mass therefore the situation is very similar to the chiral perturbation theory where the masses of the Nambu-Goldstone bosons are proportional to the current quark masses which measure the magnitude of the explicit chiral symmetry breaking. Note that whether the nonperturbative IRFP which the CT approach is anchored on exists in QCD is not yet confirmed. Among a variety of approaches, we simply refer to the positive arguments given in Refs.~\cite{Brodsky:2010ur,Horsley:2013pra,Yu:2021yvw}. Moreover, the lattice QCD indicates that in the thermal system, the scale symmetry may exists in IR region and therefore leads to the exists of massless glueballs~\cite{Alexandru:2019gdm}. It seems not strange to expect such a scale invariance may emerge in medium.

%In order to proceed, it is convenient to introduce the conformal compensator field $\chi$ which has the scale dimension 1. In the vacuum,

Following the procedure of CT~\cite{Crewther:2013vea,Cata:2018wzl}, introducing the conformal compensator field $\chi$ which has the scale dimension 1 and $\langle\theta_\mu^\mu\rangle = \langle \chi^4\rangle = f_\chi^4$, one can write the chiral-scale effective Lagrangian in terms of the scale dimensions in the chiral-scale power counting in momentum (derivative), quark mass and $\Delta\alpha=\alpha_{IR}-\alpha_s$
\be
\mathcal{O}(p^2) \sim \mathcal{O}(m_q) \sim \mathcal{O}(\Delta\alpha).
\ee

In terms of the pseudoscalar pions $U(x) = e^{2i\pi(x)/f_\pi}$ and $\chi = e^{\sigma(x)/f_\chi}$ with $\sigma$ being the dilaton field which will be identified with the lightest scalar meson $f_0(500)$, one can write the effective Lagrangian at leading chiral-scale order as
\be
{\cal L}_{\rm LO} & = & {\cal L}_{\rm inv}^{d=4} + {\cal L}_{\rm anom}^{d > 4} + {\cal L}_{\rm mass}^{d < 4}
\ee
where $d$ stands for the scale dimension and
\begin{subequations}
	\begin{eqnarray}
		{\cal L}_{\rm inv}^{d=4} & = & c_1 \frac{f_\pi^2}{4} \left( \frac{\chi}{f_\chi}\right)^2 {\rm Tr}\left( \partial_\mu U \partial^\mu U^\dagger \right) + \frac{1}{2} c_2 \partial_\mu \chi \partial^\mu \chi \nonumber\\
		& &{} + c_3 \left( \frac{\chi}{f_\chi}\right)^4, \label{eq:CTL40}\\
		{\cal L}_{\rm anom}^{d > 4} & = & (1 - c_1)\frac{f_\pi^2}{4} \left( \frac{\chi}{f_\chi}\right)^{2+\beta^\prime} {\rm Tr}\left( \partial_\mu U \partial^\mu U^\dagger \right) \nonumber\\
		& &{} + \frac{1}{2}(1 - c_2) \left( \frac{\chi}{f_\chi}\right)^{\beta^\prime} \partial_\mu \chi \partial^\mu \chi \nonumber\\
		& &{} + c_4 \left( \frac{\chi}{f_\chi}\right)^{4+\beta^\prime},\label{eq:CTLg40}\\
		{\cal L}_{\rm mass}^{d < 4} & = &{} \frac{f_\pi^2}{4} \left( \frac{\chi}{f_\chi}\right)^{3-\gamma_m} {\rm Tr}\left( \mathcal{M}^\dagger U + U^\dagger \mathcal{M} \right),\label{eq:CTLm40}
	\end{eqnarray}
\end{subequations}
where $\mathcal{M}$ stands for  the current quark matrix with $\mathcal{M} = {\rm diag}(m_\pi^2,m_\pi^2, 2m_K^2 - m_\pi^2)$,  $\gamma_m$ is the anomalous dimension of the quark mass operator $\bar{q}q$, $c_i$'s are unknown constants. It should be noted that, different from the chiral perturbation theory, $c_3$ and $c_4$ have scale-chiral order $O(p^2)$ since, as will be seen latter, they are proportional to the dilaton mass square, similarly to $\mathcal{M}$ for the pseudo-scalar Nambu-Goldstone mesons.

We next consider the dilaton potential $V(\chi)$
\be
V(\chi) & = &{} - c_3 \left( \frac{\chi}{f_\chi}\right)^4 - c_4 \left( \frac{\chi}{f_\chi}\right)^{4+\beta^\prime}.
\ee
The saddle point equation in the matter-free space yields
\begin{eqnarray}
	V(\chi) & = & {} - (4 + \beta^\prime)c\left( \frac{\chi}{f_\chi}\right)^4 + 4 c \left( \frac{\chi}{f_\chi}\right)^{4 + \beta^\prime},
	\label{eq:potentialchictLO}
\end{eqnarray}
where
\be
c ={} - \frac{1}{4}c_4 = \frac{1}{4+\beta^\prime}c_3 >0.
\label{eq:CC3C4}
\ee
We see that, with $\beta^\prime \neq 0$, that is, the dilaton potential breaks scale symmery, the dilaton potential is in the Nambu-Goldstone (NG)  mode, i.e., the minima of the potential appears at $\langle \chi \rangle = f_\chi$. However, if $\beta^\prime = 0$, $V(\chi) = 0$ and dilaton potential is scale invariant, the scale symmetry cannot spontaneous break. This simple observation illustrates that the spontaneous breaking and explicit breaking of scale symmetry are correlated and the spontaneous breaking is triggered by explicit breaking, which agrees with that unlike chiral symmetry, spontaneous breaking of scale symmetry cannot take place in the absence of explicit symmetry breaking~\cite{Freund:1968gyq}. We refer to  this as ``Freund-Nambu theorem."

Using the definition of the dilaton mass $m_\sigma$, from the dilaton potential one has
\be
c & = & \frac{m_\sigma^2f_\chi^2}{4\beta^\prime(4+\beta^\prime)}.
\ee
So that the constant $c$ and therefore $c_3$ and $c_4$ through relation \eqref{eq:CC3C4}, has chiral-scale dimension $O(p^2)$. We finally obtained the dilaton potential as
\be
V(\chi) & = & {} \frac{m_\sigma^2f_\chi^2}{4\beta^\prime(4+\beta^\prime)}\left( \frac{\chi}{f_\chi}\right)^4\left[- (4 + \beta^\prime) + 4 \left( \frac{\chi}{f_\chi}\right)^{\beta^\prime}\right].
\ee
When $\beta^\prime \ll 1$, the dilaton potential is approximated to~\cite{Goldberger:2007zk}
\begin{equation}
	V(\chi) = \frac{m_\sigma^2 f_\chi^2}{4} \left(\frac{\chi}{f_\chi}\right)^4 \left(\ln \frac{\chi}{f_\chi} -\frac{1}{4}\right).
	\label{anomaly1}
\end{equation}
This yields the scale Ward-Takahashi identity
%~\footnote{\cmh{Check this identity.}}
\be
\la\theta_\mu^\mu\ra=\la\partial_\mu D^\mu\ra ={} - \frac{m_\chi^2}{4f_\chi^2}\left\la\chi^4\right\ra
\ee
which is the partially conserved dilatonic current (PCDC) relation, the counterpart to the PCAC for the pion.
%Other low-energy theorems, some with baryons incorporated, e.g., the Goldberger-Treiman relation, can be derived.

Along the reasoning of CT, one can set up a systematic higher-order expansion and write down the higher order terms~\cite{Li:2016uzn,Cata:2019edh}. In the general Lagrangian, there are so many unknown parameters that it is difficult to give any prediction in practice, even at the leading order. However, one can make a substantial progress and arrive at a manageable form by taking the so called ``leading-order scale symmetry (LOSS)" approximation that corresponds to
\be
c_1\approx c_2\approx 1.
\label{eq:LOSSCi}
\ee
That is, in LOSS, the scale symmetry breaking---in the chiral limit---is lodged entirely in the dilaton potential $V(\chi)$. The resulting Lagrangian is
\begin{eqnarray}
	{\cal L}_{{\rm LO}}^{\chi{\rm limit}}  =  \frac{f_\pi^2}{4} \left( \frac{\chi}{f_\chi}\right)^2 {\rm Tr}\left( \partial_\mu U \partial^\mu U^\dagger \right) + \frac{1}{2} \partial_\mu \chi \partial^\mu \chi
	- V(\chi).
\nonumber\\
	\label{LOSS}
\end{eqnarray}

Whether the LOSS approximation is valid cannot be justified from the first principle. The numerical analysis shows that it works for light nuclei~\cite{Li:2017udr,Li:2018ykx} and compact star matter~\cite{Ma:2019ery} but it violates around saturation density~\cite{Ma:2020tsj}. In this contribution, without specification, we work with LOSS.

\subsection{Hidden local flavor symmetry}

To include the vector mesons into the chiral effective theory, among a variety of approaches, we use the hidden local symmetry (HLS)~\cite{Bando:1984ej,Bando:1987br,Harada:2003jx}.
%The basic idea of the HLS is summarized as follows: The chiral perturbation theory based on the manifold $G/H$ is gauge equivalent to a model having the symmetry $G_{\rm global}\times H_{\rm local}$. After dynamical breaking of the local symmetry $H_{\rm local}$, the gauge boson of the local symmetry is identified as the massive vector mesons. After integrating the massive vector mesons from the model at low energy, one can get chiral perturbation theory.
%
Explicitly, considering the chiral symmetry $G_{\rm global} = [SU(2)_L \times SU(2)_R]_{\rm global}$, following the convention of Ref.~\cite{Harada:2003jx}, we decompose the field $U(x)$ as
\be
U(x) & = & \xi_L^\dagger \xi_R(x).
\ee
Therefore, one can sandwich a local unitary transformation $H_{\rm local} = [U(2)_V ]_{\rm local}$ between this decomposition. Under transformation $G_{\rm global}\times H_{\rm local}$, $\xi_{L,R}$ transform as
\be
\xi_{L,R}(x)\mapsto\xi_{L,R}^\prime(x) =
h(x)\xi_{L,R}(x)g_{L,R}^\dag,
\label{hidden-trans}
\ee
where $h(x) \in H_{\rm local}$ and $g_{L,R}\in SU(2)_{L,R}$. The variables $\xi_{L,R}$ can be parameterized as
\be
& & \xi_{L,R}(x)=e^{i\sigma(x)/(2f_\sigma)}e^{\pm
i\pi(x)/(2f_\pi)}, \label{parameHidden}
\ee
where $\pi(x)=\pi^aX^a$ and $\sigma(x)=\sigma^\alpha S^\alpha $ with $X^a$ being the generators of the broken chiral symmetry and $S^a$ as the generators of the unbroken subgroup $H$. Note that here $\sigma$ is the Nambu-Goldstone boson which becomes the longitudinal part of gauge boson $V_\mu$ of symmetry $H_{\rm local}$ with $f_\sigma$ being its decay constant.

With quantities $\xi_{L,R}$ one can define the following two $1$-forms:
\begin{eqnarray}
\hat{\alpha}_{\parallel\mu} & = & \frac{1}{2i}(D_\mu \xi_R \cdot
\xi_R^\dag +
D_\mu \xi_L \cdot \xi_L^\dag), \nonumber\\
\hat{\alpha}_{\perp\mu} & = & \frac{1}{2i}(D_\mu \xi_R \cdot
\xi_R^\dag - D_\mu \xi_L \cdot \xi_L^\dag)
\ , \label{eq:1form}
\end{eqnarray}
where the covariant derivative is defined as $D_\mu \xi_{R,L} = (\partial_\mu - i V_\mu)\xi_{R,L}$, and both of these quantities transform as $\hat{\alpha}_{\parallel,\perp}^{\mu} \rightarrow h(x) \hat{\alpha}_{\parallel,\perp}^{\mu} h(x)^\dag$. For the gauge field $V_\mu$ we have the field strength tensor
\begin{eqnarray}
V_{\mu\nu}(x) = \partial_\mu V_\nu(x)-\partial_\nu
V_\mu(x)-i[V_\mu(x),V_\nu(x)],\label{eq:hiddentensor}
\end{eqnarray}
with the transformation $V_{\mu\nu}(x) \rightarrow h(x) V_{\mu\nu}(x) h(x)^\dag$.

In terms of the two 1-forms defined by Eq.~(\ref{eq:1form}) and field strength tensor (\ref{eq:hiddentensor}), one can construct a Lorentz invariant Lagrangian with the minimal number of derivatives as
\begin{eqnarray}
{\cal L}_{\rm HLS} & = & f_\pi^2 {\rm
Tr}[\hat{a}_{\perp\mu}\hat{a}_{\perp}^{\mu}] + f_\sigma^2{\rm
Tr}[\hat{a}_{\parallel\mu}\hat{a}_{\parallel}^{\mu}] -
\frac{1}{2g^2}{\rm Tr}[V_{\mu\nu}V^{\mu\nu}]. \nonumber\\
\label{eq:lagrhls}
\end{eqnarray}
To generate the masses of the gauge bosons therefore to identify them as the physical vector mesons $\rho$ and $\omega$, we use the Higgs mechanism and take the unitary gauge
\begin{eqnarray}
\xi_L^\dag & = & \xi_R \equiv \xi = e^{i\pi/(2f_\pi)}, \;\;\;\; U(x)
= \xi^2(x). \label{eq:hiddenbreaking}
\end{eqnarray}
Then, the gauge bosons $V_\mu$ acquires mass
\begin{eqnarray}
m_V^2 = m_\rho^2 = m_\omega^2 = g^2 f_\sigma^2,
\label{eq:massvhls}
\end{eqnarray}
which has the standard form of the gauge boson mass from the Higgs mechanism.

So far, we use $H_{\rm local} = [U(2)_V ]_{\rm local}$ therefore the rho meson and omega meson have the same mass \eqref{eq:massvhls}. This approximation works well in the matter-free space. However, in medium, it is found that this approximation breaks~\cite{Paeng:2015noa,Paeng:2011hy} and it is reasonable to take the HLS $H_{\rm local} = [SU(2)_V \times U(1)_V]_{\rm local}$.

In the HLS, considering that the masses of $\rho$ mesons are smaller than the chiral symmetry breaking scale $\Lambda_\chi$, one can make a systematic expansion including vector meson loops due to the gauge invariance~\cite{Georgi:1989gp,Georgi:1989xy} and, set up a self-consistent power counting mechanism, the essential character of effective theory~\cite{Harada:2003jx}. Since in the nuclear matter, due to the strong correlation among hadrons, the effective mass of $\rho$ meson is reduced, the convergence of the expansion is enhanced.

Based on the Wilsonian renormalization group (RG) approach, it is found that at high energy scale $f_\pi \to 0$ and $m_\rho \to m_\pi \to 0 $, i.e., there is a vector manifestation (VM) fixed point in the hidden local symmetry (HLS)~\cite{Harada:2003jx,Harada:2000kb}. It is not strange to expect that the VM appears at (super-)high density. We will see latter that this VM, although happens at the supper high density beyond the cores of massive stars, it affects the equation of state of neutron star matter in an indirect way.

In addition, in the approach to the baryonic matter using the dilaton compensated chiral effective theory, people found that there is a dilaton limit fixed point (DLFP) which states that the medium modified decay constant $f_\chi^\ast \to 0$ in theory at high density~\cite{Beane:1994ds,Paeng:2011hy}. When the DLFP is approached, the vector meson $\rho$ becomes massless and the HLS is emerged~\cite{Suzuki:2017tux}. Although the DLFP is saturated at the density beyond the core of massive neutron stars $\sim 10 n_0$, it affects the properties of the equation of state of neutron star, for example the sound velocity~\cite{Paeng:2015noa,Paeng:2017qvp,Yang:2022}.

\section{Topology change and hadron-quark continuity}

It is recognized long time ago that in the large $N_c$ limit, baryon properties share the same $N_c$ scaling as the soliton properties in the nonlinear mesonic theories~\cite{Witten:1979kh}. This gives an alternative approach to nuclear physics by using the topology properties of QCD at large $N_c$ limit, that is, regarding the baryon as a skyrmion in the Skyrme model~\cite{Skyrme:1961vq}, other than the standard EFT approach including the baryon fields as explicit degrees of freedom.

Using the skyrmion approach and regarding baryon as skyrmion, one can study the single baryon, multibaryon and nuclear matter in a unified way~\cite{Ma:2016gdd}. Since the skyrmion approach is only based on the topology structure of QCD, some qualitative conclusions obtained in the approach, such as the existence of the topology change and density dependence of some parameters in nuclear matter that will be illustrated later should be model independent.

\subsection{Baryons as topology objects and topology change}

In the nonlinear realization of chiral symmetry, the pion figures as the Nambu-Goldstone boson of the spontaneous breaking of chiral symmetry and it is expressed in the polar parameterization through $U(x)$.
%\begin{eqnarray}
%U(x) & = & \exp\left(2i\pi^a T^a/f_\pi\right),
%\label{eq:defU}
%\end{eqnarray}
%where $T^a$ is the generator of the $SU(2)$ group satisfying ${\rm Tr}(T^a T^b) = \frac{1}{2}\delta^{ab}$ and $f_\pi$ is the pion decay constant with a empirical value $f_\pi \simeq 92.4$~MeV.

Since the unitary field $U(x)$ satisfies $U(x)U(x)^\dagger = U(x)^\dagger U(x) = 1$, for any fixed time, say,  $t_0$, the matrix $U(\mathbf{x},t_0)$ defines a map from the manifold $R^3$ to the manifold $S^3$ in isospin space, that is
\begin{eqnarray}
U(\mathbf{x}, t_0): R^3 \to S^3,
\label{eq:mapU2S3}
\end{eqnarray}
for the static configuration $U(\mathbf{x}, t_0)$. At low energy limit, QCD goes to the vacuum, i.e.,
\begin{eqnarray}
U(|\mathbf{x}| \to \infty, t_0)& = & \mathbf{1},
\label{eq:configuxinfinite}
\end{eqnarray}
therefore, all the points at $|\mathbf{x}| \to \infty$ are mapped onto the north pole of $S^3$ and energy of the system is finite.

In the language of topology, maps \eqref{eq:mapU2S3} constitute the third homotopy group $\pi_3(S^{\,3}) \sim Z$ with $Z$ being the additive group of integers which accounts for the times that $S^3$ is covered by the mapping $U(\mathbf{x}, t_0)$, i.e., winding numbers. Because a change of the time coordinate can be regarded as a homotopy transformation which cannot transit between the field configurations in homotopically distinct classes, the winding number is a conserved quantity in the homotopy transformation by the unitary condition of the field $U(x)$ and condition~\eqref{eq:configuxinfinite}. In skyrmion models, the conserved winding number represent the conserved baryon number in QCD. The baryon arises as a topological soliton with the topology lodged in the chiral field $U(x)$. Therefore, in the construction of the skyrmion-type model, only the unitary condition of the field $U(x)$ and the condition \eqref{eq:configuxinfinite} are essential characteristics that should be taken into account~\cite{Ma:2016gdd}.

In the skyrmion approach, one can simulate the nuclear matter by putting skyrmions onto the crystal lattice, first put forward by Klebanov~\cite{Klebanov:1985qi}, and regarding the skyrmion matter as baryonic matter. The density effect enters when the crystal size is changed. This approach suggests a method to study the nuclear matter at densities higher than the dilute density using the topology of QCD. In practice, we do not know which crystalline the nature favors. So far, the face-centered-cubic (FCC) crystal is the known configuration  which yields the lowest energy~\cite{Kugler:1988mu,Kugler:1989uc}.

Among a variety of properties revealed in the crystal approach to dense matter, the most important one is the existence of half-skyrmion---a winding number-1/2 object---configurations at some higher density. Being topological, its presence is a robust prediction~\cite{Goldhaber:1987pb}.  Its does not depend on what degrees of freedom other than the pions are involved. What is significant is that it involves a topology change from skyrmions to half-skyrmions, which is responsible for a dramatic change in the properties of the dense matter at a density $n_{1/2}\gsim 2n_0$, a feature which has not been observed in other approaches in the literature. We will see that it plays significant roles in describing the equation of state for compressed baryonic matter relevant for massive compact stars.

\begin{figure}[ht]\centering
\includegraphics[scale=0.2]{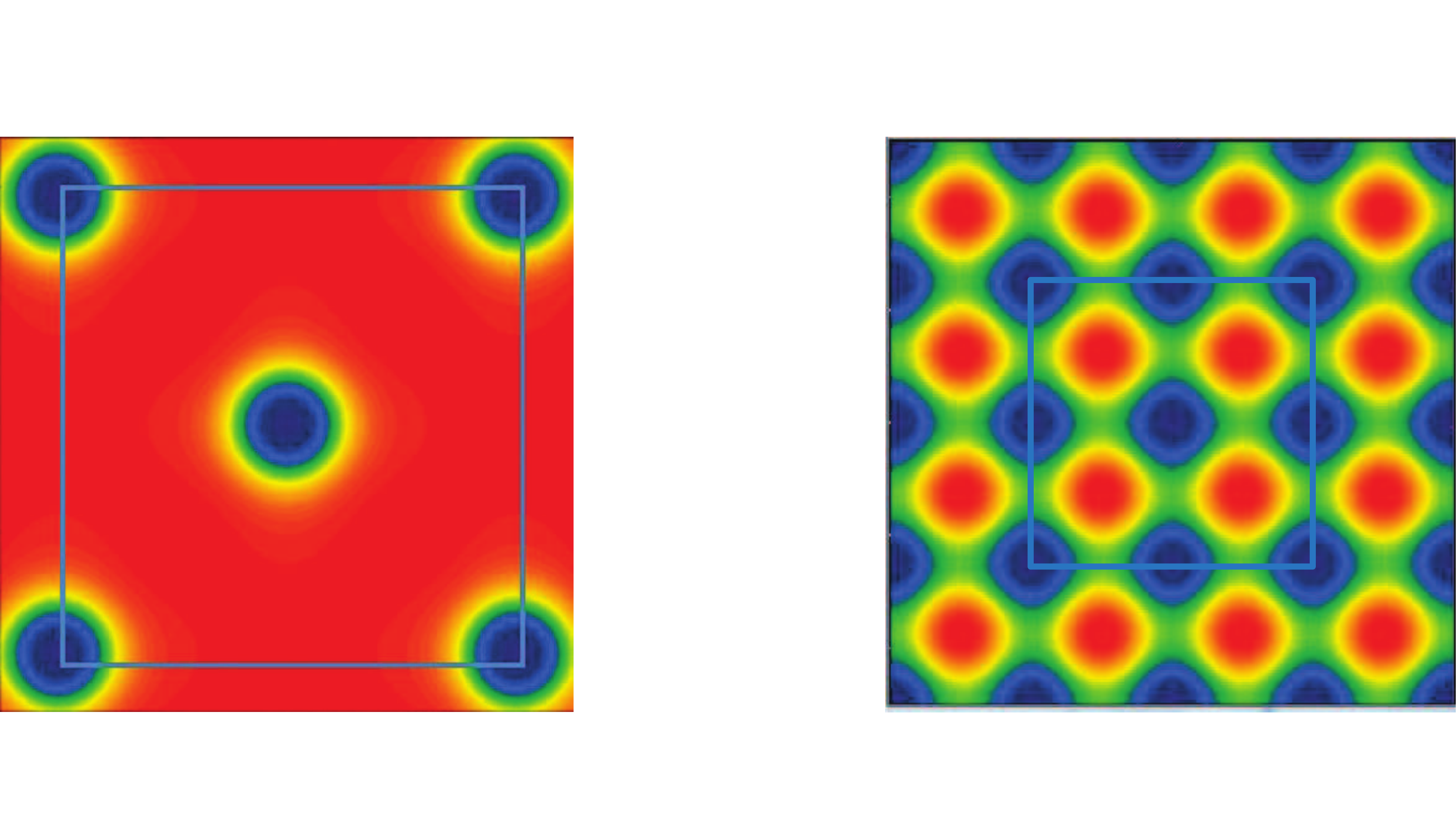}
\caption{The distribution of the baryon number density in the skyrmion (left panel) matter and half-skyrmion matter (right panel). The crystal size $2L$ is denoted by the blue quare.}
\label{fig:SkyrPhaseTran}
\end{figure}

Fig.~\ref{fig:SkyrPhaseTran} shows how the skyrmion FCC crystal configuration transforms to the half-skyrmion configuration in terms of the distribution of baryon number density. In the left panel, one can easily see that besides the corners and the center of the square where the skyrmions are originally put, the baryon number density emerges at the middles of the lines connecting the corners. That is, in the half-skyrmion phase, the vertices where the baryon number accumulates forms the CC crystal. After integration, each blue area has winding number-1/2 (for a detailed explanation, see, e.g., Ref.~\cite{Ma:2016npf}).

\begin{figure}[h]
\begin{center}
\includegraphics[width=7.5cm]{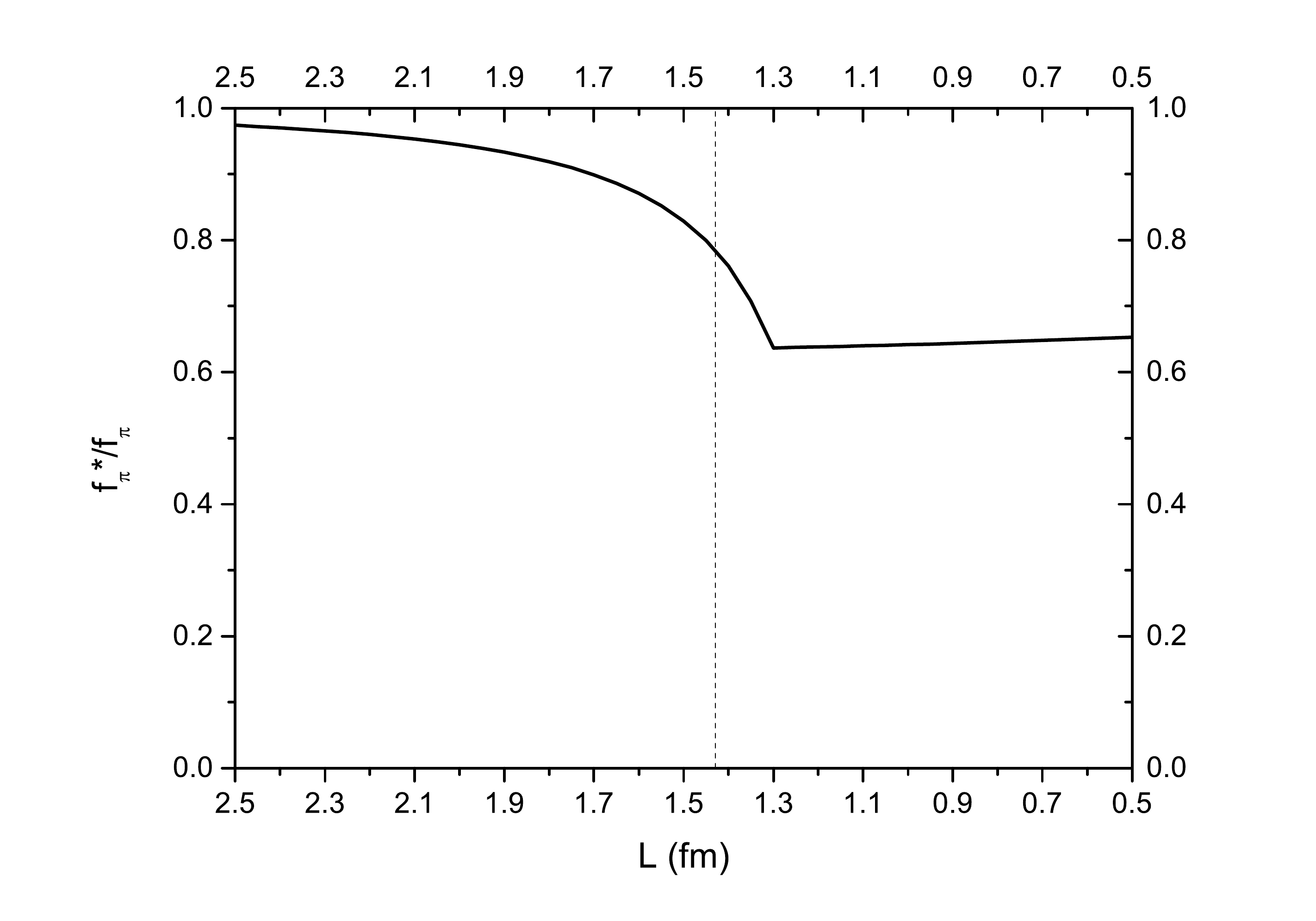}
\includegraphics[width=7.0cm]{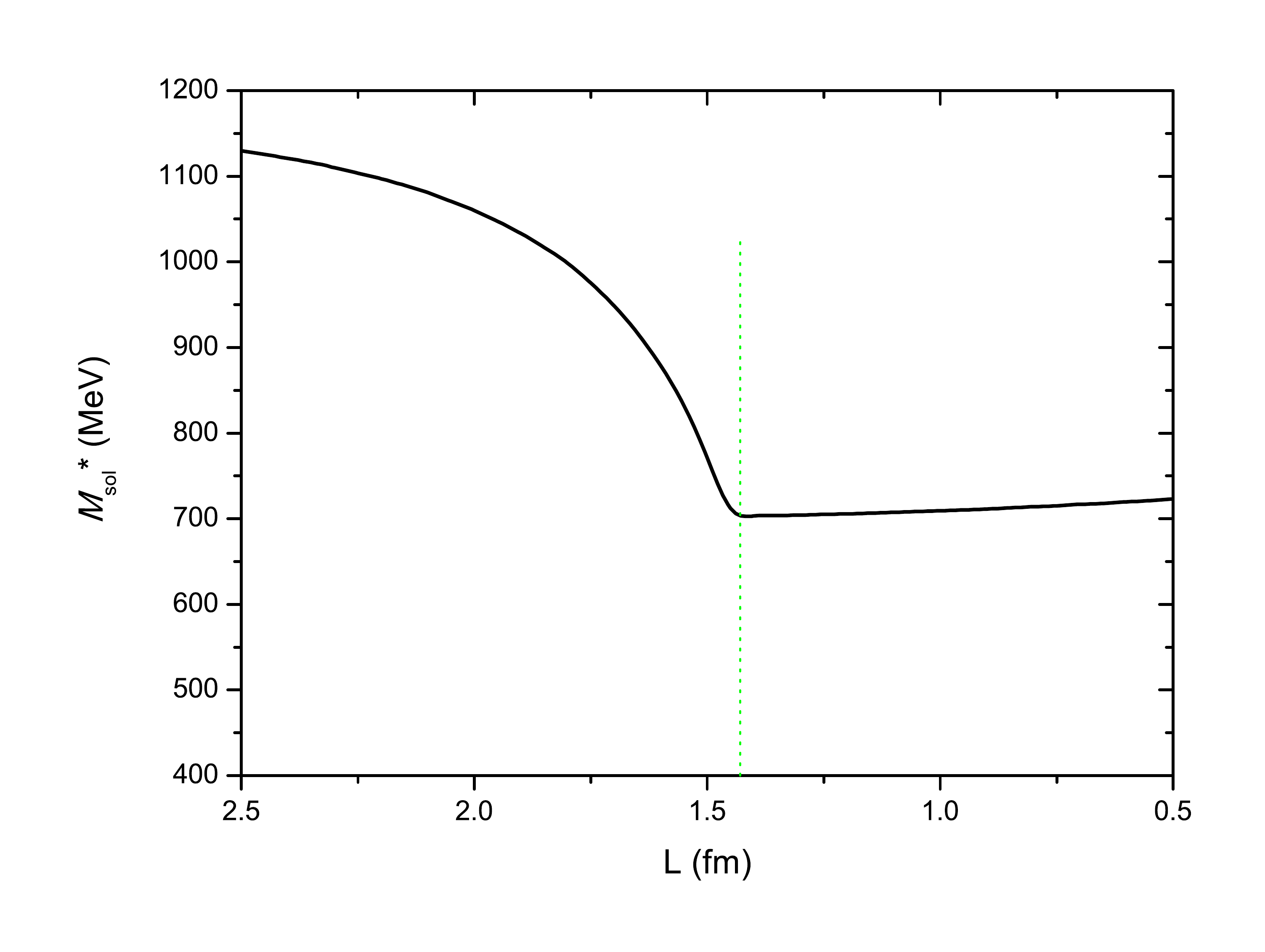}
\caption{The medium modified $f_\pi^\ast$ and $m_N^\ast$ vs. lattice size $L$ in HLS~\cite{Ma:2013vga}.}\label{fig:fpimass}
 \end{center}
\end{figure}

Due to this topology change, a variety of novel phenomena which have not been observed in the standard nucleon EFT approach emerges. Although the locations of the density $n_{1/2}$ where these phenomena start depend on model, their existences are model independent. Some of them which closely relate the present review are summarized as the following:
\begin{itemize}
  \item Quark condensate: In skyrmion matter, the space-average of the normalized quark condensate $\phi_0 = \frac{1}{2}{\rm Tr}(U)$ is
      \begin{eqnarray}
\langle\phi_0\rangle & = & \frac{1}{V}\int^{2L}_0 d^3 x \,\frac{1}{2}{\rm Tr}(U) , \label{eq:defvevphi0}
\end{eqnarray}
with $V$ being the volume of the crystal cell.  It is found that in the skyrmion matter $\langle\phi_0\rangle \neq 0$ but $\langle\phi_0\rangle \to 0$ in the half-skyrmion state. This implies that the quark condensate $\langle\bar{q}q\rangle^\ast \to 0$ in the half-skyrmion matter when space-averaged.

  \item Pion decay constant: In the skyrmion crystal approach, it is found that the medium modified pion decay constant $f_\pi^\ast$ first decreases with density until $n_{1/2}$ but after $n_{1/2}$ $f_\pi^\ast$ stays as a constant. We plot $f_\pi^\ast$ as a function of crystal size in Fig.~\ref{fig:fpimass}. This means the in the half-skyrmion matter, although the space averaged quark condensate vanishes, the chiral symmetry is not restored and it is still in the Nambu-Goldstone mode. Actually, in the half-skyrmion matter, the inhomogeneous quark condensate persists~\cite{Harada:2015lma}.

  \item Nucleon mass: By using the medium modified pion decay constant $f_\pi^\ast$, one can calculate the density dependence of nucleon mass $m_N^\ast$ and obtain the scaling relation
      \be
      \frac{m_N^\ast}{m_N} & \approx & \frac{f_\pi^\ast}{f_\pi},
      \ee
      which, as discussed later, is consistent with the Brown-Rho scaling from the LOSS~\cite{Brown:1991kk}. It is found that, as shown in Fig.~\ref{fig:fpimass}, similar to $f_\pi^\ast$, $m_N^\ast$ first decreases with density until $n_{1/2}$ after which it keeps as a constant. This is predominantly, if not entirely, due to the space-averaged quark condensate going to zero at $n_{1/2}$. Since in the half-skyrmion matter, $\langle \bar{q}q\rangle^\ast \to 0$, this observation indicates that the nucleon mass as decomposition
      \be
      m_N = m_0 + \Delta(\bar{q}q),
      \ee
      that is, there is chiral invariant part in the nucleon mass and the parity doubling of the nucleons may emerge in dense nuclear matter~\cite{Detar:1988kn,Motohiro:2015taa}.

\end{itemize}

We want to emphasize that the tendencies discussed above are robust but the location of $n_{1/2}$ is highly model dependent. So far we cannot pin down the value of $n_{1/2}$. The combination of the terrestrial experiments and astrophysical obserevations leads to the constraint, as see later, $2.0 n_0 \lsim n_{1/2} \lsim 4.0n_0$.

\subsection{Topological baryon for $N_f=1$}

As stated above that the skyrmion approach is anchored on the map~\eqref{eq:mapU2S3}. How to or if it is possible to study baryon such as $\Delta$ resonance in one-flavor QCD using the topology approach is a problem since $\pi_3(U(1))=0$. In one-flavor QCD case, the chiral effective theory is dominated by the axial $U(1)$ anomaly for the $\eta^\prime$ meson, and the soliton construction no longer applies since, for instance, the standard topological charge cannot be identified.

In 2018, Komargodski~\cite{Komargodski:2018odf} noted that the effective theory has a conserved topological current $J_{\alpha\beta\gamma}=\epsilon_{\alpha\beta\gamma\lambda}\partial^\lambda\eta^\prime/2\pi$ which are carried by $(1+2)$-dimensional charged sheets with the $\eta^\prime$ field undergoing a $2\pi$ jump across the sheet. When these sheets are finite dimensional with a boundary, they can carry massless edge excitations with baryon quantum numbers therefore can be identified with fast spinning baryons. These sheets are described by a topological field theory through a level-rank duality argument~\cite{Hsin:2016blu,Gaiotto:2017tne,Benini:2017aed}, much like in the fractional quantum Hall (FQH) effect~\cite{Tong:2016kpv}. The baryons are analogous to the gapless edge excitations in quantum Hall (QH) droplets.

By using the HLS approach, Karasik~\cite{Karasik:2020pwu} pointed that the vector mesons play the role of the Chern-Simons vector fields living on the QHD that forms the $N_f = 1$ baryon. This proposal gives a unified picture for the two types of baryons and allows them to continuously transform one to the other. Recently, Bigazzi et al proposed a string theory description of the QH sheet using the Witten-Sakai-Sugimoto model~\cite{Bigazzi:2022luo}.

\subsection{Cheshire Cat Principle and Quark-Hadron Continuity}

Based on what we discussed above and will develop below, it is found that the topology change is significant for developing the pseudo-conformal model (PCM) of dense nuclear matter, especially for the existence of the conformal sound velocity in compact star matter. Because of this topology change, there is a cusp structure in the symmetry energy $E_{sym}$~\cite{Ma:2018jze,Liu:2018wgv} which provides a simple mechanism for the putative soft-to-hard change in the EoS for compact stars at $n\sim 2n_0$ needed to account for the observed massive $\sim 2 M_\odot$. In the models that resort to hadron-quark continuity in terms of specific quark degrees of freedom that are strongly coupled, the hardening of the EoS at $n\gsim 2n_0$ is associated with ``deconfinement" of quarks~\cite{Baym:2017whm,McLerran:2018hbz}. The question is whether or how the topology change represents the ``quark deconfinement" process. Here, we give a conjecture on this issue.

We first consider one-flavor QCD in which case, as we discussed above, the topological baryon an be interpreted as the fractional Quantum Hall (FQH) droplet~\cite{Komargodski:2018odf}. In this case, the connection between the topology change and the quark deconfinement can be made by using the Cheshire Cat mechanism~\cite{Ma:2019xtx}.

\begin{figure}[h]
%\begin{center}
\includegraphics[width=7.0cm]{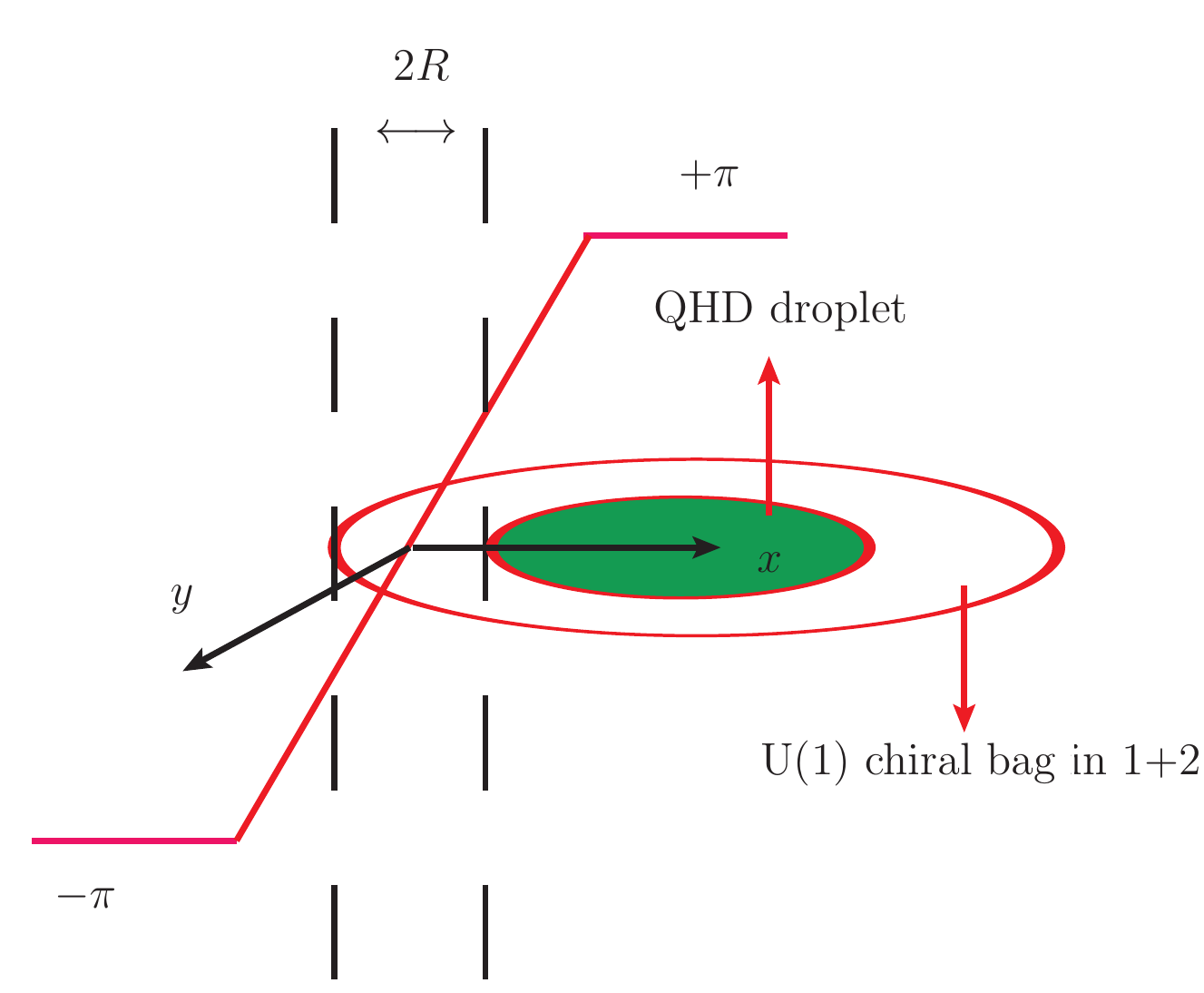}
\caption{$(1 + 2)$-dimensional chiral bag surrounding a QH droplet. ~\cite{Ma:2019xtx}.}\label{fig:droplet}
% \end{center}
\end{figure}

Explicitly, considering a $(1 + 2)$-dimensional chiral bag surrounding a QH droplet as shown in Fig.~\ref{fig:droplet}. The bag is an annulus of width $2R$ clouded by an $\eta^\prime$ with a monodromy of $2\pi$. The bag is filled in by $N_c$ quarks. In the limit of the zero bag radius, the chiral bag reduces to a vortex string with unit baryon number---the simile is left.

It is shown that a current transverse to the smile embodying the Callan-Harvey anomaly outflow~\cite{Callan:1984sa} appears. This transverse current is analogous to the Hall current of the QH effect through the emergence of an effective $U(1)$ gauge field. This $U(1)$ gauge field lives in the disk enclosed by the Cheshire Cat smile and is described by a purely topological field theory in $1 + 2$ dimensions and the emergent action of the emergent $U(1)$ gauge field is of FQH droplet. The quantum numbers of this baryon as a QH droplet follow readily from the chiral bag construction. This argument can be extended to the case where $N_f=2,3$ that we are concerned with in this review.

Now suppose the $\eta^\prime$ becomes light as is expected at high density. Then the FQH pancakes could become relevant as density increases and figure in dense matter in a form of a stack of FQH pancakes. Interactions must then induce the $N_c$ quarks with the fractional (1/$N_c$) baryon charge living on the boundary of the pancakes could tunnel between the pancakes. This could lead to sheets of fractional baryon-charged topological objects in (3+1) dimensions. In fact in recent analysis of dense matter using skyrmion crystal approach, one finds certain configurations unstable at low density but stabilized at high density of sheets with half-baryon charged objects called ``lasagnes"~\cite{Park:2019bmi} and  also  with $1/q$-charged baryons in tube configurations with baryons living on the surface of the tube~\cite{Canfora:2018rdz}. In addition, it is recently found that this QHD sheet exists in the string theory description of single-flavor QCD~\cite{Bigazzi:2022luo}. Anyway, it seems not impossible that the layers of FQH droplets in (3+1) dimensions give rise to deconfined quasiparticles dual to quarks of fractional charges, e.g, half-skyrmions and this FQH droplet may be explored in superdense compact star matter~\cite{Rho:2022meo}. Such deconfinement can take place in the presence of domain walls as in some condensed matter systems~\cite{Sulejmanpasic:2016uwq} and the half-skyrmions probed in the density regime $n> n_{1/2}$ could be deconfined as in the N\'eel-VBS deconfined quantum critical transition~\cite{Senthil:2003eed,Sulejmanpasic:2016uwq}.

\section{Generalized nuclear effective field theory}

Equipped by the discussion of the hidden symmetries, following the procedure discussed above, one can write down the generalized nuclear effective field theory (G$n$EFT). Here, for simplicity, we only consider the leading order scale symmetry (LOSS). In this limit, the effective Lagrangian is expressed as
\be
{\cal L}_{GnEFT} & = & {\cal L}_{GnEFT}^{M} + {\cal L}_{GnEFT}^{B} - V(\chi)
\label{eq:GnEFT}
\ee
where
\be
{\cal L}_{GnEFT}^{M} & = & f_\pi^2 \left(\frac{\chi}{f_\sigma}\right)^2{\rm Tr}\left[\hat{\alpha}_{\perp \mu}\hat{\alpha}_{\perp}^{\mu}\right] \nonumber\\
& &{} + af_\pi^2 \left(\frac{\chi}{f_\sigma}\right)^2{\rm Tr}\left[\hat{\alpha}_{\parallel \mu}\hat{\alpha}_{\parallel}^{\mu}\right] + \frac{1}{2}\partial_\mu \chi \partial^\mu \chi + \cdots,\nonumber\\
{\cal L}_{GnEFT}^{B} & = & \bar{N}i\gamma_\mu D^\mu N - \frac{\chi}{f_\sigma}m_0\bar{N}N \nonumber\\
& & {} + g_A \bar{N}\gamma_\mu \gamma_5 \hat{\alpha}_{\perp}^{\mu} N + g_V \bar{N}\gamma_\mu \hat{\alpha}_{\parallel}^{\mu} N ,\nonumber\\
V(\chi) & = & h_5 \left(\frac{\chi}{f_\sigma}\right)^4 + h_6 \left(\frac{\chi}{f_\sigma}\right)^{4+\beta^\prime},
\ee
with $N$ being the iso-doublet of the baryon fields. Using the saddle-point equations and in terms of the dilaton mass $m_\sigma$, the dilaton potential is reexpressed as
\begin{eqnarray}
	V(\chi) & = & {} \frac{m_\sigma^2f_\chi^2}{\beta^\prime(4+\beta^\prime)}\left( \frac{\chi}{f_\chi}\right)^4\left[ \left( \frac{\chi}{f_\chi}\right)^{\beta^\prime} - \left(1 + \frac{\beta^\prime}{4}\right)\right].
\end{eqnarray}
Lagrangian~\eqref{eq:GnEFT} is the starting point of the PCM of the compact star matter.

\subsection{Generalized Brown-Rho Scaling}

From G$n$EFT, one can derive the generalized BR scaling~\cite{Brown:1991kk} which mimics the medium modified hadron properties at LOSS. The most general scaling including the corrects to LOSS can be found in Ref.~\cite{Li:2017hqe}. At LOSS, one finds
\be
\frac{f_\pi^\ast}{f_\pi} & = & \frac{m_V^\ast}{m_V} = \frac{m_N^\ast}{m_N} =\Phi(n), \quad \frac{m_\sigma^\ast}{m_\sigma} = \Phi(n)^{1+\beta^\prime},
\label{eq:NewBRS}
\ee
where $\Phi(n) = \langle \chi \rangle^\ast/\langle \chi \rangle$.

Note that, different from the BR scaling originally proposed in Ref.~\cite{Brown:1991kk}, here, the density scaling of the sigma mass depends on $\beta^\prime$. When $\beta^\prime \ll 1$, the dilaton potential reduces to the logarithm form~\cite{Li:2016uzn} and scaling of sigma mass becomes $m_\sigma^\ast/m_\sigma \to \Phi(n)$, the form suggested in Ref.~\cite{Brown:1991kk}.

So far, we do not have any priority to fix $\beta^\prime$ without ambiguity. What we learned is that when using chiral-scale EFT to dense skyrmion matter $1\lesssim|\beta^\prime|\lesssim 3.5$~\cite{Ma:2016nki,Shao:2022njr}. This magnitude is consistent with the phenomenological Lagrangian approach to nuclear matter where the six-point interaction of the sigma meson---roughly $\beta^\prime=2$ in the present framework---is found significant~\cite{Motohiro:2015taa}.

It should be noted that the scaling relation~\eqref{eq:NewBRS} is obtained from the LOSS. In the following explicit calculation, to fit the nuclear matter properties around saturation density $n_0$, we should fine tuning the scaling parameters which is attributed to the corrections to LOSS.

%\subsection{Dichotomy problem}

\subsection{Quenching of $g_A$ in nuclei transition}

Let us put the discussion of compact star matter aside for a moment. We show how the scale symmetry manifests in nuclei by looking at the $g_A$ quench problem in the nuclear Gamow-Teller transitions, that is, the axial coupling constant $g_A^{\rm free} = 1.276$ in the vacuum should be taken as $g_A^{\rm eff} \to 1 $ in the nuclear Gamow-Teller transitions~\cite{Wilkinson:1973zz,Suhonen:2017krv,Engel:2016xgb}. Some results are reviewed in Ref.~\cite{Rho:2022vju} recently.

In G$n$EFT, the axial current relevant to the nuclear Gamow-Teller transitions is expressed as
\be
q_{\rm SSB}g_A\bar{\psi}\tau^+\gamma_\mu\gamma_5 \psi
\label{eq:WeakCurrent}
\ee
where
\be
q_{\rm SSB} & = & c_A + (1-c_A)\Phi^{\beta^\prime}.
\ee
In the LOSS, $c_A=1$. So that, the value $c_A <1$ accounts for the breaking of LOSS and reflects the patterns of the scale symmetry in nuclei system. Using \eqref{eq:WeakCurrent}, the quench factor is finally expressed as~\cite{Ma:2020tsj}
\be
q_{\rm GnEFT}^{\rm ESPM} & = & q_{\rm SSB}\times q_{\rm SNC},
\label{eq:qGnEFT}
\ee
where $q_{\rm SNC}$ accounts for strong nuclear many-body correlations. By using the Fermi-liquid fixed point theory~\cite{Friman:1996qc}, one can work out $q_{\rm SNC}$~\cite{Li:2017udr}. With the value $\Phi(n_0)\simeq 0.8$~\cite{Kienle:2004hq}, one obtains~\cite{Ma:2020tsj}
\be
q_{\rm SNC}\simeq 0.79
\label{eq:qSNC}
\ee

In nuclei upto $A\sim 60$, $g_A^{\rm eff}$ in shell model comes out to be ~\cite{Suhonen:2017krv,Engel:2016xgb}
\be
g_A^{\rm eff} =q_{\rm light}g_A^{\rm free} = 0.98-1.18
\ee
with $g_A^{\rm free} = 1.276$. In the range $q_{\rm light}=0.76 - 0.93$ implied by this equation, let us pick what gives $g_A^{\rm eff} \simeq 1$
\be
q_{\rm light} \simeq 0.78.
\ee
With respect to \eqref{eq:qSNC}, one concludes that $q_{\rm SSB}\simeq1$ in Eq.~\eqref{eq:qSNC}. This indicates that $c_A =1$, LOSS work well for the light nuclei system.

However, as the mass number of nuclei goes up above $A \sim 60$, the scenario is different. A more stringent recent experiment from RIKEN on the superallowed GT decay of the doubly magic nucleus $^{100}$Sn~\cite{Lubos:2019nik} yields~\cite{Ma:2020tsj}
\be
q_{\rm RIKEN}^{\rm ESPN} = 0.46-0.55.
\ee
This means that in the heavy nuclei system, only~\eqref{eq:qSNC} is not enough to account for the quench factor. This discrepancy can be interpreted by assuming $c_A \simeq 0.15$ and $\beta^\prime \simeq 2.5$---the same values that resolve the HWZ problem in Refs.~\cite{Ma:2016nki,Shao:2022njr}. This choice gives
\be
q_{\rm SSB} = 0.64
\ee
which leads to
\be
q_{GnEFT}^{\rm ESPM} = q_{\rm SSB} \times q_{\rm SNC} = 0.64 \times 0.79  \simeq 0.5
\ee
that consists with the RIKEN data well.

What we learned from this analysis can be summarized as follows: At very low energy and density, at the unitarity limit (in the framework of pionless EFT), conformal symmetry emerges in light nuclei and in
the EOS of baryonic matter~\cite{vanKolck:2019qea,Tews:2016jhi}. At the normal nuclear matter density, on the contrary, such symmetry is evidently absent, but at high density approaching the dilaton-limit
fixed point (DLFP), as we will see later, the symmetry reappears. This tells us how the scale symmetry manifests in nuclear system.

\section{Equation of state of nuclear matter}

Let us come back to the nuclear matter properties by using the G$n$EFT with generalized BR scaling. Since the G$n$EFT includes, in addition to the Nambu-Goldstone bosons pions, the effects from the hadron resonances $\sigma, \rho$ and $\omega$, the obtained equation of state is expected to applicable to the core of massive stars, i.e., $\sim 10 n_0$. To take the meson fluctuation effects into account, we use the $V_{low k}$ approach~\cite{Bogner:2003wn}. Therefore, the density effect come from both the intrinsic density dependence inherits from the BR scaling and hadron correlations.

%In the explicit calculation, we implicate the medium modified hadron properties to the G$n$EFT constructed above by using the Brown-Rho scaling~\cite{Brown:1991kk}.
Considering that the density dependence of the medium modified hadron properties are categorized into two regions due to the topology change delimited by density $n_{1/2}$, we denote the region $n<n_{1/2}$ as R-I and region $n>n_{1/2}$ as R-II. The density scaling of the medium modified hadron properties are summarized as follows:
\begin{itemize}
  \item R-I: In this region, the scaling function $\Phi$ in the master formulism \eqref{eq:NewBRS} decreases with density. Without first principle information on the explicit form of $\Phi$, we parameterize it as
      \be
      \Phi_I=\frac{1}{1+c_I\frac{n}{n_0}}
      \ee
      with $c_I$ being a constant. With respect to the nuclear matter properties~\cite{Ma:2021nuf} and the measured pion decay constant~\cite{Kienle:2004hq}, the range of $c_I$ is found to be
      \be
      c_I\approx 0.13-0.20 . \label{range}
      \ee
      In practice, to reproduce the nuclear matter properties around saturation density, it is easy to imagine that there should be fine-tuning within the range~\eqref{range}.
      %This reflects the fine-tuning nature required for ground-state properties of nuclear matter, be that EFT or phenomenological.

  \item R-II: Due to the topology change at $n_{1/2} \gsim 2n_0$, the scaling behaviours of some parameters in R-II are drastically different from that in R-I. The existence of such a topology change is one of the most robust inputs from skyrmion matter. The scaling behaviours of the parameters are quite involved.
      \begin{itemize}
        \item $g_\rho$ and $\rho$ mass: The hidden local gauge coupling $g_\rho$ related to the $\rho$ mass through the KSRF relation. Combined with the vector manifestation(VM) fixed-point structure of HLS leads to that for $n > n_{1/2}$ the coupling $g_\rho$ should drop to zero toward the putative VM fixed point $n_{VM}$. We take the simple form~\cite{Yang:2022}
            \be
            \frac{g^*_{\rho NN}}{{g_{\rho NN}}}=\begin{cases}
		1-0.1\frac{n}{n_0}, & for \quad n\in (n_{1/2}, 3.5 n_0) \\
		0.65-0.04\frac{(n-3.5n_0)}{n_0}, & for \quad n\in (3.5n_0, n_{VM})
				\end{cases}\nonumber\\
            \ee
            which gives $m_\rho^\ast/m_\rho = g^*_{\rho NN}/g_{\rho NN} \to 0$ at $n_{\rm VM} \approx 20n_0$. Where $n_{\rm VM}$ is located is not known in QCD. In compact stars, whether it is $\sim 6n_0$ or $\gsim 20 n_0$ does not make noticeable differences with one possible exception, namely, the star sound velocity as we will see below.
        \item Nucleon mass: As we learn from the 1/2-skyrmion phase that the parity doubling emerges giving rise to the chiral-invariant mass $m_0$ and the pion decay constant $f_\pi^\ast$ becomes density invariant. In the chiral-scale effective theory, they both locked to the dilaton condensate $f_\chi^\ast$.  Therefore we have
            \be
            \frac{m_N^\ast}{m_N}\approx \frac{f_\chi^\ast}{f_\chi}\approx \frac{f_\pi^\ast}{f_\pi} \equiv \kappa\sim (0.6-0.9).\label{kappa}
            \ee
        \item Dilaton mass: The dilaton mass is also proportional to the dilaton condensate which follows from the partially conserved dilatation current (PCDC)~\cite{Crewther:2013vea}, we then have
            \be
            \frac{m_\sigma^\ast}{m_\sigma}\approx\kappa.
            \ee
      \item $\omega$ meson: The nuclear matter density dependences of the $\omega$ meson properties are subtle. Using the HLS, the $\omega$ mass is locked to the hidden gauge coupling constant. Since the $U(2)$ HLS which works well in R-I breaks in R-II~\cite{Paeng:2011hy,Paeng:2015noa}, some sort of fine-tuning is needed in the density-scaling of $\omega$ mass and hidden gauge coupling constant. We take it as
          \be
          \frac{m_\omega^\ast}{m_\omega}\approx \kappa\frac{g_\omega^\ast}{g_\omega} = \kappa\Phi_\omega(n).
          \ee
          In the numerical calculation, we take
          \be
          \Phi_\omega\equiv \frac{g_\omega^\ast}{g_\omega}\approx 1-d\frac{n-n_{1/2}}{n_0}
          \ee
          with $d\approx 0.05$.
      \end{itemize}

\end{itemize}

After the above discussions, one can make a numerical calculation of the equation of state of the nuclear matter once the vacuum values of the parameters are fixed. It is found that, with the only parameter $c_I$, all the nuclear matter properties $\leq 2n_0 \lesssim n_{1/2}$ can be well reproduced~\cite{Ma:2021nuf}.

The density $n_{1/2}$ where the topology change happens and how the R-I and R-II are delineated changes density dependence of the hadron properties drastically and therefore impact the EoS in a qualitative way. However, as we discussed above, the location of the topology change is model dependent so that we cannot pin down its value theoretically. With respect to constraints from the various astrophysical observations so far available, the maximum mass of neutron star and the gravity-wave data, we constrain $n_{1/2}$ as $2.0n_0 < n_{1/2}<4.0n_0$~\cite{Ma:2018xjw}.

%\subsection{Equation of state before topology change}

\subsection{Vector manifestation}

%Based on the Wilsonian renormalization group (RG) approach, it is found that at high energy scale $f_\pi \to 0, m_\rho \to m_\pi \to 0$, i.e., there is a vector manifestation (VM) fixed point in the hidden local symmetry (HLS)~\cite{Harada:2000kb,Harada:2003jx}. By the KSRF relation, at VM, the HLS gauge coupling $g_\rho$ goes to zero therefore the $\rho$ decouples from the theory. It is reasonable to expect that the VM appears at (super-)high density denoted as $n_{\rm vm}$ if we regard density as energy scale.

Where the vector manifestation fixed point  $n_{\rm vm}$ is located is known neither theoretically nor empirically. While most of the global properties of compact stars do not seem to depend much on where $n_{\rm vm}$ lies since its value is above the possible central density of massive compact stars, it seems that it affects the sound velocity of compact star matter in an indirect way.

\begin{figure}[h]%\centering
\includegraphics[width=7.0cm]{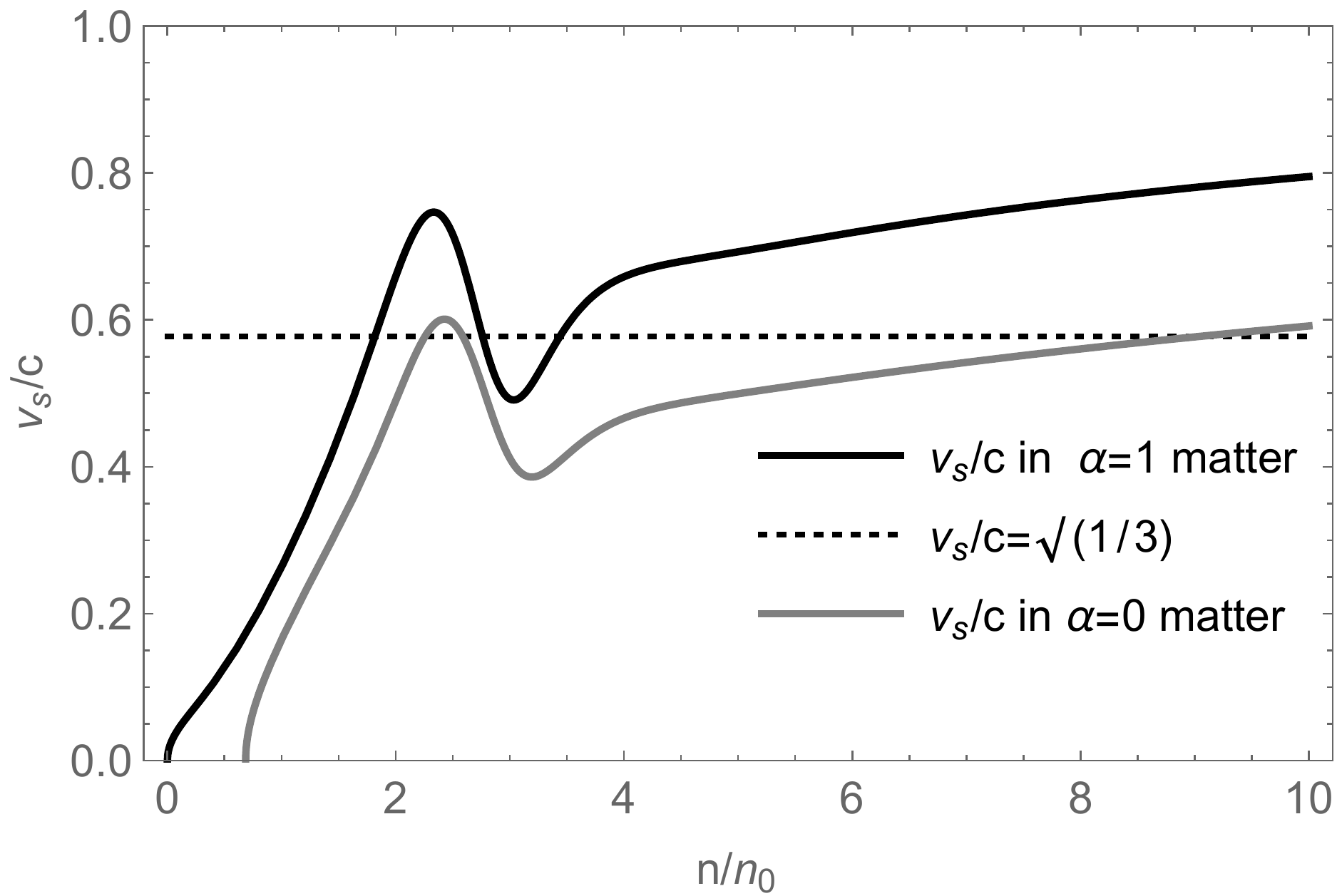}
\includegraphics[width=7.0cm]{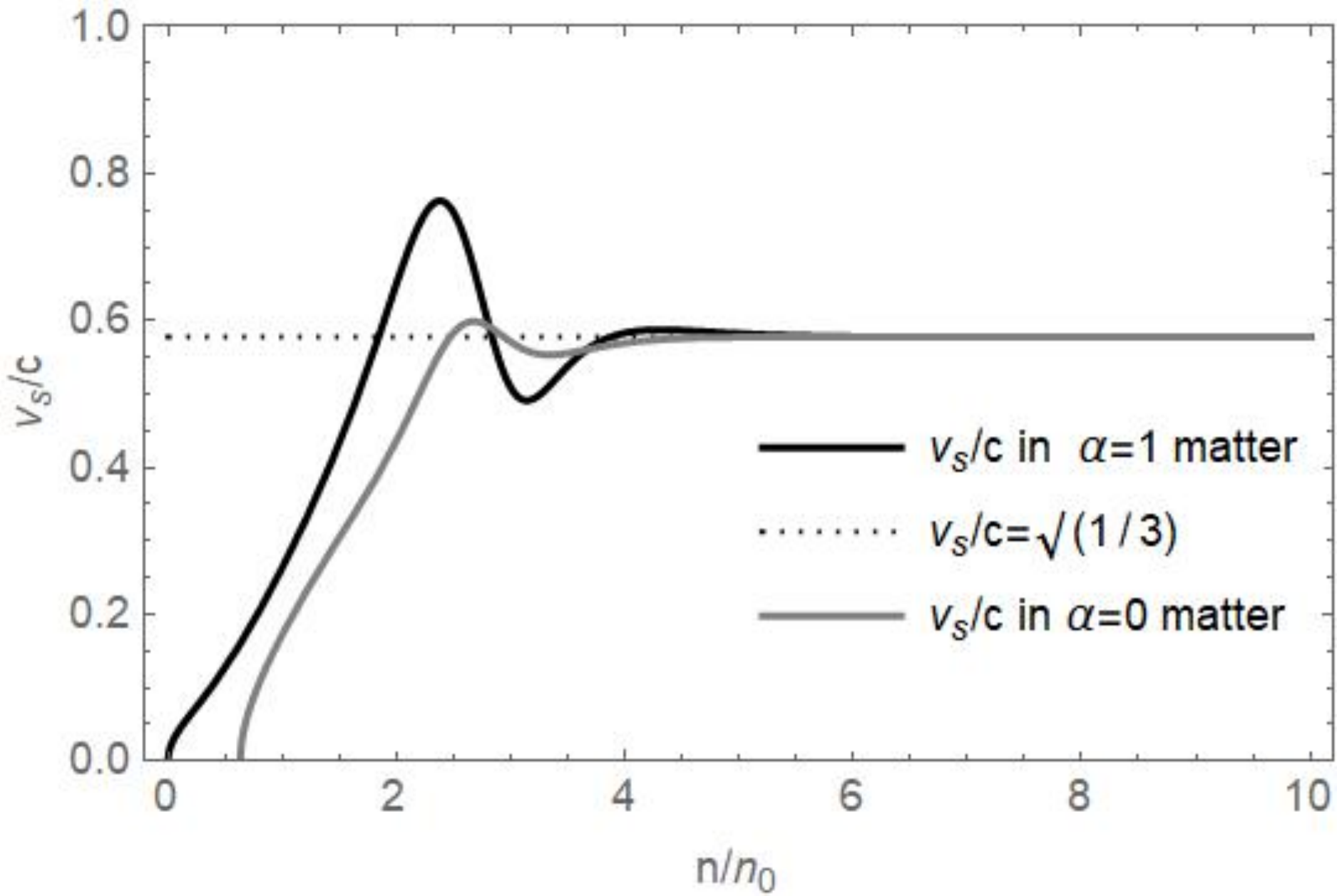}
\caption{Sound velocity for $n_{\rm vm}=6.75n_0$ (upper panel) and $20n_0$ (lower panel), both computed in $V_{lowk}$ RG with $n_{1/2}=2.5n_0$~\cite{Yang:2022}.
}
\label{Vs}
\end{figure}

Here, to show the effect of $n_{\rm vm}$ on the sound velocity, we fix the typical value $n_{1/2} = 2.5 n_0$. We choose $n_{\rm VM}= 6.75 n_0$ and $ 20 n_0 $.  The lower value of the density is about the central density of massive stars,  and the upper value represents an ``asymptotic density" where perturbative QCD is expected to be applicable. The $n_{\rm VM}$ dependences of the sound velocity are plotted in Fig.~\ref{Vs}.

From Fig.~\ref{Vs} one can easily see that the location of $n_{\rm vm}$ drastically affects the behaviour of the sound velocity. When $n_{\rm vm}$ is big, the sound velocity converges to the ``conformal velocity" $v_s^2\approx 1/3$ after $n_{1/2}$. However, for a smaller $n_{\rm vm}$, e.g., $n_{\rm vm} \approx 7n_0$, it increases steadily after $n_{1/2}$ and overshoots conformal velocity. We will see later that the conformal sound velocity is locked to the (pseudo-)conformality of the matter.

\subsection{Pseudoconformal structure}

Now, let us understand what is the implication of the conformal velocity in nuclear matter.

In the matter system, the sound velocity is defined by
\be
v_s^2=\frac{\partial P(n)}{\partial n}/\frac{\partial\epsilon}{\partial n},
\ee
where $\epsilon$ and $P$ are, respectively, the energy density and the pressure density. We then have
\be
\int \frac{\partial P(n)}{\partial n}dn = \int v_s^2\frac{\partial\epsilon}{\partial n} dn - \frac{1}{3}C_0
\ee
with $C_0$ being a constant independent of density. For a constant sound velocity, one obtains
\be
P(n) & = & v_s^2 \epsilon(n) - \frac{1}{3}C_0.
\ee
And for $v_s^2/c^2 = 1/3$, we obtain
\be
\epsilon(n) - 3 P(n) & = & C_0.
\ee
For an ideal liquid system where the nuclear matter is assumed to work, the trace of the energy-momentum tensor (TEMT) is expressed as
\be
\theta_\mu^\mu = \epsilon - 3 P.
\ee
Therefore, when $v_s^2/c^2 = 1/3$, one has
\be
\langle\theta_\mu^\mu\rangle = C_0,
\ee
which is a density independent quantity. When $C_0=0$, the TEMT vanishes so that the scale symmetry in dense matter is restored. The system with this property can be regarded as that where both the scale symmetry and chiral symmetry are restored, quarks are deconfined and perturbative QCD applies. Since this scenario happens at the density much beyond that in the core of massive stars, we are not interested in it.

\begin{figure}[h]
\begin{center}
\includegraphics[width=7.0cm]{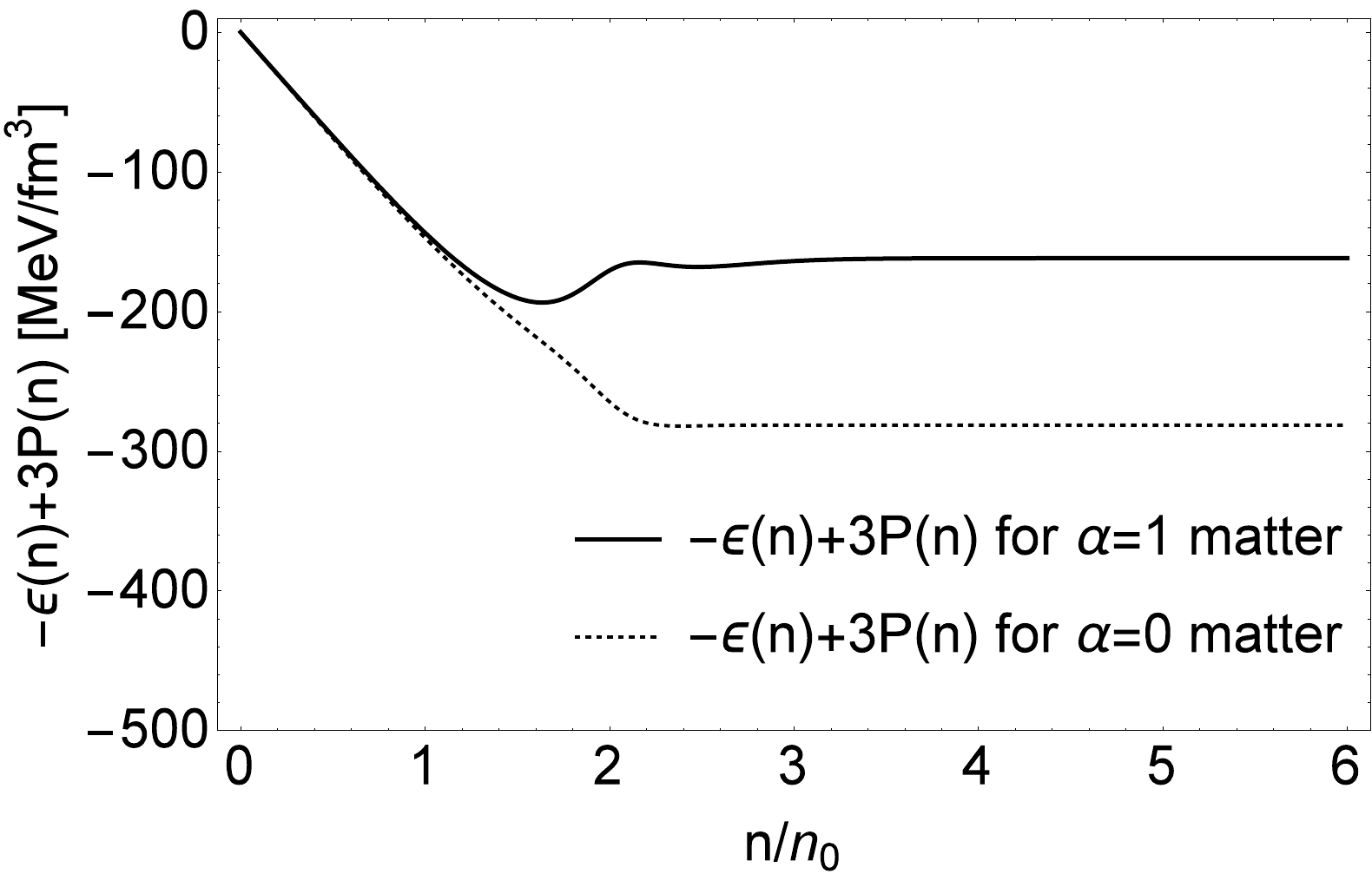}\;\;
\includegraphics[width=7.0cm]{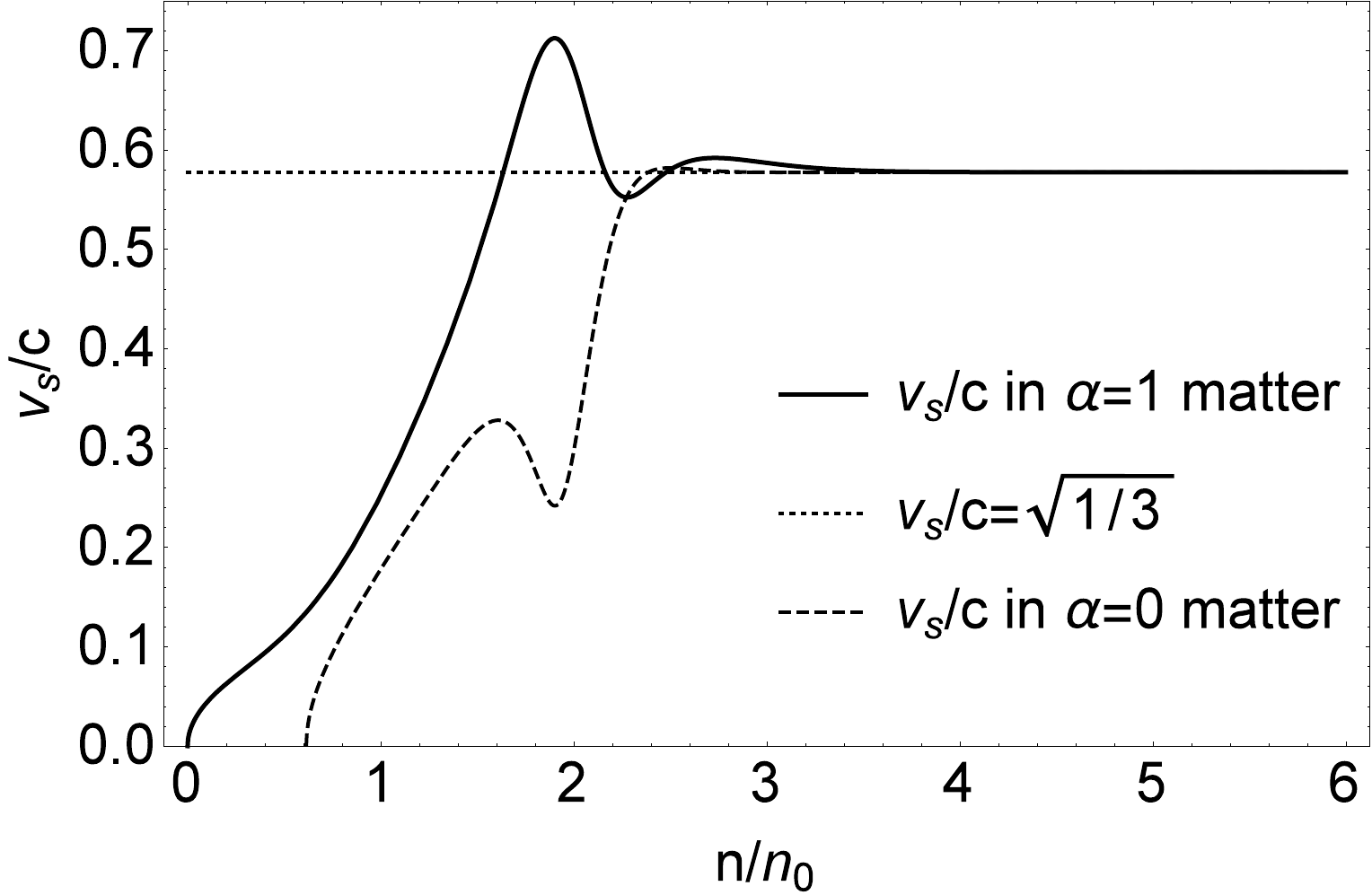}
\caption{$\la\theta_\mu^\mu\ra$ (upper panel) and $v_s$ (lower panel) vs. density for $\alpha=0$ (nuclear matter) and $\alpha=1$ (neutron matter) in $V_{lowk}$ RG for $n_{1/2}=2 n_0$ and $v_{\rm vm}=25 n_0$.
 }\label{TEMT}
 \end{center}
\end{figure}

Let us focus on the scenario $C_0 \neq 0$, that is, the TEMT is a density independent quantity. This scenario does happen in the chiral-scale EFT approach to nuclear matter. In the mean field approach it is shown that, going toward the DLFP, the TEMT $\la\theta_\mu^\mu\ra$ is a function of only the dilaton condensate~\cite{Paeng:2011hy}. Now if the condensate goes to a constant $\sim m_0$ due to the emergence of parity-doubling which is found in the skyrmion crystal approach to nuclear matter~\cite{Ma:2013ooa}, then the $\la\theta_\mu^\mu\ra$ will be independent of density. This chain of reasoning is confirmed in the full $V_{lowk}$ RG formalism specifically for the case of $n_{1/2} = 2 n_0$. In Fig.~\ref{TEMT} is shown the TEMT (left panel) that gives the conformal velocity for $n\gsim 3n_0$ (right panel).

One can easily see that, the scenario $\langle \theta_\mu^\mu \rangle = C_0 \neq 0$ means that, the sound velocity saturated the conformal limit but the conformal symmetry is not restored. We call this matter as pseudoconformal matter in which the nucleon has an effective constant mass and pion decay constant is not zero.

\begin{figure}[htbp]
\includegraphics[width=7.0cm]{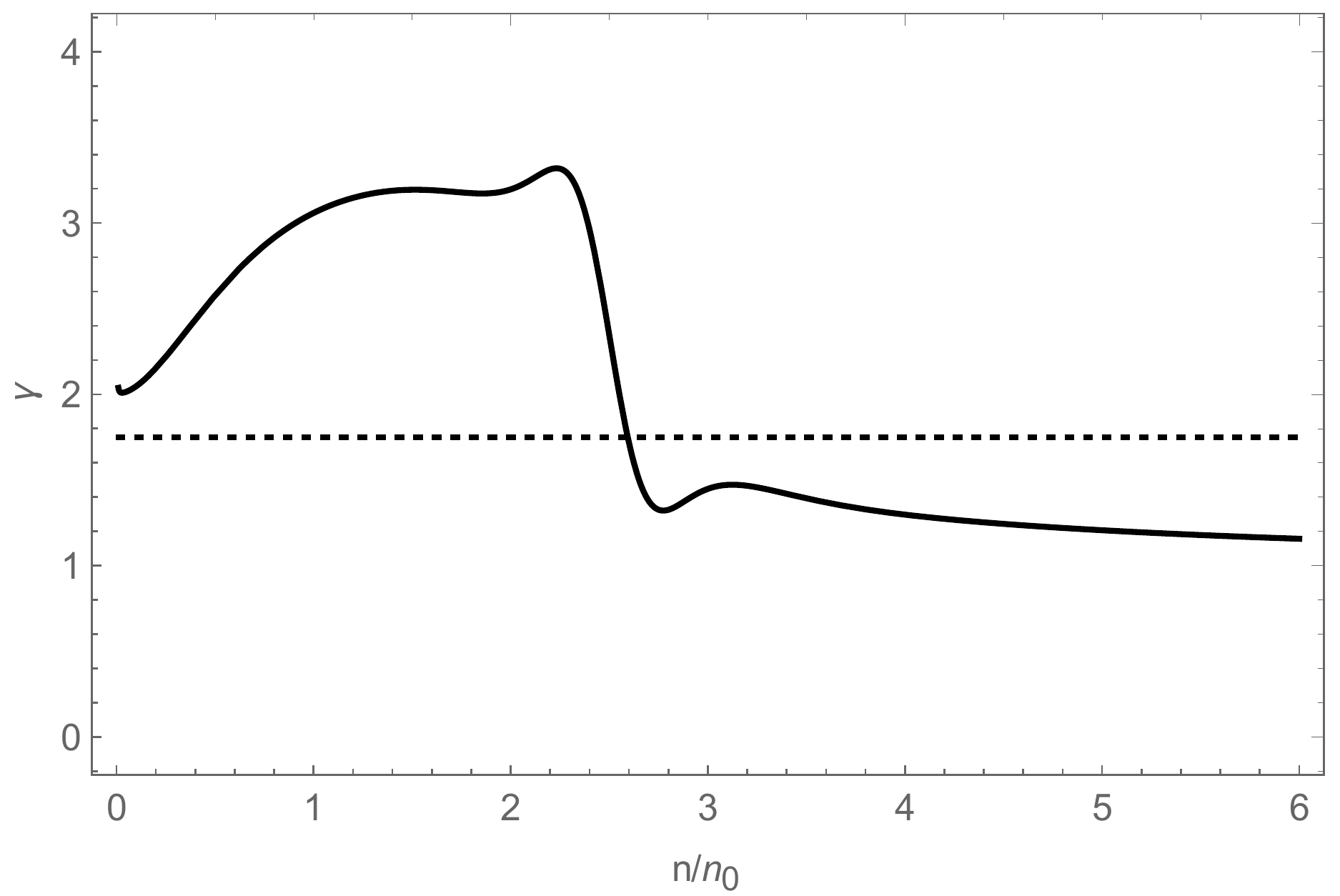}
\caption{Density dependence of the polytropic index $\gamma= {d\ln P}/{d\ln \epsilon}$ in neutron matter from the pseudo-conformal model.}
\label{fig:gamma}
\end{figure}

A recent analysis combining astrophysical observations and model independent theoretical {\it ab initio} calculations~\cite{Annala:2019puf} shows that in the core of massive stars the sound velocity approaches the comformal limit $v_s^2/c^2 \to 1/3$ and the polytropic index takes the value $\gamma < 1.75$ --- the value close to the minimal one obtained in hadronic models. Therefore the core of the massive stars is populated by  ``deconfined" quarks. An explicit calculation shows that the polytropic index $\gamma < 1.75$ in the PCM (see Fig.~\ref{fig:gamma}) but we are still in the confined phased. Therefore the smallness of the polytropic index and conformal velocity cannot be regarded as sufficient criterions for the appearance of the deconfined quark.

\begin{figure}[htbp]
%\begin{center}
%\vskip 0.3cm
\includegraphics[width=7.0cm]{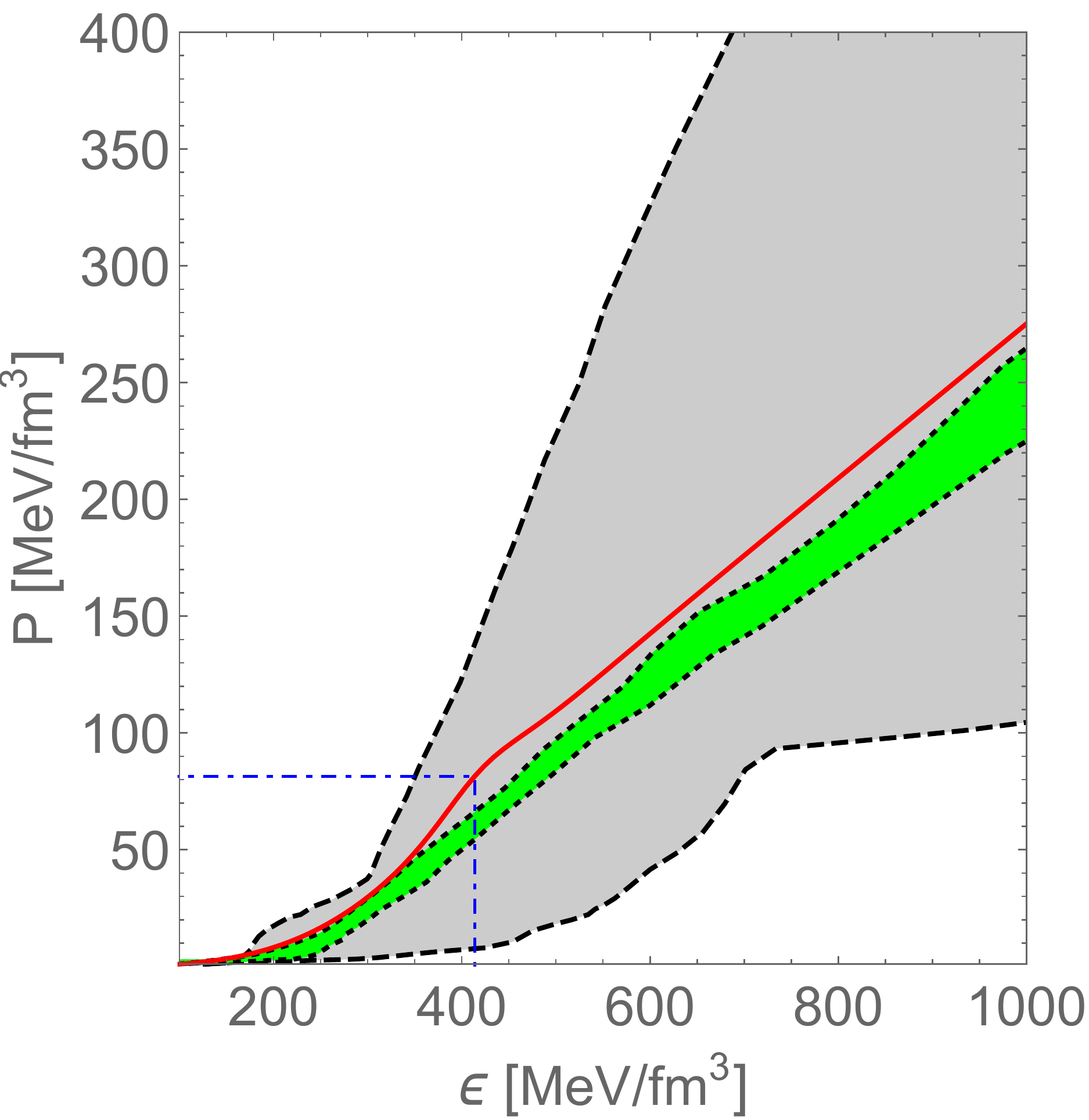}
%\end{center}
\caption{Comparison of $(P/\epsilon)$  between the PCM  velocity and the band generated with the
sound velocity interpolation method used in~\cite{Annala:2019puf}. The gray band is  from the causality and the green band from the conformality. The red line is the PCM prediction. The dash-dotted line indicates the location of the topology change.}
\label{fig:EoS}
\end{figure}

Finally we compare in Fig.~\ref{fig:EoS} our prediction for $P/\epsilon$ with the conformality band obtained by the sound velocity interpolation method ~\cite{Annala:2019puf}. We see that our prediction is close to, and parallel with, the conformality band, but most significantly, it lies above this band. The parallelism and location of our prediction come from the fact that in PCM the trace of the energy-momentum tensor is a position constant. The predicted results of GnEFT as a whole resemble  the ``deconfined" quark structure of \cite{Annala:2019puf}. There are, however, basic differences between the two. First of all, in our theory, conformality is broken, though perhaps only slightly at high density, in the system. Most important of all, the confined half-skyrmion fermion in the half-skyrmion phase is not deconfined. It is a quasiparticle of fractional baryon charge, neither purely baryonic nor purely quarkonic. In fact it can be anyonic lying on a (2+1) dimensional sheet~\cite{Ma:2020nih}.

\subsection{Equation of state}

We next compute the equation of state of the pseudoconformal dense nuclear matter and compare it to the constraint from the astrophysical observation and gravitational wave detection. We also vary the last parameter in the model $n_{1/2}$ to see its effect on the EoS.

We should say that, whatever the topology change density $n_{1/2} \gsim 2n_0$ is, the properties of ordinary  nuclear matter are fixed as stated already. In addition, we assume and actually numerically checked that for $n_{1/2}\geq 2n_0$, slightly above that transition density, the sound velocity must be $v_s^2/c^2\approx 1/3$.

It turns out that the feature of the EoS at $n> n_{1/2}$ can be captured by a simple two-parameter formula for the energy per particle
\be
E/A= - m_N +X^\alpha  x^{1/3} + Y^\alpha x^{-1}
\label{PC-RII}
\ee
where $X$ and $Y$ are parameters to be fixed and $\alpha=(N-Z)/(N+Z)$ and $x\equiv n/n_0$. From \eqref{PC-RII}, one concludes that the sound velocity satisies
\be
\frac{v_s^2}{c^2}=\frac{1}{3},
\label{pc-sound}
\ee
independently of $X^\alpha$ and $Y^\alpha$.

What we refer to as the pseudo-conformal model for the EoS is then $E/A$ given by the union of that given by $V_{lowk}$ in R-I ($ n<n_{1/2}$) and that given by Eq.~(\ref{PC-RII}) in R-II ( $n\geq n_{1/2}$)  with the parameters $X^\alpha$ and $Y^\alpha$ fixed by the continuity at $n=n_{1/2}$ of the chemical potential and pressure
\be
\mu_I=\mu_{II},\ P_I=P_{II}\ \ {\rm at} \ \ n=n_{1/2}.
\label{matchingE}
\ee
This formulation is found to work very well for both $\alpha=0$ and $1$ in the entire range of densities appropriate for massive compact stars, say up to $n\sim (6 - 7)n_0$, for the case $n_{1/2}\gsim 2n_0$ where the full $V_{lowk}$RG calculation is available~\cite{Paeng:2017qvp}.
%We apply this PCM formalism for the cases where $n_{1/2}> 2n_0$.

\begin{figure}[!h]
 \begin{center}
   \includegraphics[width=7cm]{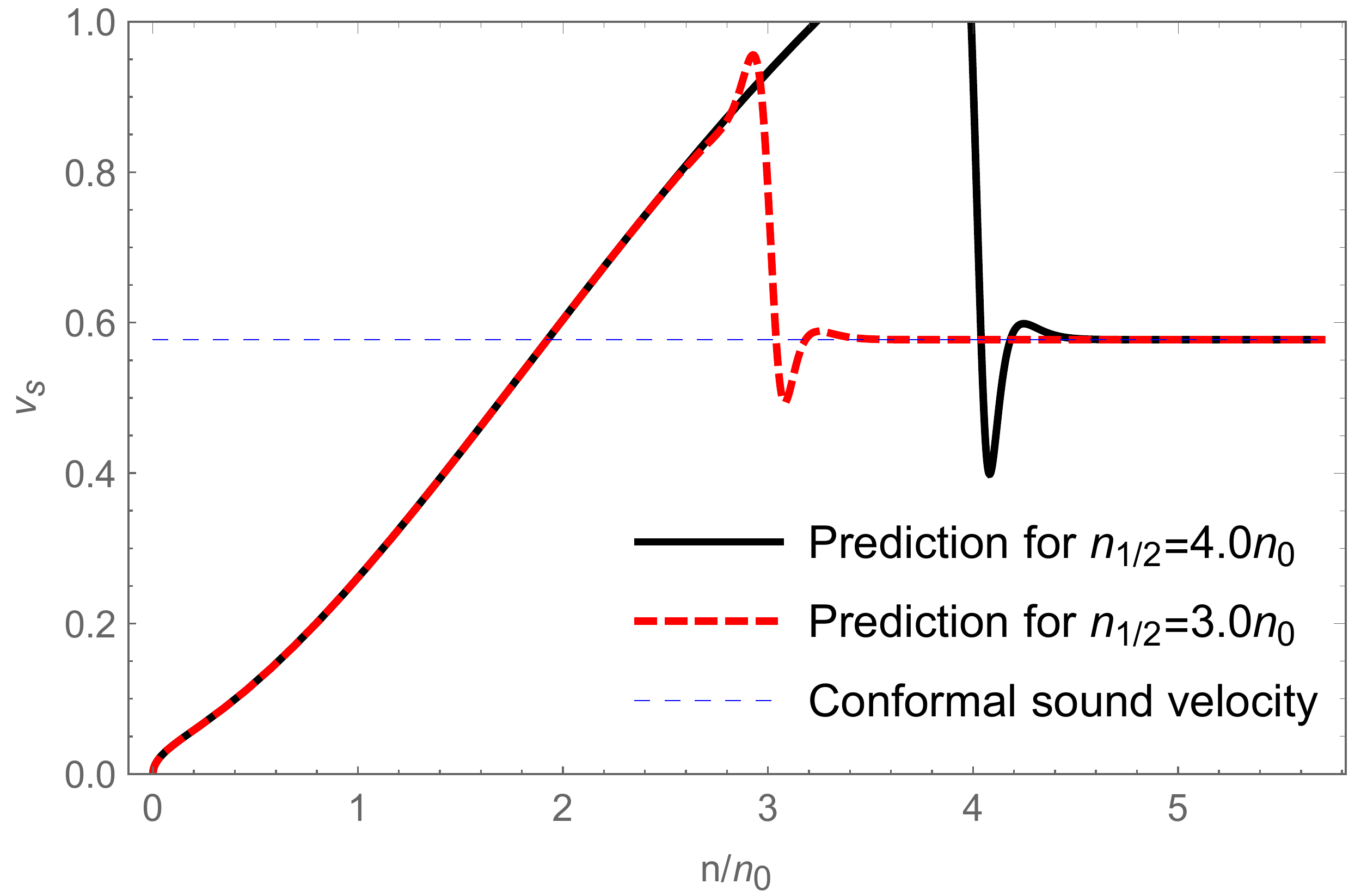}
  \end{center}
%  \vskip -1cm
 \caption{Sound velocity as a function of density in neutron matter with different $n_{1/2}$.
 }
\label{Vs2-4}
\end{figure}

We plot the sound velocity in Fig.~\ref{Vs2-4} by varying $n_{1/2}$. From this figure one can easily see that after $n_{1/2}$, the PC sound velocity $v_s^2/c^2=1/3$ emerges which indicates the emergence of pseudoconformal symmetry. It is clear from Fig.~\ref{Vs2-4} that the sound velocity for the case of $n_{1/2}=4 n_0$  violates the causality bound $v_s^2/c^2 < 1$.  The spike structure could very well be an artifact of the sharp connection made at the boundary. What is however physical is the rapid increase of the sound speed at the transition point signaling the changeover of the degrees of freedom~\cite{Hippert:2021gfs} and the derivative contribution from the trace anomaly~\cite{Fujimoto:2022ohj,Marczenko:2022jhl}. Significantly, this allows us to set the constraint for $n_{1/2}$
\be
2 n_0 \lsim  n_{1/2} \lsim  4 n_0.\label{boundhalf}
\ee
What is important of this constraint is that the emergence of the conformal sound velocity is an order of magnitude lower than the asymptotic density $\gsim 50 n_0$ perturbative QCD predicts. This is signals the precocious emergence of pseudo-conformality in compact stars. A recent detailed analysis of currently available data in the quarkyonic madel does confirm the onset density of $v_c^2\approx 1/3$ at $\sim 4 n_0$~\cite{Zhao:2020dvu,Kapusta:2021ney,Margueron:2021dtx}

 \begin{figure}[h]
 \begin{center}
   \includegraphics[width=7cm]{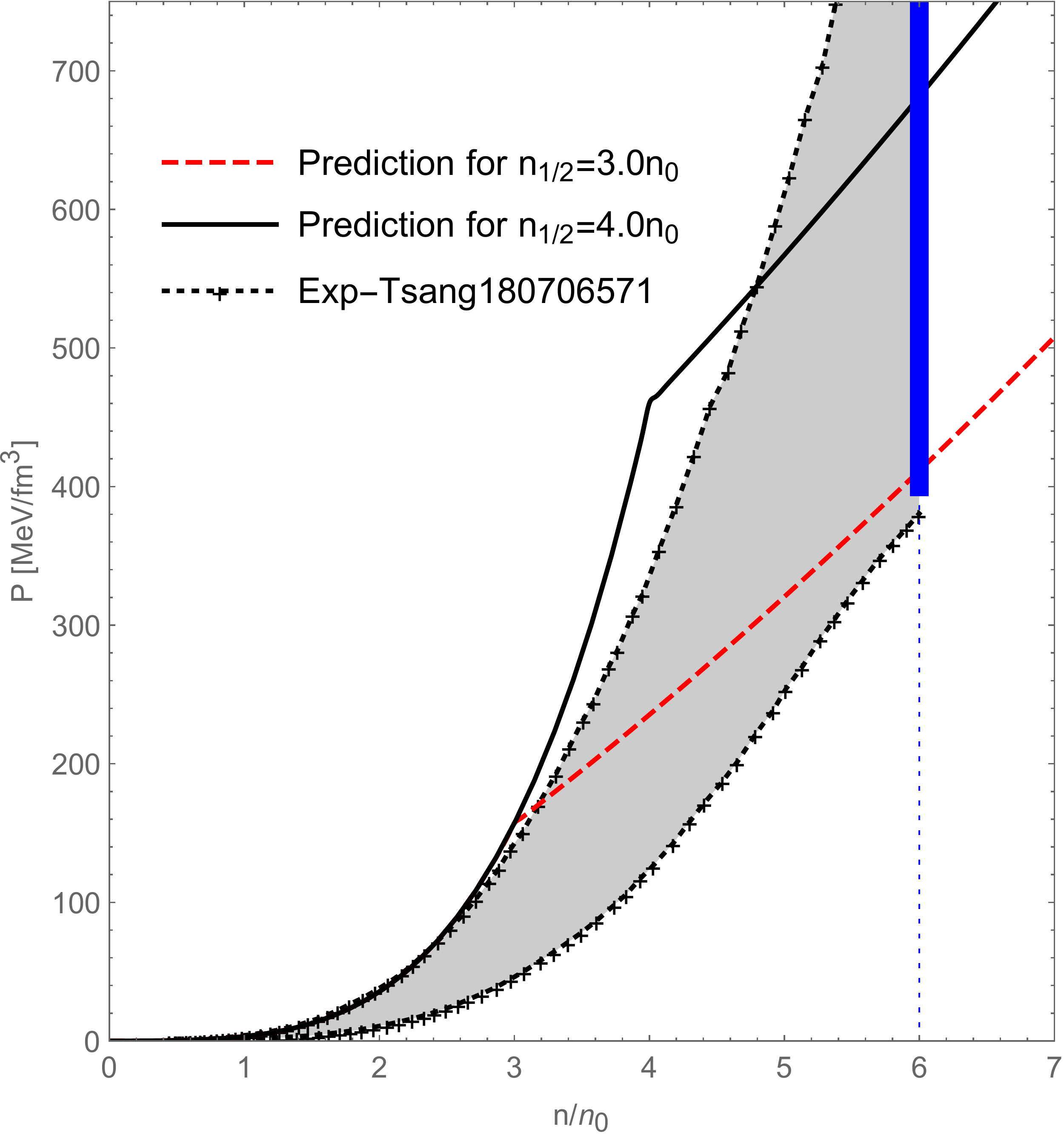}
  \end{center}
%  \vskip -0.6cm
 \caption{Predicted pressure for neutron matter ($\alpha=1$) vs density compared  with the available experimental bound (shaded) given by Ref.~\cite{Tsang:2018kqj} and the bound at $6n_0$ given by the blue band.}
\label{Tsang}
\end{figure}

Plotted below in Fig.~\ref{Tsang} is the predicted pressure $P$ for $n_{1/2}/n_0= 3,4$ compared with the presently available heavy-ion data~\cite{Tsang:2018kqj}.
The case of $n_{1/2}=4 n_0$, while consistent with the bound at $n\sim 6n_0$,  goes outside of the  presently available experimental bound at $n\sim 4n_0$. This may again be an artifact of the sharp matching, but that it violates the causality bound seems to put it in tension with Nature. Nonetheless it may  be too hasty to rule out the threshold density $n_{1/2}=4n_0$.

\section{Star properties and gravitational waves}

The final topic we want to discuss is to confront the PCM with the star properties and the gravitational wave detections for the purpose to show the rationality of the PCM.

\begin{figure}[h]
 \begin{center}
   \includegraphics[width=7.0cm]{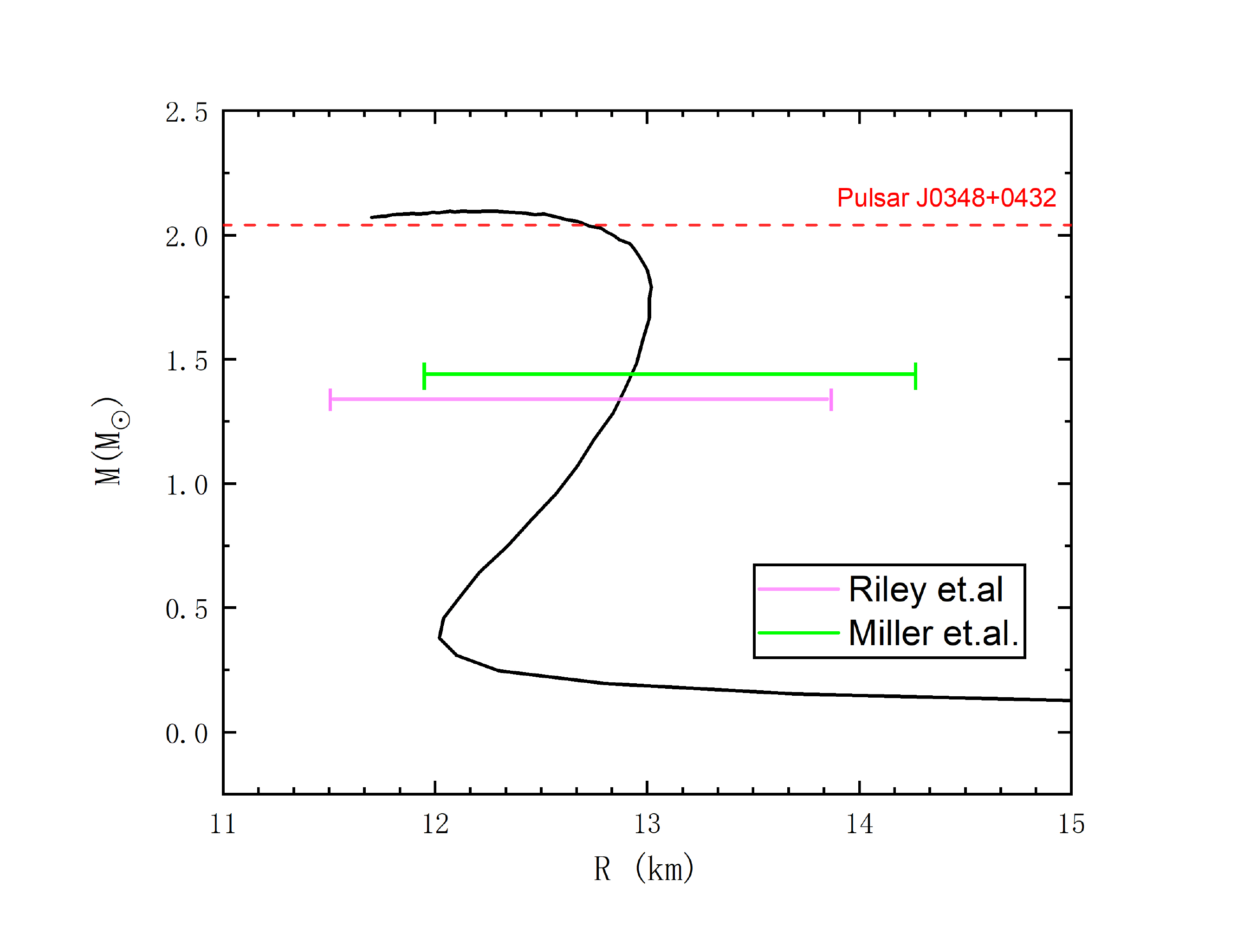}
 \end{center}
 \caption{M-R relation from PCM with observed mass of pulsar J0348+0432\cite{Antoniadis:2013pzd} and radius constraints\cite{Raaijmakers:2019qny}\cite{Miller:2019cac} from NICER.}
\label{star-mass}
%\end{center}
\end{figure}

The solution of the TOV equation with the pressures of leptons in beta equilibrium duly taken into account as in Ref.~\cite{Paeng:2017qvp} yields the results for the star mass $M$ vs. the radius $R$. It is found that the maximum mass comes out to be roughly  $2.04M_\odot \sim 2.23 M_\odot$ for $2.0 \leq n_{1/2}/n_0 \leq 4.0$ which is consistent with the present astrophysical observation~\cite{Demorest:2010bx,Antoniadis:2013pzd,NANOGrav:2019jur}. We plot in Fig.~\ref{star-mass} for the M-R relation with typical value $n_{1/2} = 2.5 n_0$. One can easily see that the present calculation is consistent with the observations.

Next, let us turn to how our theory fares with what came out of the LIGO/Virgo gravitational observations. The quantities that we will consider are the dimensionless tidal deformability $\Lambda_i$ for the star $M_i$ and $\tilde{\Lambda}$ defined  by
\begin{eqnarray}
\tilde{\Lambda} & = &  \frac{16}{13}\frac{(M_1 + 12 M_2)M_1^4 \Lambda_1 + (M_2 + 12 M_1)M_2^4 \Lambda_2}{(M_1 + M_2)^{5}}
\nonumber\\
\label{eq:tildeL}
\end{eqnarray}
for $M_1$ and $M_2$ constrained to the well-measured ``chirp mass"
\begin{eqnarray}
{\cal M} & = & \frac{(M_1 M_2)^{3/5}}{(M_1 + M_2)^{1/5}} = 1.188 M_\odot .
\label{eq:chirpmass}
\ee

To confront the LIGO/Virgo data, we plot our prediction for $\Lambda_1$ vs. $\Lambda_2$ in Fig.~\ref{L1vsL2}. As it stands, our prediction is compatible with the LIGO/Virgo constraint for $n_{1/2}\gsim 2n_0$. Although there seems to be some tension with the pressure, the result for $n_{1/2}=4 n_0$ is of quality comparable to that of $n_{1/2}=3 n_0$.

\begin{figure}[h]
 \begin{center}
   \includegraphics[width=6cm]{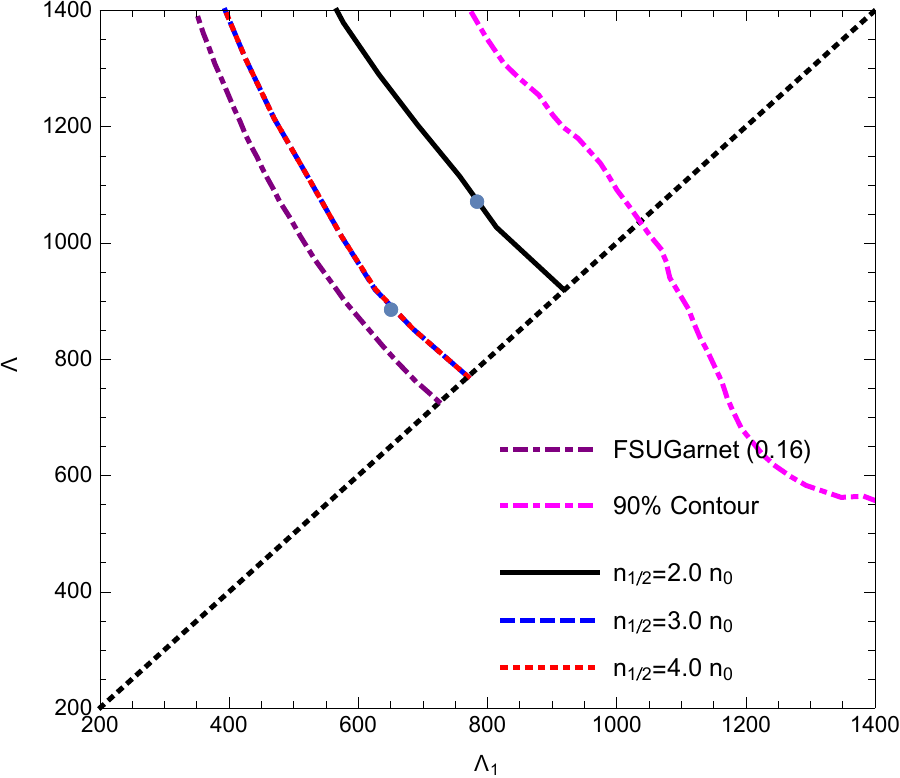}
  \end{center}
 \vskip -0.6cm
 \caption{Tidal deformabiliitiess $\Lambda_1$ and $\Lambda_2$ associated with the high-mass $M_1$ and low mass $M_2$ components of the binary neutron star system GW170817 with chirp
mass $1.188M_\odot$. The constraint from GW170817 at the 90\% probability contour is also indicated. We quote  ``FSUGarnet (0.16)"~\cite{Fattoyev:2017jql} as a presently available ``state-of-art" theoretical prediction.}
\label{L1vsL2}
\end{figure}

\section{Summary and perspective}

We reviewed in this work the possible emergent symmetries and topology change in dense compact star matter. The information of the medium modified hadron properties obtained from the skyrmion crystal approach, in addition to the presumed emergent scale and flavor symmetry, inspired the construction of the pseudoconformal model of dense nuclear matter relevant to compact stars.

In the pseudoconformal model, a peculiar feature that has not been observed by any other models previously is that in compact star matter the trace of the energy-momentum tensor is a nonzero density independent quantity and therefore, induce the precious appearance of the conformal limit $v_s^2/c^2 = 1/3$, in stark contrast to what was widely accepted in the society~\cite{Tews:2018kmu}. That is, there is a pseudoconformal symmetry in the compact star matter. The predictions of the pseudoconformal model are consistent with all the constraints from all the terrestrial experimental and astrophysical observations.

We finally devote ourselves to the possible extensions and revisions of the model.

The idea of the chiral-scale EFT \'a la Crewther and Tunstall which G$n$EFT is based on is anchored on three-flavor QCD. Therefore $f_0(500)$ can be taken as the same footing as the pseudoscalar mesons pion and kaon. However, in the present approach, we only included the up and down quarks and ignored the strange degrees of freedom for simplicity. So that it is interesting to extend the present framework to see the effect of strangeness on compact star matter~\cite{Weise:2019mou}. In addition, it is also interesting to extend the present approach to include the $\Delta$ baryon~\cite{Takeda:2017mrm,Marczenko:2021uaj}.

Another issue should be addressed in the future is to include the corrections to the LOSS applied so far such that, in addition to the mass parameters, the coupling constants also have IDD. This procedure may change the stiffness of the EOS and the tidal the deformability of the compact star. Meanwhile, the sound velocity after the topology change may deviate from the conformal limit because of the explicit breaking of the conformal limit. It should be noted that since the correction from the explicit breaking of the conformal limit is taken as a perturbative one, the global picture of the compact star discussed is intact.

Finally, it is interesting to pin down the density at which the hidden scale and local flavor symmetry emerge. This is encoded in the IDDs of the hadron parameters such as pion decay constant, dilaton decay constant, rho-N-N coupling and rho meson mass. By checking the effect of the location of the emergent symmetries on the star properties as well ths waveforms of the gravitational waves, one can also extract the information of the emergent symmetries and so phase structure of QCD at low temperature~\cite{Yang:2020ucv}.

%The last issue we want to discuss in this review is on the strangness in the present framework. The scale-chiral effective theory which the pseudoconformal relies on is based on three flavor QCD such that $f_0(500)$ can be taken as the same footing as the pseudoscalar mesons pion and kaon.

\acknowledgments

The work of Y.~L. M. was supported in part by National Science Foundation of China (NSFC) under Grant No. 11875147 and No. 12147103 and National Key R\&D Program of China (2021YFC2202900).

\bibliography{MDPIRef}

%merlin.mbs apsrev4-1.bst 2010-07-25 4.21a (PWD, AO, DPC) hacked
%Control: key (0)
%Control: author (8) initials jnrlst
%Control: editor formatted (1) identically to author
%Control: production of article title (-1) disabled
%Control: page (0) single
%Control: year (1) truncated
%Control: production of eprint (0) enabled
\begin{thebibliography}{120}%
\makeatletter
\providecommand \@ifxundefined [1]{%
 \@ifx{#1\undefined}
}%
\providecommand \@ifnum [1]{%
 \ifnum #1\expandafter \@firstoftwo
 \else \expandafter \@secondoftwo
 \fi
}%
\providecommand \@ifx [1]{%
 \ifx #1\expandafter \@firstoftwo
 \else \expandafter \@secondoftwo
 \fi
}%
\providecommand \natexlab [1]{#1}%
\providecommand \enquote  [1]{``#1''}%
\providecommand \bibnamefont  [1]{#1}%
\providecommand \bibfnamefont [1]{#1}%
\providecommand \citenamefont [1]{#1}%
\providecommand \href@noop [0]{\@secondoftwo}%
\providecommand \href [0]{\begingroup \@sanitize@url \@href}%
\providecommand \@href[1]{\@@startlink{#1}\@@href}%
\providecommand \@@href[1]{\endgroup#1\@@endlink}%
\providecommand \@sanitize@url [0]{\catcode `\\12\catcode `\$12\catcode
  `\&12\catcode `\#12\catcode `\^12\catcode `\_12\catcode `\%12\relax}%
\providecommand \@@startlink[1]{}%
\providecommand \@@endlink[0]{}%
\providecommand \url  [0]{\begingroup\@sanitize@url \@url }%
\providecommand \@url [1]{\endgroup\@href {#1}{\urlprefix }}%
\providecommand \urlprefix  [0]{URL }%
\providecommand \Eprint [0]{\href }%
\providecommand \doibase [0]{http://dx.doi.org/}%
\providecommand \selectlanguage [0]{\@gobble}%
\providecommand \bibinfo  [0]{\@secondoftwo}%
\providecommand \bibfield  [0]{\@secondoftwo}%
\providecommand \translation [1]{[#1]}%
\providecommand \BibitemOpen [0]{}%
\providecommand \bibitemStop [0]{}%
\providecommand \bibitemNoStop [0]{.\EOS\space}%
\providecommand \EOS [0]{\spacefactor3000\relax}%
\providecommand \BibitemShut  [1]{\csname bibitem#1\endcsname}%
\let\auto@bib@innerbib\@empty
%</preamble>
\bibitem [{\citenamefont {Brown}\ and\ \citenamefont
  {Rho}(2002)}]{Brown:2001nh}%
  \BibitemOpen
  \bibfield  {author} {\bibinfo {author} {\bibfnamefont {G.~E.}\ \bibnamefont
  {Brown}}\ and\ \bibinfo {author} {\bibfnamefont {M.}~\bibnamefont {Rho}},\
  }\href {\doibase 10.1016/S0370-1573(01)00084-9} {\bibfield  {journal}
  {\bibinfo  {journal} {Phys. Rept.}\ }\textbf {\bibinfo {volume} {363}},\
  \bibinfo {pages} {85} (\bibinfo {year} {2002})},\ \Eprint
  {http://arxiv.org/abs/hep-ph/0103102} {arXiv:hep-ph/0103102} \BibitemShut
  {NoStop}%
\bibitem [{\citenamefont {Holt}\ \emph {et~al.}(2016)\citenamefont {Holt},
  \citenamefont {Rho},\ and\ \citenamefont {Weise}}]{Holt:2014hma}%
  \BibitemOpen
  \bibfield  {author} {\bibinfo {author} {\bibfnamefont {J.~W.}\ \bibnamefont
  {Holt}}, \bibinfo {author} {\bibfnamefont {M.}~\bibnamefont {Rho}}, \ and\
  \bibinfo {author} {\bibfnamefont {W.}~\bibnamefont {Weise}},\ }\href
  {\doibase 10.1016/j.physrep.2015.10.011} {\bibfield  {journal} {\bibinfo
  {journal} {Phys. Rept.}\ }\textbf {\bibinfo {volume} {621}},\ \bibinfo
  {pages} {2} (\bibinfo {year} {2016})},\ \Eprint
  {http://arxiv.org/abs/1411.6681} {arXiv:1411.6681 [nucl-th]} \BibitemShut
  {NoStop}%
\bibitem [{\citenamefont {Drews}\ and\ \citenamefont
  {Weise}(2017)}]{Drews:2016wpi}%
  \BibitemOpen
  \bibfield  {author} {\bibinfo {author} {\bibfnamefont {M.}~\bibnamefont
  {Drews}}\ and\ \bibinfo {author} {\bibfnamefont {W.}~\bibnamefont {Weise}},\
  }\href {\doibase 10.1016/j.ppnp.2016.10.002} {\bibfield  {journal} {\bibinfo
  {journal} {Prog. Part. Nucl. Phys.}\ }\textbf {\bibinfo {volume} {93}},\
  \bibinfo {pages} {69} (\bibinfo {year} {2017})},\ \Eprint
  {http://arxiv.org/abs/1610.07568} {arXiv:1610.07568 [nucl-th]} \BibitemShut
  {NoStop}%
\bibitem [{\citenamefont {Baym}\ \emph {et~al.}(2018)\citenamefont {Baym},
  \citenamefont {Hatsuda}, \citenamefont {Kojo}, \citenamefont {Powell},
  \citenamefont {Song},\ and\ \citenamefont {Takatsuka}}]{Baym:2017whm}%
  \BibitemOpen
  \bibfield  {author} {\bibinfo {author} {\bibfnamefont {G.}~\bibnamefont
  {Baym}}, \bibinfo {author} {\bibfnamefont {T.}~\bibnamefont {Hatsuda}},
  \bibinfo {author} {\bibfnamefont {T.}~\bibnamefont {Kojo}}, \bibinfo {author}
  {\bibfnamefont {P.~D.}\ \bibnamefont {Powell}}, \bibinfo {author}
  {\bibfnamefont {Y.}~\bibnamefont {Song}}, \ and\ \bibinfo {author}
  {\bibfnamefont {T.}~\bibnamefont {Takatsuka}},\ }\href {\doibase
  10.1088/1361-6633/aaae14} {\bibfield  {journal} {\bibinfo  {journal} {Rept.
  Prog. Phys.}\ }\textbf {\bibinfo {volume} {81}},\ \bibinfo {pages} {056902}
  (\bibinfo {year} {2018})},\ \Eprint {http://arxiv.org/abs/1707.04966}
  {arXiv:1707.04966 [astro-ph.HE]} \BibitemShut {NoStop}%
\bibitem [{\citenamefont {Ma}\ and\ \citenamefont
  {Rho}(2020{\natexlab{a}})}]{Ma:2019ery}%
  \BibitemOpen
  \bibfield  {author} {\bibinfo {author} {\bibfnamefont {Y.-L.}\ \bibnamefont
  {Ma}}\ and\ \bibinfo {author} {\bibfnamefont {M.}~\bibnamefont {Rho}},\
  }\href {\doibase 10.1016/j.ppnp.2020.103791} {\bibfield  {journal} {\bibinfo
  {journal} {Prog. Part. Nucl. Phys.}\ }\textbf {\bibinfo {volume} {113}},\
  \bibinfo {pages} {103791} (\bibinfo {year} {2020}{\natexlab{a}})},\ \Eprint
  {http://arxiv.org/abs/1909.05889} {arXiv:1909.05889 [nucl-th]} \BibitemShut
  {NoStop}%
\bibitem [{\citenamefont {Li}\ \emph {et~al.}(2019)\citenamefont {Li},
  \citenamefont {Krastev}, \citenamefont {Wen},\ and\ \citenamefont
  {Zhang}}]{Li:2019xxz}%
  \BibitemOpen
  \bibfield  {author} {\bibinfo {author} {\bibfnamefont {B.-A.}\ \bibnamefont
  {Li}}, \bibinfo {author} {\bibfnamefont {P.~G.}\ \bibnamefont {Krastev}},
  \bibinfo {author} {\bibfnamefont {D.-H.}\ \bibnamefont {Wen}}, \ and\
  \bibinfo {author} {\bibfnamefont {N.-B.}\ \bibnamefont {Zhang}},\ }\href
  {\doibase 10.1140/epja/i2019-12780-8} {\bibfield  {journal} {\bibinfo
  {journal} {Eur. Phys. J. A}\ }\textbf {\bibinfo {volume} {55}},\ \bibinfo
  {pages} {117} (\bibinfo {year} {2019})},\ \Eprint
  {http://arxiv.org/abs/1905.13175} {arXiv:1905.13175 [nucl-th]} \BibitemShut
  {NoStop}%
\bibitem [{\citenamefont {Ma}\ and\ \citenamefont
  {Rho}(2021{\natexlab{a}})}]{Ma:2020nih}%
  \BibitemOpen
  \bibfield  {author} {\bibinfo {author} {\bibfnamefont {Y.-L.}\ \bibnamefont
  {Ma}}\ and\ \bibinfo {author} {\bibfnamefont {M.}~\bibnamefont {Rho}},\
  }\href {\doibase 10.3390/sym13101888} {\bibfield  {journal} {\bibinfo
  {journal} {Symmetry}\ }\textbf {\bibinfo {volume} {13}},\ \bibinfo {pages}
  {1888} (\bibinfo {year} {2021}{\natexlab{a}})},\ \Eprint
  {http://arxiv.org/abs/2009.09219} {arXiv:2009.09219 [nucl-th]} \BibitemShut
  {NoStop}%
\bibitem [{\citenamefont {Lovato}\ \emph {et~al.}(2022)\citenamefont {Lovato}
  \emph {et~al.}}]{Lovato:2022vgq}%
  \BibitemOpen
  \bibfield  {author} {\bibinfo {author} {\bibfnamefont {A.}~\bibnamefont
  {Lovato}} \emph {et~al.},\ }\href@noop {} {\  (\bibinfo {year} {2022})},\
  \Eprint {http://arxiv.org/abs/2211.02224} {arXiv:2211.02224 [nucl-th]}
  \BibitemShut {NoStop}%
\bibitem [{\citenamefont {Paeng}\ \emph {et~al.}(2016)\citenamefont {Paeng},
  \citenamefont {Kuo}, \citenamefont {Lee},\ and\ \citenamefont
  {Rho}}]{Paeng:2015noa}%
  \BibitemOpen
  \bibfield  {author} {\bibinfo {author} {\bibfnamefont {W.-G.}\ \bibnamefont
  {Paeng}}, \bibinfo {author} {\bibfnamefont {T.~T.~S.}\ \bibnamefont {Kuo}},
  \bibinfo {author} {\bibfnamefont {H.~K.}\ \bibnamefont {Lee}}, \ and\
  \bibinfo {author} {\bibfnamefont {M.}~\bibnamefont {Rho}},\ }\href {\doibase
  10.1103/PhysRevC.93.055203} {\bibfield  {journal} {\bibinfo  {journal} {Phys.
  Rev. C}\ }\textbf {\bibinfo {volume} {93}},\ \bibinfo {pages} {055203}
  (\bibinfo {year} {2016})},\ \Eprint {http://arxiv.org/abs/1508.05210}
  {arXiv:1508.05210 [hep-ph]} \BibitemShut {NoStop}%
\bibitem [{\citenamefont {Paeng}\ \emph {et~al.}(2017)\citenamefont {Paeng},
  \citenamefont {Kuo}, \citenamefont {Lee}, \citenamefont {Ma},\ and\
  \citenamefont {Rho}}]{Paeng:2017qvp}%
  \BibitemOpen
  \bibfield  {author} {\bibinfo {author} {\bibfnamefont {W.-G.}\ \bibnamefont
  {Paeng}}, \bibinfo {author} {\bibfnamefont {T.~T.~S.}\ \bibnamefont {Kuo}},
  \bibinfo {author} {\bibfnamefont {H.~K.}\ \bibnamefont {Lee}}, \bibinfo
  {author} {\bibfnamefont {Y.-L.}\ \bibnamefont {Ma}}, \ and\ \bibinfo {author}
  {\bibfnamefont {M.}~\bibnamefont {Rho}},\ }\href {\doibase
  10.1103/PhysRevD.96.014031} {\bibfield  {journal} {\bibinfo  {journal} {Phys.
  Rev. D}\ }\textbf {\bibinfo {volume} {96}},\ \bibinfo {pages} {014031}
  (\bibinfo {year} {2017})},\ \Eprint {http://arxiv.org/abs/1704.02775}
  {arXiv:1704.02775 [nucl-th]} \BibitemShut {NoStop}%
\bibitem [{\citenamefont {Ma}\ \emph {et~al.}(2019{\natexlab{a}})\citenamefont
  {Ma}, \citenamefont {Lee}, \citenamefont {Paeng},\ and\ \citenamefont
  {Rho}}]{Ma:2018jze}%
  \BibitemOpen
  \bibfield  {author} {\bibinfo {author} {\bibfnamefont {Y.-L.}\ \bibnamefont
  {Ma}}, \bibinfo {author} {\bibfnamefont {H.~K.}\ \bibnamefont {Lee}},
  \bibinfo {author} {\bibfnamefont {W.-G.}\ \bibnamefont {Paeng}}, \ and\
  \bibinfo {author} {\bibfnamefont {M.}~\bibnamefont {Rho}},\ }\href {\doibase
  10.1007/s11433-019-9399-5} {\bibfield  {journal} {\bibinfo  {journal} {Sci.
  China Phys. Mech. Astron.}\ }\textbf {\bibinfo {volume} {62}},\ \bibinfo
  {pages} {112011} (\bibinfo {year} {2019}{\natexlab{a}})},\ \Eprint
  {http://arxiv.org/abs/1804.00305} {arXiv:1804.00305 [nucl-th]} \BibitemShut
  {NoStop}%
\bibitem [{\citenamefont {Ma}\ and\ \citenamefont
  {Rho}(2019{\natexlab{a}})}]{Ma:2018xjw}%
  \BibitemOpen
  \bibfield  {author} {\bibinfo {author} {\bibfnamefont {Y.-L.}\ \bibnamefont
  {Ma}}\ and\ \bibinfo {author} {\bibfnamefont {M.}~\bibnamefont {Rho}},\
  }\href {\doibase 10.1103/PhysRevD.99.014034} {\bibfield  {journal} {\bibinfo
  {journal} {Phys. Rev. D}\ }\textbf {\bibinfo {volume} {99}},\ \bibinfo
  {pages} {014034} (\bibinfo {year} {2019}{\natexlab{a}})},\ \Eprint
  {http://arxiv.org/abs/1810.06062} {arXiv:1810.06062 [nucl-th]} \BibitemShut
  {NoStop}%
\bibitem [{\citenamefont {Ma}\ and\ \citenamefont
  {Rho}(2019{\natexlab{b}})}]{Ma:2018qkg}%
  \BibitemOpen
  \bibfield  {author} {\bibinfo {author} {\bibfnamefont {Y.-L.}\ \bibnamefont
  {Ma}}\ and\ \bibinfo {author} {\bibfnamefont {M.}~\bibnamefont {Rho}},\
  }\href {\doibase 10.1103/PhysRevD.100.114003} {\bibfield  {journal} {\bibinfo
   {journal} {Phys. Rev. D}\ }\textbf {\bibinfo {volume} {100}},\ \bibinfo
  {pages} {114003} (\bibinfo {year} {2019}{\natexlab{b}})},\ \Eprint
  {http://arxiv.org/abs/1811.07071} {arXiv:1811.07071 [nucl-th]} \BibitemShut
  {NoStop}%
\bibitem [{\citenamefont {Ma}\ and\ \citenamefont
  {Rho}(2020{\natexlab{b}})}]{Ma:2020hno}%
  \BibitemOpen
  \bibfield  {author} {\bibinfo {author} {\bibfnamefont {Y.-L.}\ \bibnamefont
  {Ma}}\ and\ \bibinfo {author} {\bibfnamefont {M.}~\bibnamefont {Rho}},\
  }\href@noop {} {\  (\bibinfo {year} {2020}{\natexlab{b}})},\ \Eprint
  {http://arxiv.org/abs/2006.14173} {arXiv:2006.14173 [nucl-th]} \BibitemShut
  {NoStop}%
\bibitem [{\citenamefont {Rho}\ and\ \citenamefont {Ma}(2021)}]{Rho:2021zwm}%
  \BibitemOpen
  \bibfield  {author} {\bibinfo {author} {\bibfnamefont {M.}~\bibnamefont
  {Rho}}\ and\ \bibinfo {author} {\bibfnamefont {Y.-L.}\ \bibnamefont {Ma}},\
  }\href {\doibase 10.1142/S0217732321300123} {\bibfield  {journal} {\bibinfo
  {journal} {Mod. Phys. Lett. A}\ }\textbf {\bibinfo {volume} {36}},\ \bibinfo
  {pages} {2130012} (\bibinfo {year} {2021})},\ \Eprint
  {http://arxiv.org/abs/2101.07121} {arXiv:2101.07121 [nucl-th]} \BibitemShut
  {NoStop}%
\bibitem [{\citenamefont {Ma}\ and\ \citenamefont
  {Rho}(2021{\natexlab{b}})}]{Ma:2021nuf}%
  \BibitemOpen
  \bibfield  {author} {\bibinfo {author} {\bibfnamefont {Y.-L.}\ \bibnamefont
  {Ma}}\ and\ \bibinfo {author} {\bibfnamefont {M.}~\bibnamefont {Rho}},\
  }\href {\doibase 10.1007/s43673-021-00016-1} {\bibfield  {journal} {\bibinfo
  {journal} {AAPPS Bull.}\ }\textbf {\bibinfo {volume} {31}},\ \bibinfo {pages}
  {16} (\bibinfo {year} {2021}{\natexlab{b}})},\ \Eprint
  {http://arxiv.org/abs/2103.00744} {arXiv:2103.00744 [nucl-th]} \BibitemShut
  {NoStop}%
\bibitem [{\citenamefont {Lee}\ \emph {et~al.}(2022)\citenamefont {Lee},
  \citenamefont {Ma}, \citenamefont {Paeng},\ and\ \citenamefont
  {Rho}}]{Lee:2021hrw}%
  \BibitemOpen
  \bibfield  {author} {\bibinfo {author} {\bibfnamefont {H.~K.}\ \bibnamefont
  {Lee}}, \bibinfo {author} {\bibfnamefont {Y.-L.}\ \bibnamefont {Ma}},
  \bibinfo {author} {\bibfnamefont {W.-G.}\ \bibnamefont {Paeng}}, \ and\
  \bibinfo {author} {\bibfnamefont {M.}~\bibnamefont {Rho}},\ }\href {\doibase
  10.1142/S0217732322300038} {\bibfield  {journal} {\bibinfo  {journal} {Mod.
  Phys. Lett. A}\ }\textbf {\bibinfo {volume} {37}},\ \bibinfo {pages}
  {2230003} (\bibinfo {year} {2022})},\ \Eprint
  {http://arxiv.org/abs/2107.01879} {arXiv:2107.01879 [nucl-th]} \BibitemShut
  {NoStop}%
\bibitem [{\citenamefont {Rho}(2022{\natexlab{a}})}]{Rho:2022uns}%
  \BibitemOpen
  \bibfield  {author} {\bibinfo {author} {\bibfnamefont {M.}~\bibnamefont
  {Rho}},\ }\href {\doibase 10.3390/sym14050994} {\bibfield  {journal}
  {\bibinfo  {journal} {Symmetry}\ }\textbf {\bibinfo {volume} {14}},\ \bibinfo
  {pages} {994} (\bibinfo {year} {2022}{\natexlab{a}})},\ \Eprint
  {http://arxiv.org/abs/2203.06998} {arXiv:2203.06998 [nucl-th]} \BibitemShut
  {NoStop}%
\bibitem [{\citenamefont {Bedaque}\ and\ \citenamefont
  {Steiner}(2015)}]{Bedaque:2014sqa}%
  \BibitemOpen
  \bibfield  {author} {\bibinfo {author} {\bibfnamefont {P.}~\bibnamefont
  {Bedaque}}\ and\ \bibinfo {author} {\bibfnamefont {A.~W.}\ \bibnamefont
  {Steiner}},\ }\href {\doibase 10.1103/PhysRevLett.114.031103} {\bibfield
  {journal} {\bibinfo  {journal} {Phys. Rev. Lett.}\ }\textbf {\bibinfo
  {volume} {114}},\ \bibinfo {pages} {031103} (\bibinfo {year} {2015})},\
  \Eprint {http://arxiv.org/abs/1408.5116} {arXiv:1408.5116 [nucl-th]}
  \BibitemShut {NoStop}%
\bibitem [{\citenamefont {Tews}\ \emph {et~al.}(2018)\citenamefont {Tews},
  \citenamefont {Carlson}, \citenamefont {Gandolfi},\ and\ \citenamefont
  {Reddy}}]{Tews:2018kmu}%
  \BibitemOpen
  \bibfield  {author} {\bibinfo {author} {\bibfnamefont {I.}~\bibnamefont
  {Tews}}, \bibinfo {author} {\bibfnamefont {J.}~\bibnamefont {Carlson}},
  \bibinfo {author} {\bibfnamefont {S.}~\bibnamefont {Gandolfi}}, \ and\
  \bibinfo {author} {\bibfnamefont {S.}~\bibnamefont {Reddy}},\ }\href
  {\doibase 10.3847/1538-4357/aac267} {\bibfield  {journal} {\bibinfo
  {journal} {Astrophys. J.}\ }\textbf {\bibinfo {volume} {860}},\ \bibinfo
  {pages} {149} (\bibinfo {year} {2018})},\ \Eprint
  {http://arxiv.org/abs/1801.01923} {arXiv:1801.01923 [nucl-th]} \BibitemShut
  {NoStop}%
\bibitem [{\citenamefont {Moustakidis}\ \emph {et~al.}(2017)\citenamefont
  {Moustakidis}, \citenamefont {Gaitanos}, \citenamefont {Margaritis},\ and\
  \citenamefont {Lalazissis}}]{Moustakidis:2016sab}%
  \BibitemOpen
  \bibfield  {author} {\bibinfo {author} {\bibfnamefont {C.~C.}\ \bibnamefont
  {Moustakidis}}, \bibinfo {author} {\bibfnamefont {T.}~\bibnamefont
  {Gaitanos}}, \bibinfo {author} {\bibfnamefont {C.}~\bibnamefont
  {Margaritis}}, \ and\ \bibinfo {author} {\bibfnamefont {G.~A.}\ \bibnamefont
  {Lalazissis}},\ }\href {\doibase 10.1103/PhysRevC.95.045801} {\bibfield
  {journal} {\bibinfo  {journal} {Phys. Rev. C}\ }\textbf {\bibinfo {volume}
  {95}},\ \bibinfo {pages} {045801} (\bibinfo {year} {2017})},\ \bibinfo {note}
  {[Erratum: Phys.Rev.C 95, 059904 (2017)]},\ \Eprint
  {http://arxiv.org/abs/1608.00344} {arXiv:1608.00344 [nucl-th]} \BibitemShut
  {NoStop}%
\bibitem [{\citenamefont {Alsing}\ \emph {et~al.}(2018)\citenamefont {Alsing},
  \citenamefont {Silva},\ and\ \citenamefont {Berti}}]{Alsing:2017bbc}%
  \BibitemOpen
  \bibfield  {author} {\bibinfo {author} {\bibfnamefont {J.}~\bibnamefont
  {Alsing}}, \bibinfo {author} {\bibfnamefont {H.~O.}\ \bibnamefont {Silva}}, \
  and\ \bibinfo {author} {\bibfnamefont {E.}~\bibnamefont {Berti}},\ }\href
  {\doibase 10.1093/mnras/sty1065} {\bibfield  {journal} {\bibinfo  {journal}
  {Mon. Not. Roy. Astron. Soc.}\ }\textbf {\bibinfo {volume} {478}},\ \bibinfo
  {pages} {1377} (\bibinfo {year} {2018})},\ \Eprint
  {http://arxiv.org/abs/1709.07889} {arXiv:1709.07889 [astro-ph.HE]}
  \BibitemShut {NoStop}%
\bibitem [{\citenamefont {McLerran}\ and\ \citenamefont
  {Reddy}(2019)}]{McLerran:2018hbz}%
  \BibitemOpen
  \bibfield  {author} {\bibinfo {author} {\bibfnamefont {L.}~\bibnamefont
  {McLerran}}\ and\ \bibinfo {author} {\bibfnamefont {S.}~\bibnamefont
  {Reddy}},\ }\href {\doibase 10.1103/PhysRevLett.122.122701} {\bibfield
  {journal} {\bibinfo  {journal} {Phys. Rev. Lett.}\ }\textbf {\bibinfo
  {volume} {122}},\ \bibinfo {pages} {122701} (\bibinfo {year} {2019})},\
  \Eprint {http://arxiv.org/abs/1811.12503} {arXiv:1811.12503 [nucl-th]}
  \BibitemShut {NoStop}%
\bibitem [{\citenamefont {Jeong}\ \emph {et~al.}(2020)\citenamefont {Jeong},
  \citenamefont {McLerran},\ and\ \citenamefont {Sen}}]{Jeong:2019lhv}%
  \BibitemOpen
  \bibfield  {author} {\bibinfo {author} {\bibfnamefont {K.~S.}\ \bibnamefont
  {Jeong}}, \bibinfo {author} {\bibfnamefont {L.}~\bibnamefont {McLerran}}, \
  and\ \bibinfo {author} {\bibfnamefont {S.}~\bibnamefont {Sen}},\ }\href
  {\doibase 10.1103/PhysRevC.101.035201} {\bibfield  {journal} {\bibinfo
  {journal} {Phys. Rev. C}\ }\textbf {\bibinfo {volume} {101}},\ \bibinfo
  {pages} {035201} (\bibinfo {year} {2020})},\ \Eprint
  {http://arxiv.org/abs/1908.04799} {arXiv:1908.04799 [nucl-th]} \BibitemShut
  {NoStop}%
\bibitem [{\citenamefont {Kapusta}\ and\ \citenamefont
  {Welle}(2021)}]{Kapusta:2021ney}%
  \BibitemOpen
  \bibfield  {author} {\bibinfo {author} {\bibfnamefont {J.~I.}\ \bibnamefont
  {Kapusta}}\ and\ \bibinfo {author} {\bibfnamefont {T.}~\bibnamefont
  {Welle}},\ }\href {\doibase 10.1103/PhysRevC.104.L012801} {\bibfield
  {journal} {\bibinfo  {journal} {Phys. Rev. C}\ }\textbf {\bibinfo {volume}
  {104}},\ \bibinfo {pages} {L012801} (\bibinfo {year} {2021})},\ \Eprint
  {http://arxiv.org/abs/2103.16633} {arXiv:2103.16633 [nucl-th]} \BibitemShut
  {NoStop}%
\bibitem [{\citenamefont {Zhao}\ and\ \citenamefont
  {Lattimer}(2020)}]{Zhao:2020dvu}%
  \BibitemOpen
  \bibfield  {author} {\bibinfo {author} {\bibfnamefont {T.}~\bibnamefont
  {Zhao}}\ and\ \bibinfo {author} {\bibfnamefont {J.~M.}\ \bibnamefont
  {Lattimer}},\ }\href {\doibase 10.1103/PhysRevD.102.023021} {\bibfield
  {journal} {\bibinfo  {journal} {Phys. Rev. D}\ }\textbf {\bibinfo {volume}
  {102}},\ \bibinfo {pages} {023021} (\bibinfo {year} {2020})},\ \Eprint
  {http://arxiv.org/abs/2004.08293} {arXiv:2004.08293 [astro-ph.HE]}
  \BibitemShut {NoStop}%
\bibitem [{\citenamefont {Margueron}\ \emph {et~al.}(2021)\citenamefont
  {Margueron}, \citenamefont {Hansen}, \citenamefont {Proust},\ and\
  \citenamefont {Chanfray}}]{Margueron:2021dtx}%
  \BibitemOpen
  \bibfield  {author} {\bibinfo {author} {\bibfnamefont {J.}~\bibnamefont
  {Margueron}}, \bibinfo {author} {\bibfnamefont {H.}~\bibnamefont {Hansen}},
  \bibinfo {author} {\bibfnamefont {P.}~\bibnamefont {Proust}}, \ and\ \bibinfo
  {author} {\bibfnamefont {G.}~\bibnamefont {Chanfray}},\ }\href {\doibase
  10.1103/PhysRevC.104.055803} {\bibfield  {journal} {\bibinfo  {journal}
  {Phys. Rev. C}\ }\textbf {\bibinfo {volume} {104}},\ \bibinfo {pages}
  {055803} (\bibinfo {year} {2021})},\ \Eprint
  {http://arxiv.org/abs/2103.10209} {arXiv:2103.10209 [nucl-th]} \BibitemShut
  {NoStop}%
\bibitem [{\citenamefont {Fujimoto}\ \emph {et~al.}(2022)\citenamefont
  {Fujimoto}, \citenamefont {Fukushima}, \citenamefont {McLerran},\ and\
  \citenamefont {Praszalowicz}}]{Fujimoto:2022ohj}%
  \BibitemOpen
  \bibfield  {author} {\bibinfo {author} {\bibfnamefont {Y.}~\bibnamefont
  {Fujimoto}}, \bibinfo {author} {\bibfnamefont {K.}~\bibnamefont {Fukushima}},
  \bibinfo {author} {\bibfnamefont {L.~D.}\ \bibnamefont {McLerran}}, \ and\
  \bibinfo {author} {\bibfnamefont {M.}~\bibnamefont {Praszalowicz}},\
  }\href@noop {} {\  (\bibinfo {year} {2022})},\ \Eprint
  {http://arxiv.org/abs/2207.06753} {arXiv:2207.06753 [nucl-th]} \BibitemShut
  {NoStop}%
\bibitem [{\citenamefont {Marczenko}\ \emph
  {et~al.}(2022{\natexlab{a}})\citenamefont {Marczenko}, \citenamefont
  {McLerran}, \citenamefont {Redlich},\ and\ \citenamefont
  {Sasaki}}]{Marczenko:2022jhl}%
  \BibitemOpen
  \bibfield  {author} {\bibinfo {author} {\bibfnamefont {M.}~\bibnamefont
  {Marczenko}}, \bibinfo {author} {\bibfnamefont {L.}~\bibnamefont {McLerran}},
  \bibinfo {author} {\bibfnamefont {K.}~\bibnamefont {Redlich}}, \ and\
  \bibinfo {author} {\bibfnamefont {C.}~\bibnamefont {Sasaki}},\ }\href@noop {}
  {\  (\bibinfo {year} {2022}{\natexlab{a}})},\ \Eprint
  {http://arxiv.org/abs/2207.13059} {arXiv:2207.13059 [nucl-th]} \BibitemShut
  {NoStop}%
\bibitem [{\citenamefont {Bando}\ \emph {et~al.}(1985)\citenamefont {Bando},
  \citenamefont {Kugo}, \citenamefont {Uehara}, \citenamefont {Yamawaki},\ and\
  \citenamefont {Yanagida}}]{Bando:1984ej}%
  \BibitemOpen
  \bibfield  {author} {\bibinfo {author} {\bibfnamefont {M.}~\bibnamefont
  {Bando}}, \bibinfo {author} {\bibfnamefont {T.}~\bibnamefont {Kugo}},
  \bibinfo {author} {\bibfnamefont {S.}~\bibnamefont {Uehara}}, \bibinfo
  {author} {\bibfnamefont {K.}~\bibnamefont {Yamawaki}}, \ and\ \bibinfo
  {author} {\bibfnamefont {T.}~\bibnamefont {Yanagida}},\ }\href {\doibase
  10.1103/PhysRevLett.54.1215} {\bibfield  {journal} {\bibinfo  {journal}
  {Phys. Rev. Lett.}\ }\textbf {\bibinfo {volume} {54}},\ \bibinfo {pages}
  {1215} (\bibinfo {year} {1985})}\BibitemShut {NoStop}%
\bibitem [{\citenamefont {Bando}\ \emph {et~al.}(1988)\citenamefont {Bando},
  \citenamefont {Kugo},\ and\ \citenamefont {Yamawaki}}]{Bando:1987br}%
  \BibitemOpen
  \bibfield  {author} {\bibinfo {author} {\bibfnamefont {M.}~\bibnamefont
  {Bando}}, \bibinfo {author} {\bibfnamefont {T.}~\bibnamefont {Kugo}}, \ and\
  \bibinfo {author} {\bibfnamefont {K.}~\bibnamefont {Yamawaki}},\ }\href
  {\doibase 10.1016/0370-1573(88)90019-1} {\bibfield  {journal} {\bibinfo
  {journal} {Phys. Rept.}\ }\textbf {\bibinfo {volume} {164}},\ \bibinfo
  {pages} {217} (\bibinfo {year} {1988})}\BibitemShut {NoStop}%
\bibitem [{\citenamefont {Harada}\ and\ \citenamefont
  {Yamawaki}(2003)}]{Harada:2003jx}%
  \BibitemOpen
  \bibfield  {author} {\bibinfo {author} {\bibfnamefont {M.}~\bibnamefont
  {Harada}}\ and\ \bibinfo {author} {\bibfnamefont {K.}~\bibnamefont
  {Yamawaki}},\ }\href {\doibase 10.1016/S0370-1573(03)00139-X} {\bibfield
  {journal} {\bibinfo  {journal} {Phys. Rept.}\ }\textbf {\bibinfo {volume}
  {381}},\ \bibinfo {pages} {1} (\bibinfo {year} {2003})},\ \Eprint
  {http://arxiv.org/abs/hep-ph/0302103} {arXiv:hep-ph/0302103} \BibitemShut
  {NoStop}%
\bibitem [{\citenamefont {Crewther}\ and\ \citenamefont
  {Tunstall}(2015)}]{Crewther:2013vea}%
  \BibitemOpen
  \bibfield  {author} {\bibinfo {author} {\bibfnamefont {R.~J.}\ \bibnamefont
  {Crewther}}\ and\ \bibinfo {author} {\bibfnamefont {L.~C.}\ \bibnamefont
  {Tunstall}},\ }\href {\doibase 10.1103/PhysRevD.91.034016} {\bibfield
  {journal} {\bibinfo  {journal} {Phys. Rev. D}\ }\textbf {\bibinfo {volume}
  {91}},\ \bibinfo {pages} {034016} (\bibinfo {year} {2015})},\ \Eprint
  {http://arxiv.org/abs/1312.3319} {arXiv:1312.3319 [hep-ph]} \BibitemShut
  {NoStop}%
\bibitem [{\citenamefont {Cat\`a}\ \emph {et~al.}(2019)\citenamefont {Cat\`a},
  \citenamefont {Crewther},\ and\ \citenamefont {Tunstall}}]{Cata:2018wzl}%
  \BibitemOpen
  \bibfield  {author} {\bibinfo {author} {\bibfnamefont {O.}~\bibnamefont
  {Cat\`a}}, \bibinfo {author} {\bibfnamefont {R.~J.}\ \bibnamefont
  {Crewther}}, \ and\ \bibinfo {author} {\bibfnamefont {L.~C.}\ \bibnamefont
  {Tunstall}},\ }\href {\doibase 10.1103/PhysRevD.100.095007} {\bibfield
  {journal} {\bibinfo  {journal} {Phys. Rev. D}\ }\textbf {\bibinfo {volume}
  {100}},\ \bibinfo {pages} {095007} (\bibinfo {year} {2019})},\ \Eprint
  {http://arxiv.org/abs/1803.08513} {arXiv:1803.08513 [hep-ph]} \BibitemShut
  {NoStop}%
\bibitem [{\citenamefont {Crewther}(2020)}]{Crewther:2020tgd}%
  \BibitemOpen
  \bibfield  {author} {\bibinfo {author} {\bibfnamefont {R.~J.}\ \bibnamefont
  {Crewther}},\ }\href {\doibase 10.3390/universe6070096} {\bibfield  {journal}
  {\bibinfo  {journal} {Universe}\ }\textbf {\bibinfo {volume} {6}},\ \bibinfo
  {pages} {96} (\bibinfo {year} {2020})},\ \Eprint
  {http://arxiv.org/abs/2003.11259} {arXiv:2003.11259 [hep-ph]} \BibitemShut
  {NoStop}%
\bibitem [{\citenamefont {Skyrme}(1961)}]{Skyrme:1961vq}%
  \BibitemOpen
  \bibfield  {author} {\bibinfo {author} {\bibfnamefont {T.~H.~R.}\
  \bibnamefont {Skyrme}},\ }\href {\doibase 10.1098/rspa.1961.0018} {\bibfield
  {journal} {\bibinfo  {journal} {Proc. Roy. Soc. Lond. A}\ }\textbf {\bibinfo
  {volume} {260}},\ \bibinfo {pages} {127} (\bibinfo {year}
  {1961})}\BibitemShut {NoStop}%
\bibitem [{\citenamefont {Kugler}\ and\ \citenamefont
  {Shtrikman}(1988)}]{Kugler:1988mu}%
  \BibitemOpen
  \bibfield  {author} {\bibinfo {author} {\bibfnamefont {M.}~\bibnamefont
  {Kugler}}\ and\ \bibinfo {author} {\bibfnamefont {S.}~\bibnamefont
  {Shtrikman}},\ }\href {\doibase 10.1016/0370-2693(88)90653-3} {\bibfield
  {journal} {\bibinfo  {journal} {Phys. Lett. B}\ }\textbf {\bibinfo {volume}
  {208}},\ \bibinfo {pages} {491} (\bibinfo {year} {1988})}\BibitemShut
  {NoStop}%
\bibitem [{\citenamefont {Kugler}\ and\ \citenamefont
  {Shtrikman}(1989)}]{Kugler:1989uc}%
  \BibitemOpen
  \bibfield  {author} {\bibinfo {author} {\bibfnamefont {M.}~\bibnamefont
  {Kugler}}\ and\ \bibinfo {author} {\bibfnamefont {S.}~\bibnamefont
  {Shtrikman}},\ }\href {\doibase 10.1103/PhysRevD.40.3421} {\bibfield
  {journal} {\bibinfo  {journal} {Phys. Rev. D}\ }\textbf {\bibinfo {volume}
  {40}},\ \bibinfo {pages} {3421} (\bibinfo {year} {1989})}\BibitemShut
  {NoStop}%
\bibitem [{\citenamefont {Lee}\ \emph {et~al.}(2003)\citenamefont {Lee},
  \citenamefont {Park}, \citenamefont {Rho},\ and\ \citenamefont
  {Vento}}]{Lee:2003eg}%
  \BibitemOpen
  \bibfield  {author} {\bibinfo {author} {\bibfnamefont {H.-J.}\ \bibnamefont
  {Lee}}, \bibinfo {author} {\bibfnamefont {B.-Y.}\ \bibnamefont {Park}},
  \bibinfo {author} {\bibfnamefont {M.}~\bibnamefont {Rho}}, \ and\ \bibinfo
  {author} {\bibfnamefont {V.}~\bibnamefont {Vento}},\ }\href {\doibase
  10.1016/S0375-9474(03)01626-9} {\bibfield  {journal} {\bibinfo  {journal}
  {Nucl. Phys. A}\ }\textbf {\bibinfo {volume} {726}},\ \bibinfo {pages} {69}
  (\bibinfo {year} {2003})},\ \Eprint {http://arxiv.org/abs/hep-ph/0304066}
  {arXiv:hep-ph/0304066} \BibitemShut {NoStop}%
\bibitem [{\citenamefont {Park}\ \emph {et~al.}(2004)\citenamefont {Park},
  \citenamefont {Rho},\ and\ \citenamefont {Vento}}]{Park:2003sd}%
  \BibitemOpen
  \bibfield  {author} {\bibinfo {author} {\bibfnamefont {B.-Y.}\ \bibnamefont
  {Park}}, \bibinfo {author} {\bibfnamefont {M.}~\bibnamefont {Rho}}, \ and\
  \bibinfo {author} {\bibfnamefont {V.}~\bibnamefont {Vento}},\ }\href
  {\doibase 10.1016/j.nuclphysa.2004.01.131} {\bibfield  {journal} {\bibinfo
  {journal} {Nucl. Phys. A}\ }\textbf {\bibinfo {volume} {736}},\ \bibinfo
  {pages} {129} (\bibinfo {year} {2004})},\ \Eprint
  {http://arxiv.org/abs/hep-ph/0310087} {arXiv:hep-ph/0310087} \BibitemShut
  {NoStop}%
\bibitem [{\citenamefont {Park}\ \emph {et~al.}(2008)\citenamefont {Park},
  \citenamefont {Rho},\ and\ \citenamefont {Vento}}]{Park:2008zg}%
  \BibitemOpen
  \bibfield  {author} {\bibinfo {author} {\bibfnamefont {B.-Y.}\ \bibnamefont
  {Park}}, \bibinfo {author} {\bibfnamefont {M.}~\bibnamefont {Rho}}, \ and\
  \bibinfo {author} {\bibfnamefont {V.}~\bibnamefont {Vento}},\ }\href
  {\doibase 10.1016/j.nuclphysa.2008.03.015} {\bibfield  {journal} {\bibinfo
  {journal} {Nucl. Phys. A}\ }\textbf {\bibinfo {volume} {807}},\ \bibinfo
  {pages} {28} (\bibinfo {year} {2008})},\ \Eprint
  {http://arxiv.org/abs/0801.1374} {arXiv:0801.1374 [hep-ph]} \BibitemShut
  {NoStop}%
\bibitem [{\citenamefont {Ma}\ \emph {et~al.}(2013)\citenamefont {Ma},
  \citenamefont {Harada}, \citenamefont {Lee}, \citenamefont {Oh},
  \citenamefont {Park},\ and\ \citenamefont {Rho}}]{Ma:2013ooa}%
  \BibitemOpen
  \bibfield  {author} {\bibinfo {author} {\bibfnamefont {Y.-L.}\ \bibnamefont
  {Ma}}, \bibinfo {author} {\bibfnamefont {M.}~\bibnamefont {Harada}}, \bibinfo
  {author} {\bibfnamefont {H.~K.}\ \bibnamefont {Lee}}, \bibinfo {author}
  {\bibfnamefont {Y.}~\bibnamefont {Oh}}, \bibinfo {author} {\bibfnamefont
  {B.-Y.}\ \bibnamefont {Park}}, \ and\ \bibinfo {author} {\bibfnamefont
  {M.}~\bibnamefont {Rho}},\ }\href {\doibase 10.1103/PhysRevD.88.014016}
  {\bibfield  {journal} {\bibinfo  {journal} {Phys. Rev. D}\ }\textbf {\bibinfo
  {volume} {88}},\ \bibinfo {pages} {014016} (\bibinfo {year} {2013})},\
  \bibinfo {note} {[Erratum: Phys.Rev.D 88, 079904 (2013)]},\ \Eprint
  {http://arxiv.org/abs/1304.5638} {arXiv:1304.5638 [hep-ph]} \BibitemShut
  {NoStop}%
\bibitem [{\citenamefont {Ma}\ \emph {et~al.}(2014{\natexlab{a}})\citenamefont
  {Ma}, \citenamefont {Harada}, \citenamefont {Lee}, \citenamefont {Oh},
  \citenamefont {Park},\ and\ \citenamefont {Rho}}]{Ma:2013ela}%
  \BibitemOpen
  \bibfield  {author} {\bibinfo {author} {\bibfnamefont {Y.-L.}\ \bibnamefont
  {Ma}}, \bibinfo {author} {\bibfnamefont {M.}~\bibnamefont {Harada}}, \bibinfo
  {author} {\bibfnamefont {H.~K.}\ \bibnamefont {Lee}}, \bibinfo {author}
  {\bibfnamefont {Y.}~\bibnamefont {Oh}}, \bibinfo {author} {\bibfnamefont
  {B.-Y.}\ \bibnamefont {Park}}, \ and\ \bibinfo {author} {\bibfnamefont
  {M.}~\bibnamefont {Rho}},\ }\href {\doibase 10.1103/PhysRevD.90.034015}
  {\bibfield  {journal} {\bibinfo  {journal} {Phys. Rev. D}\ }\textbf {\bibinfo
  {volume} {90}},\ \bibinfo {pages} {034015} (\bibinfo {year}
  {2014}{\natexlab{a}})},\ \Eprint {http://arxiv.org/abs/1308.6476}
  {arXiv:1308.6476 [hep-ph]} \BibitemShut {NoStop}%
\bibitem [{\citenamefont {Ma}\ and\ \citenamefont {Rho}(2017)}]{Ma:2016gdd}%
  \BibitemOpen
  \bibfield  {author} {\bibinfo {author} {\bibfnamefont {Y.-L.}\ \bibnamefont
  {Ma}}\ and\ \bibinfo {author} {\bibfnamefont {M.}~\bibnamefont {Rho}},\
  }\href {\doibase 10.1007/s11433-016-0497-2} {\bibfield  {journal} {\bibinfo
  {journal} {Sci. China Phys. Mech. Astron.}\ }\textbf {\bibinfo {volume}
  {60}},\ \bibinfo {pages} {032001} (\bibinfo {year} {2017})},\ \Eprint
  {http://arxiv.org/abs/1612.06600} {arXiv:1612.06600 [nucl-th]} \BibitemShut
  {NoStop}%
\bibitem [{\citenamefont {Shao}\ and\ \citenamefont {Ma}(2022)}]{Shao:2022njr}%
  \BibitemOpen
  \bibfield  {author} {\bibinfo {author} {\bibfnamefont {L.-Q.}\ \bibnamefont
  {Shao}}\ and\ \bibinfo {author} {\bibfnamefont {Y.-L.}\ \bibnamefont {Ma}},\
  }\href {\doibase 10.1103/PhysRevD.106.014014} {\bibfield  {journal} {\bibinfo
   {journal} {Phys. Rev. D}\ }\textbf {\bibinfo {volume} {106}},\ \bibinfo
  {pages} {014014} (\bibinfo {year} {2022})},\ \Eprint
  {http://arxiv.org/abs/2202.09957} {arXiv:2202.09957 [nucl-th]} \BibitemShut
  {NoStop}%
\bibitem [{\citenamefont {Brown}\ and\ \citenamefont
  {Rho}(1991)}]{Brown:1991kk}%
  \BibitemOpen
  \bibfield  {author} {\bibinfo {author} {\bibfnamefont {G.~E.}\ \bibnamefont
  {Brown}}\ and\ \bibinfo {author} {\bibfnamefont {M.}~\bibnamefont {Rho}},\
  }\href {\doibase 10.1103/PhysRevLett.66.2720} {\bibfield  {journal} {\bibinfo
   {journal} {Phys. Rev. Lett.}\ }\textbf {\bibinfo {volume} {66}},\ \bibinfo
  {pages} {2720} (\bibinfo {year} {1991})}\BibitemShut {NoStop}%
\bibitem [{\citenamefont {Bogner}\ \emph {et~al.}(2003)\citenamefont {Bogner},
  \citenamefont {Kuo},\ and\ \citenamefont {Schwenk}}]{Bogner:2003wn}%
  \BibitemOpen
  \bibfield  {author} {\bibinfo {author} {\bibfnamefont {S.~K.}\ \bibnamefont
  {Bogner}}, \bibinfo {author} {\bibfnamefont {T.~T.~S.}\ \bibnamefont {Kuo}},
  \ and\ \bibinfo {author} {\bibfnamefont {A.}~\bibnamefont {Schwenk}},\ }\href
  {\doibase 10.1016/j.physrep.2003.07.001} {\bibfield  {journal} {\bibinfo
  {journal} {Phys. Rept.}\ }\textbf {\bibinfo {volume} {386}},\ \bibinfo
  {pages} {1} (\bibinfo {year} {2003})},\ \Eprint
  {http://arxiv.org/abs/nucl-th/0305035} {arXiv:nucl-th/0305035} \BibitemShut
  {NoStop}%
\bibitem [{\citenamefont {Serot}\ and\ \citenamefont
  {Walecka}(1986)}]{Serot:1984ey}%
  \BibitemOpen
  \bibfield  {author} {\bibinfo {author} {\bibfnamefont {B.~D.}\ \bibnamefont
  {Serot}}\ and\ \bibinfo {author} {\bibfnamefont {J.~D.}\ \bibnamefont
  {Walecka}},\ }\href@noop {} {\bibfield  {journal} {\bibinfo  {journal} {Adv.
  Nucl. Phys.}\ }\textbf {\bibinfo {volume} {16}},\ \bibinfo {pages} {1}
  (\bibinfo {year} {1986})}\BibitemShut {NoStop}%
\bibitem [{\citenamefont {Li}\ \emph {et~al.}(2022)\citenamefont {Li},
  \citenamefont {Cai}, \citenamefont {Zhou}, \citenamefont {Jiang},\ and\
  \citenamefont {Chen}}]{Li:2022okx}%
  \BibitemOpen
  \bibfield  {author} {\bibinfo {author} {\bibfnamefont {F.}~\bibnamefont
  {Li}}, \bibinfo {author} {\bibfnamefont {B.-J.}\ \bibnamefont {Cai}},
  \bibinfo {author} {\bibfnamefont {Y.}~\bibnamefont {Zhou}}, \bibinfo {author}
  {\bibfnamefont {W.-Z.}\ \bibnamefont {Jiang}}, \ and\ \bibinfo {author}
  {\bibfnamefont {L.-W.}\ \bibnamefont {Chen}},\ }\href {\doibase
  10.3847/1538-4357/ac5e2a} {\bibfield  {journal} {\bibinfo  {journal}
  {Astrophys. J.}\ }\textbf {\bibinfo {volume} {929}},\ \bibinfo {pages} {183}
  (\bibinfo {year} {2022})},\ \Eprint {http://arxiv.org/abs/2202.08705}
  {arXiv:2202.08705 [nucl-th]} \BibitemShut {NoStop}%
\bibitem [{\citenamefont {Miyatsu}\ \emph {et~al.}(2022)\citenamefont
  {Miyatsu}, \citenamefont {Cheoun},\ and\ \citenamefont
  {Saito}}]{Miyatsu:2022wuy}%
  \BibitemOpen
  \bibfield  {author} {\bibinfo {author} {\bibfnamefont {T.}~\bibnamefont
  {Miyatsu}}, \bibinfo {author} {\bibfnamefont {M.-K.}\ \bibnamefont {Cheoun}},
  \ and\ \bibinfo {author} {\bibfnamefont {K.}~\bibnamefont {Saito}},\ }\href
  {\doibase 10.3847/1538-4357/ac5f40} {\bibfield  {journal} {\bibinfo
  {journal} {Astrophys. J.}\ }\textbf {\bibinfo {volume} {929}},\ \bibinfo
  {pages} {82} (\bibinfo {year} {2022})},\ \Eprint
  {http://arxiv.org/abs/2202.06468} {arXiv:2202.06468 [nucl-th]} \BibitemShut
  {NoStop}%
\bibitem [{\citenamefont {Workman}(2022)}]{Workman:2022ynf}%
  \BibitemOpen
  \bibfield  {author} {\bibinfo {author} {\bibfnamefont {R.~L.}\ \bibnamefont
  {Workman}} (\bibinfo {collaboration} {Particle Data Group}),\ }\href
  {\doibase 10.1093/ptep/ptac097} {\bibfield  {journal} {\bibinfo  {journal}
  {PTEP}\ }\textbf {\bibinfo {volume} {2022}},\ \bibinfo {pages} {083C01}
  (\bibinfo {year} {2022})}\BibitemShut {NoStop}%
\bibitem [{\citenamefont {Isham}\ \emph {et~al.}(1970)\citenamefont {Isham},
  \citenamefont {Salam},\ and\ \citenamefont {Strathdee}}]{Isham:1970xz}%
  \BibitemOpen
  \bibfield  {author} {\bibinfo {author} {\bibfnamefont {C.~J.}\ \bibnamefont
  {Isham}}, \bibinfo {author} {\bibfnamefont {A.}~\bibnamefont {Salam}}, \ and\
  \bibinfo {author} {\bibfnamefont {J.~A.}\ \bibnamefont {Strathdee}},\ }\href
  {\doibase 10.1103/PhysRevD.2.685} {\bibfield  {journal} {\bibinfo  {journal}
  {Phys. Rev. D}\ }\textbf {\bibinfo {volume} {2}},\ \bibinfo {pages} {685}
  (\bibinfo {year} {1970})}\BibitemShut {NoStop}%
\bibitem [{\citenamefont {Ellis}(1970)}]{Ellis:1970yd}%
  \BibitemOpen
  \bibfield  {author} {\bibinfo {author} {\bibfnamefont {J.~R.}\ \bibnamefont
  {Ellis}},\ }\href {\doibase 10.1016/0550-3213(70)90422-0} {\bibfield
  {journal} {\bibinfo  {journal} {Nucl. Phys. B}\ }\textbf {\bibinfo {volume}
  {22}},\ \bibinfo {pages} {478} (\bibinfo {year} {1970})},\ \bibinfo {note}
  {[Erratum: Nucl.Phys.B 25, 639--639 (1971)]}\BibitemShut {NoStop}%
\bibitem [{\citenamefont {Schechter}(1980)}]{Schechter:1980ak}%
  \BibitemOpen
  \bibfield  {author} {\bibinfo {author} {\bibfnamefont {J.}~\bibnamefont
  {Schechter}},\ }\href {\doibase 10.1103/PhysRevD.21.3393} {\bibfield
  {journal} {\bibinfo  {journal} {Phys. Rev. D}\ }\textbf {\bibinfo {volume}
  {21}},\ \bibinfo {pages} {3393} (\bibinfo {year} {1980})}\BibitemShut
  {NoStop}%
\bibitem [{\citenamefont {Golterman}\ and\ \citenamefont
  {Shamir}(2016)}]{Golterman:2016lsd}%
  \BibitemOpen
  \bibfield  {author} {\bibinfo {author} {\bibfnamefont {M.}~\bibnamefont
  {Golterman}}\ and\ \bibinfo {author} {\bibfnamefont {Y.}~\bibnamefont
  {Shamir}},\ }\href {\doibase 10.1103/PhysRevD.94.054502} {\bibfield
  {journal} {\bibinfo  {journal} {Phys. Rev. D}\ }\textbf {\bibinfo {volume}
  {94}},\ \bibinfo {pages} {054502} (\bibinfo {year} {2016})},\ \Eprint
  {http://arxiv.org/abs/1603.04575} {arXiv:1603.04575 [hep-ph]} \BibitemShut
  {NoStop}%
\bibitem [{\citenamefont {Brodsky}\ \emph {et~al.}(2010)\citenamefont
  {Brodsky}, \citenamefont {de~Teramond},\ and\ \citenamefont
  {Deur}}]{Brodsky:2010ur}%
  \BibitemOpen
  \bibfield  {author} {\bibinfo {author} {\bibfnamefont {S.~J.}\ \bibnamefont
  {Brodsky}}, \bibinfo {author} {\bibfnamefont {G.~F.}\ \bibnamefont
  {de~Teramond}}, \ and\ \bibinfo {author} {\bibfnamefont {A.}~\bibnamefont
  {Deur}},\ }\href {\doibase 10.1103/PhysRevD.81.096010} {\bibfield  {journal}
  {\bibinfo  {journal} {Phys. Rev. D}\ }\textbf {\bibinfo {volume} {81}},\
  \bibinfo {pages} {096010} (\bibinfo {year} {2010})},\ \Eprint
  {http://arxiv.org/abs/1002.3948} {arXiv:1002.3948 [hep-ph]} \BibitemShut
  {NoStop}%
\bibitem [{\citenamefont {Horsley}\ \emph {et~al.}(2014)\citenamefont
  {Horsley}, \citenamefont {Perlt}, \citenamefont {Rakow}, \citenamefont
  {Schierholz},\ and\ \citenamefont {Schiller}}]{Horsley:2013pra}%
  \BibitemOpen
  \bibfield  {author} {\bibinfo {author} {\bibfnamefont {R.}~\bibnamefont
  {Horsley}}, \bibinfo {author} {\bibfnamefont {H.}~\bibnamefont {Perlt}},
  \bibinfo {author} {\bibfnamefont {P.~E.~L.}\ \bibnamefont {Rakow}}, \bibinfo
  {author} {\bibfnamefont {G.}~\bibnamefont {Schierholz}}, \ and\ \bibinfo
  {author} {\bibfnamefont {A.}~\bibnamefont {Schiller}},\ }\href {\doibase
  10.1016/j.physletb.2013.11.012} {\bibfield  {journal} {\bibinfo  {journal}
  {Phys. Lett. B}\ }\textbf {\bibinfo {volume} {728}},\ \bibinfo {pages} {1}
  (\bibinfo {year} {2014})},\ \Eprint {http://arxiv.org/abs/1309.4311}
  {arXiv:1309.4311 [hep-lat]} \BibitemShut {NoStop}%
\bibitem [{\citenamefont {Yu}\ \emph {et~al.}(2022)\citenamefont {Yu},
  \citenamefont {Zhou}, \citenamefont {Huang}, \citenamefont {Shen},\ and\
  \citenamefont {Wu}}]{Yu:2021yvw}%
  \BibitemOpen
  \bibfield  {author} {\bibinfo {author} {\bibfnamefont {Q.}~\bibnamefont
  {Yu}}, \bibinfo {author} {\bibfnamefont {H.}~\bibnamefont {Zhou}}, \bibinfo
  {author} {\bibfnamefont {X.-D.}\ \bibnamefont {Huang}}, \bibinfo {author}
  {\bibfnamefont {J.-M.}\ \bibnamefont {Shen}}, \ and\ \bibinfo {author}
  {\bibfnamefont {X.-G.}\ \bibnamefont {Wu}},\ }\href {\doibase
  10.1088/0256-307X/39/7/071201} {\bibfield  {journal} {\bibinfo  {journal}
  {Chin. Phys. Lett.}\ }\textbf {\bibinfo {volume} {39}},\ \bibinfo {pages}
  {071201} (\bibinfo {year} {2022})},\ \Eprint
  {http://arxiv.org/abs/2112.01200} {arXiv:2112.01200 [hep-ph]} \BibitemShut
  {NoStop}%
\bibitem [{\citenamefont {Alexandru}\ and\ \citenamefont
  {Horv\'ath}(2019)}]{Alexandru:2019gdm}%
  \BibitemOpen
  \bibfield  {author} {\bibinfo {author} {\bibfnamefont {A.}~\bibnamefont
  {Alexandru}}\ and\ \bibinfo {author} {\bibfnamefont {I.}~\bibnamefont
  {Horv\'ath}},\ }\href {\doibase 10.1103/PhysRevD.100.094507} {\bibfield
  {journal} {\bibinfo  {journal} {Phys. Rev. D}\ }\textbf {\bibinfo {volume}
  {100}},\ \bibinfo {pages} {094507} (\bibinfo {year} {2019})},\ \Eprint
  {http://arxiv.org/abs/1906.08047} {arXiv:1906.08047 [hep-lat]} \BibitemShut
  {NoStop}%
\bibitem [{\citenamefont {Freund}\ and\ \citenamefont
  {Nambu}(1968)}]{Freund:1968gyq}%
  \BibitemOpen
  \bibfield  {author} {\bibinfo {author} {\bibfnamefont {P.~G.~O.}\
  \bibnamefont {Freund}}\ and\ \bibinfo {author} {\bibfnamefont
  {Y.}~\bibnamefont {Nambu}},\ }\href {\doibase 10.1103/PhysRev.174.1741}
  {\bibfield  {journal} {\bibinfo  {journal} {Phys. Rev.}\ }\textbf {\bibinfo
  {volume} {174}},\ \bibinfo {pages} {1741} (\bibinfo {year}
  {1968})}\BibitemShut {NoStop}%
\bibitem [{\citenamefont {Goldberger}\ \emph {et~al.}(2008)\citenamefont
  {Goldberger}, \citenamefont {Grinstein},\ and\ \citenamefont
  {Skiba}}]{Goldberger:2007zk}%
  \BibitemOpen
  \bibfield  {author} {\bibinfo {author} {\bibfnamefont {W.~D.}\ \bibnamefont
  {Goldberger}}, \bibinfo {author} {\bibfnamefont {B.}~\bibnamefont
  {Grinstein}}, \ and\ \bibinfo {author} {\bibfnamefont {W.}~\bibnamefont
  {Skiba}},\ }\href {\doibase 10.1103/PhysRevLett.100.111802} {\bibfield
  {journal} {\bibinfo  {journal} {Phys. Rev. Lett.}\ }\textbf {\bibinfo
  {volume} {100}},\ \bibinfo {pages} {111802} (\bibinfo {year} {2008})},\
  \Eprint {http://arxiv.org/abs/0708.1463} {arXiv:0708.1463 [hep-ph]}
  \BibitemShut {NoStop}%
\bibitem [{\citenamefont {Li}\ \emph {et~al.}(2017)\citenamefont {Li},
  \citenamefont {Ma},\ and\ \citenamefont {Rho}}]{Li:2016uzn}%
  \BibitemOpen
  \bibfield  {author} {\bibinfo {author} {\bibfnamefont {Y.-L.}\ \bibnamefont
  {Li}}, \bibinfo {author} {\bibfnamefont {Y.-L.}\ \bibnamefont {Ma}}, \ and\
  \bibinfo {author} {\bibfnamefont {M.}~\bibnamefont {Rho}},\ }\href {\doibase
  10.1103/PhysRevD.95.114011} {\bibfield  {journal} {\bibinfo  {journal} {Phys.
  Rev. D}\ }\textbf {\bibinfo {volume} {95}},\ \bibinfo {pages} {114011}
  (\bibinfo {year} {2017})},\ \Eprint {http://arxiv.org/abs/1609.07014}
  {arXiv:1609.07014 [hep-ph]} \BibitemShut {NoStop}%
\bibitem [{\citenamefont {Cat\`a}\ and\ \citenamefont
  {M\"uller}(2020)}]{Cata:2019edh}%
  \BibitemOpen
  \bibfield  {author} {\bibinfo {author} {\bibfnamefont {O.}~\bibnamefont
  {Cat\`a}}\ and\ \bibinfo {author} {\bibfnamefont {C.}~\bibnamefont
  {M\"uller}},\ }\href {\doibase 10.1016/j.nuclphysb.2020.114938} {\bibfield
  {journal} {\bibinfo  {journal} {Nucl. Phys. B}\ }\textbf {\bibinfo {volume}
  {952}},\ \bibinfo {pages} {114938} (\bibinfo {year} {2020})},\ \Eprint
  {http://arxiv.org/abs/1906.01879} {arXiv:1906.01879 [hep-ph]} \BibitemShut
  {NoStop}%
\bibitem [{\citenamefont {Li}\ \emph {et~al.}(2018{\natexlab{a}})\citenamefont
  {Li}, \citenamefont {Ma},\ and\ \citenamefont {Rho}}]{Li:2017udr}%
  \BibitemOpen
  \bibfield  {author} {\bibinfo {author} {\bibfnamefont {Y.-L.}\ \bibnamefont
  {Li}}, \bibinfo {author} {\bibfnamefont {Y.-L.}\ \bibnamefont {Ma}}, \ and\
  \bibinfo {author} {\bibfnamefont {M.}~\bibnamefont {Rho}},\ }\href {\doibase
  10.1088/1674-1137/42/9/094102} {\bibfield  {journal} {\bibinfo  {journal}
  {Chin. Phys. C}\ }\textbf {\bibinfo {volume} {42}},\ \bibinfo {pages}
  {094102} (\bibinfo {year} {2018}{\natexlab{a}})},\ \Eprint
  {http://arxiv.org/abs/1710.02840} {arXiv:1710.02840 [nucl-th]} \BibitemShut
  {NoStop}%
\bibitem [{\citenamefont {Li}\ \emph {et~al.}(2018{\natexlab{b}})\citenamefont
  {Li}, \citenamefont {Ma},\ and\ \citenamefont {Rho}}]{Li:2018ykx}%
  \BibitemOpen
  \bibfield  {author} {\bibinfo {author} {\bibfnamefont {Y.-L.}\ \bibnamefont
  {Li}}, \bibinfo {author} {\bibfnamefont {Y.-L.}\ \bibnamefont {Ma}}, \ and\
  \bibinfo {author} {\bibfnamefont {M.}~\bibnamefont {Rho}},\ }\href {\doibase
  10.1103/PhysRevC.98.044318} {\bibfield  {journal} {\bibinfo  {journal} {Phys.
  Rev. C}\ }\textbf {\bibinfo {volume} {98}},\ \bibinfo {pages} {044318}
  (\bibinfo {year} {2018}{\natexlab{b}})},\ \Eprint
  {http://arxiv.org/abs/1804.00310} {arXiv:1804.00310 [nucl-th]} \BibitemShut
  {NoStop}%
\bibitem [{\citenamefont {Ma}\ and\ \citenamefont
  {Rho}(2020{\natexlab{c}})}]{Ma:2020tsj}%
  \BibitemOpen
  \bibfield  {author} {\bibinfo {author} {\bibfnamefont {Y.-L.}\ \bibnamefont
  {Ma}}\ and\ \bibinfo {author} {\bibfnamefont {M.}~\bibnamefont {Rho}},\
  }\href {\doibase 10.1103/PhysRevLett.125.142501} {\bibfield  {journal}
  {\bibinfo  {journal} {Phys. Rev. Lett.}\ }\textbf {\bibinfo {volume} {125}},\
  \bibinfo {pages} {142501} (\bibinfo {year} {2020}{\natexlab{c}})},\ \Eprint
  {http://arxiv.org/abs/2002.03310} {arXiv:2002.03310 [nucl-th]} \BibitemShut
  {NoStop}%
\bibitem [{\citenamefont {Paeng}\ \emph {et~al.}(2012)\citenamefont {Paeng},
  \citenamefont {Lee}, \citenamefont {Rho},\ and\ \citenamefont
  {Sasaki}}]{Paeng:2011hy}%
  \BibitemOpen
  \bibfield  {author} {\bibinfo {author} {\bibfnamefont {W.-G.}\ \bibnamefont
  {Paeng}}, \bibinfo {author} {\bibfnamefont {H.~K.}\ \bibnamefont {Lee}},
  \bibinfo {author} {\bibfnamefont {M.}~\bibnamefont {Rho}}, \ and\ \bibinfo
  {author} {\bibfnamefont {C.}~\bibnamefont {Sasaki}},\ }\href {\doibase
  10.1103/PhysRevD.85.054022} {\bibfield  {journal} {\bibinfo  {journal} {Phys.
  Rev. D}\ }\textbf {\bibinfo {volume} {85}},\ \bibinfo {pages} {054022}
  (\bibinfo {year} {2012})},\ \Eprint {http://arxiv.org/abs/1109.5431}
  {arXiv:1109.5431 [hep-ph]} \BibitemShut {NoStop}%
\bibitem [{\citenamefont {Georgi}(1989)}]{Georgi:1989gp}%
  \BibitemOpen
  \bibfield  {author} {\bibinfo {author} {\bibfnamefont {H.}~\bibnamefont
  {Georgi}},\ }\href {\doibase 10.1103/PhysRevLett.63.1917} {\bibfield
  {journal} {\bibinfo  {journal} {Phys. Rev. Lett.}\ }\textbf {\bibinfo
  {volume} {63}},\ \bibinfo {pages} {1917} (\bibinfo {year}
  {1989})}\BibitemShut {NoStop}%
\bibitem [{\citenamefont {Georgi}(1990)}]{Georgi:1989xy}%
  \BibitemOpen
  \bibfield  {author} {\bibinfo {author} {\bibfnamefont {H.}~\bibnamefont
  {Georgi}},\ }\href {\doibase 10.1016/0550-3213(90)90210-5} {\bibfield
  {journal} {\bibinfo  {journal} {Nucl. Phys. B}\ }\textbf {\bibinfo {volume}
  {331}},\ \bibinfo {pages} {311} (\bibinfo {year} {1990})}\BibitemShut
  {NoStop}%
\bibitem [{\citenamefont {Harada}\ and\ \citenamefont
  {Yamawaki}(2001)}]{Harada:2000kb}%
  \BibitemOpen
  \bibfield  {author} {\bibinfo {author} {\bibfnamefont {M.}~\bibnamefont
  {Harada}}\ and\ \bibinfo {author} {\bibfnamefont {K.}~\bibnamefont
  {Yamawaki}},\ }\href {\doibase 10.1103/PhysRevLett.86.757} {\bibfield
  {journal} {\bibinfo  {journal} {Phys. Rev. Lett.}\ }\textbf {\bibinfo
  {volume} {86}},\ \bibinfo {pages} {757} (\bibinfo {year} {2001})},\ \Eprint
  {http://arxiv.org/abs/hep-ph/0010207} {arXiv:hep-ph/0010207} \BibitemShut
  {NoStop}%
\bibitem [{\citenamefont {Beane}\ and\ \citenamefont {van
  Kolck}(1994)}]{Beane:1994ds}%
  \BibitemOpen
  \bibfield  {author} {\bibinfo {author} {\bibfnamefont {S.~R.}\ \bibnamefont
  {Beane}}\ and\ \bibinfo {author} {\bibfnamefont {U.}~\bibnamefont {van
  Kolck}},\ }\href {\doibase 10.1016/0370-2693(94)90441-3} {\bibfield
  {journal} {\bibinfo  {journal} {Phys. Lett. B}\ }\textbf {\bibinfo {volume}
  {328}},\ \bibinfo {pages} {137} (\bibinfo {year} {1994})},\ \Eprint
  {http://arxiv.org/abs/hep-ph/9401218} {arXiv:hep-ph/9401218} \BibitemShut
  {NoStop}%
\bibitem [{\citenamefont {Suzuki}(2017)}]{Suzuki:2017tux}%
  \BibitemOpen
  \bibfield  {author} {\bibinfo {author} {\bibfnamefont {M.}~\bibnamefont
  {Suzuki}},\ }\href {\doibase 10.1103/PhysRevD.96.065010} {\bibfield
  {journal} {\bibinfo  {journal} {Phys. Rev. D}\ }\textbf {\bibinfo {volume}
  {96}},\ \bibinfo {pages} {065010} (\bibinfo {year} {2017})},\ \Eprint
  {http://arxiv.org/abs/1707.01589} {arXiv:1707.01589 [hep-ph]} \BibitemShut
  {NoStop}%
\bibitem [{\citenamefont {Yang}\ and\ \citenamefont {Ma}()}]{Yang:2022}%
  \BibitemOpen
  \bibfield  {author} {\bibinfo {author} {\bibfnamefont {W.-C.}\ \bibnamefont
  {Yang}}\ and\ \bibinfo {author} {\bibfnamefont {Y.-L.}\ \bibnamefont {Ma}},\
  }\href@noop {} {\ }\Eprint {http://arxiv.org/abs/In preparation} {In
  preparation} \BibitemShut {NoStop}%
\bibitem [{\citenamefont {Witten}(1979)}]{Witten:1979kh}%
  \BibitemOpen
  \bibfield  {author} {\bibinfo {author} {\bibfnamefont {E.}~\bibnamefont
  {Witten}},\ }\href {\doibase 10.1016/0550-3213(79)90232-3} {\bibfield
  {journal} {\bibinfo  {journal} {Nucl. Phys. B}\ }\textbf {\bibinfo {volume}
  {160}},\ \bibinfo {pages} {57} (\bibinfo {year} {1979})}\BibitemShut
  {NoStop}%
\bibitem [{\citenamefont {Klebanov}(1985)}]{Klebanov:1985qi}%
  \BibitemOpen
  \bibfield  {author} {\bibinfo {author} {\bibfnamefont {I.~R.}\ \bibnamefont
  {Klebanov}},\ }\href {\doibase 10.1016/0550-3213(85)90068-9} {\bibfield
  {journal} {\bibinfo  {journal} {Nucl. Phys. B}\ }\textbf {\bibinfo {volume}
  {262}},\ \bibinfo {pages} {133} (\bibinfo {year} {1985})}\BibitemShut
  {NoStop}%
\bibitem [{\citenamefont {Goldhaber}\ and\ \citenamefont
  {Manton}(1987)}]{Goldhaber:1987pb}%
  \BibitemOpen
  \bibfield  {author} {\bibinfo {author} {\bibfnamefont {A.~S.}\ \bibnamefont
  {Goldhaber}}\ and\ \bibinfo {author} {\bibfnamefont {N.~S.}\ \bibnamefont
  {Manton}},\ }\href {\doibase 10.1016/0370-2693(87)91502-4} {\bibfield
  {journal} {\bibinfo  {journal} {Phys. Lett. B}\ }\textbf {\bibinfo {volume}
  {198}},\ \bibinfo {pages} {231} (\bibinfo {year} {1987})}\BibitemShut
  {NoStop}%
\bibitem [{\citenamefont {Ma}\ and\ \citenamefont {Harada}(2016)}]{Ma:2016npf}%
  \BibitemOpen
  \bibfield  {author} {\bibinfo {author} {\bibfnamefont {Y.-L.}\ \bibnamefont
  {Ma}}\ and\ \bibinfo {author} {\bibfnamefont {M.}~\bibnamefont {Harada}},\
  }\href@noop {} {\  (\bibinfo {year} {2016})},\ \Eprint
  {http://arxiv.org/abs/1604.04850} {arXiv:1604.04850 [hep-ph]} \BibitemShut
  {NoStop}%
\bibitem [{\citenamefont {Harada}\ \emph {et~al.}(2015)\citenamefont {Harada},
  \citenamefont {Lee}, \citenamefont {Ma},\ and\ \citenamefont
  {Rho}}]{Harada:2015lma}%
  \BibitemOpen
  \bibfield  {author} {\bibinfo {author} {\bibfnamefont {M.}~\bibnamefont
  {Harada}}, \bibinfo {author} {\bibfnamefont {H.~K.}\ \bibnamefont {Lee}},
  \bibinfo {author} {\bibfnamefont {Y.-L.}\ \bibnamefont {Ma}}, \ and\ \bibinfo
  {author} {\bibfnamefont {M.}~\bibnamefont {Rho}},\ }\href {\doibase
  10.1103/PhysRevD.91.096011} {\bibfield  {journal} {\bibinfo  {journal} {Phys.
  Rev. D}\ }\textbf {\bibinfo {volume} {91}},\ \bibinfo {pages} {096011}
  (\bibinfo {year} {2015})},\ \Eprint {http://arxiv.org/abs/1502.02508}
  {arXiv:1502.02508 [hep-ph]} \BibitemShut {NoStop}%
\bibitem [{\citenamefont {Detar}\ and\ \citenamefont
  {Kunihiro}(1989)}]{Detar:1988kn}%
  \BibitemOpen
  \bibfield  {author} {\bibinfo {author} {\bibfnamefont {C.~E.}\ \bibnamefont
  {Detar}}\ and\ \bibinfo {author} {\bibfnamefont {T.}~\bibnamefont
  {Kunihiro}},\ }\href {\doibase 10.1103/PhysRevD.39.2805} {\bibfield
  {journal} {\bibinfo  {journal} {Phys. Rev. D}\ }\textbf {\bibinfo {volume}
  {39}},\ \bibinfo {pages} {2805} (\bibinfo {year} {1989})}\BibitemShut
  {NoStop}%
\bibitem [{\citenamefont {Motohiro}\ \emph {et~al.}(2015)\citenamefont
  {Motohiro}, \citenamefont {Kim},\ and\ \citenamefont
  {Harada}}]{Motohiro:2015taa}%
  \BibitemOpen
  \bibfield  {author} {\bibinfo {author} {\bibfnamefont {Y.}~\bibnamefont
  {Motohiro}}, \bibinfo {author} {\bibfnamefont {Y.}~\bibnamefont {Kim}}, \
  and\ \bibinfo {author} {\bibfnamefont {M.}~\bibnamefont {Harada}},\ }\href
  {\doibase 10.1103/PhysRevC.92.025201} {\bibfield  {journal} {\bibinfo
  {journal} {Phys. Rev. C}\ }\textbf {\bibinfo {volume} {92}},\ \bibinfo
  {pages} {025201} (\bibinfo {year} {2015})},\ \bibinfo {note} {[Erratum:
  Phys.Rev.C 95, 059903 (2017)]},\ \Eprint {http://arxiv.org/abs/1505.00988}
  {arXiv:1505.00988 [nucl-th]} \BibitemShut {NoStop}%
\bibitem [{\citenamefont {Ma}\ \emph {et~al.}(2014{\natexlab{b}})\citenamefont
  {Ma}, \citenamefont {Harada}, \citenamefont {Lee}, \citenamefont {Oh},\ and\
  \citenamefont {Rho}}]{Ma:2013vga}%
  \BibitemOpen
  \bibfield  {author} {\bibinfo {author} {\bibfnamefont {Y.-L.}\ \bibnamefont
  {Ma}}, \bibinfo {author} {\bibfnamefont {M.}~\bibnamefont {Harada}}, \bibinfo
  {author} {\bibfnamefont {H.~K.}\ \bibnamefont {Lee}}, \bibinfo {author}
  {\bibfnamefont {Y.}~\bibnamefont {Oh}}, \ and\ \bibinfo {author}
  {\bibfnamefont {M.}~\bibnamefont {Rho}},\ }\href {\doibase
  10.1142/S2010194514602385} {\bibfield  {journal} {\bibinfo  {journal} {Int.
  J. Mod. Phys. Conf. Ser.}\ }\textbf {\bibinfo {volume} {29}},\ \bibinfo
  {pages} {1460238} (\bibinfo {year} {2014}{\natexlab{b}})},\ \Eprint
  {http://arxiv.org/abs/1312.2290} {arXiv:1312.2290 [hep-ph]} \BibitemShut
  {NoStop}%
\bibitem [{\citenamefont {Komargodski}(2018)}]{Komargodski:2018odf}%
  \BibitemOpen
  \bibfield  {author} {\bibinfo {author} {\bibfnamefont {Z.}~\bibnamefont
  {Komargodski}},\ }\href@noop {} {\  (\bibinfo {year} {2018})},\ \Eprint
  {http://arxiv.org/abs/1812.09253} {arXiv:1812.09253 [hep-th]} \BibitemShut
  {NoStop}%
\bibitem [{\citenamefont {Hsin}\ and\ \citenamefont
  {Seiberg}(2016)}]{Hsin:2016blu}%
  \BibitemOpen
  \bibfield  {author} {\bibinfo {author} {\bibfnamefont {P.-S.}\ \bibnamefont
  {Hsin}}\ and\ \bibinfo {author} {\bibfnamefont {N.}~\bibnamefont {Seiberg}},\
  }\href {\doibase 10.1007/JHEP09(2016)095} {\bibfield  {journal} {\bibinfo
  {journal} {JHEP}\ }\textbf {\bibinfo {volume} {09}},\ \bibinfo {pages} {095}
  (\bibinfo {year} {2016})},\ \Eprint {http://arxiv.org/abs/1607.07457}
  {arXiv:1607.07457 [hep-th]} \BibitemShut {NoStop}%
\bibitem [{\citenamefont {Gaiotto}\ \emph {et~al.}(2018)\citenamefont
  {Gaiotto}, \citenamefont {Komargodski},\ and\ \citenamefont
  {Seiberg}}]{Gaiotto:2017tne}%
  \BibitemOpen
  \bibfield  {author} {\bibinfo {author} {\bibfnamefont {D.}~\bibnamefont
  {Gaiotto}}, \bibinfo {author} {\bibfnamefont {Z.}~\bibnamefont
  {Komargodski}}, \ and\ \bibinfo {author} {\bibfnamefont {N.}~\bibnamefont
  {Seiberg}},\ }\href {\doibase 10.1007/JHEP01(2018)110} {\bibfield  {journal}
  {\bibinfo  {journal} {JHEP}\ }\textbf {\bibinfo {volume} {01}},\ \bibinfo
  {pages} {110} (\bibinfo {year} {2018})},\ \Eprint
  {http://arxiv.org/abs/1708.06806} {arXiv:1708.06806 [hep-th]} \BibitemShut
  {NoStop}%
\bibitem [{\citenamefont {Benini}(2018)}]{Benini:2017aed}%
  \BibitemOpen
  \bibfield  {author} {\bibinfo {author} {\bibfnamefont {F.}~\bibnamefont
  {Benini}},\ }\href {\doibase 10.1007/JHEP02(2018)068} {\bibfield  {journal}
  {\bibinfo  {journal} {JHEP}\ }\textbf {\bibinfo {volume} {02}},\ \bibinfo
  {pages} {068} (\bibinfo {year} {2018})},\ \Eprint
  {http://arxiv.org/abs/1712.00020} {arXiv:1712.00020 [hep-th]} \BibitemShut
  {NoStop}%
\bibitem [{\citenamefont {Tong}(2016)}]{Tong:2016kpv}%
  \BibitemOpen
  \bibfield  {author} {\bibinfo {author} {\bibfnamefont {D.}~\bibnamefont
  {Tong}}\ }(\bibinfo {year} {2016})\ \Eprint {http://arxiv.org/abs/1606.06687}
  {arXiv:1606.06687 [hep-th]} \BibitemShut {NoStop}%
\bibitem [{\citenamefont {Karasik}(2020)}]{Karasik:2020pwu}%
  \BibitemOpen
  \bibfield  {author} {\bibinfo {author} {\bibfnamefont {A.}~\bibnamefont
  {Karasik}},\ }\href {\doibase 10.21468/SciPostPhys.9.1.008} {\bibfield
  {journal} {\bibinfo  {journal} {SciPost Phys.}\ }\textbf {\bibinfo {volume}
  {9}},\ \bibinfo {pages} {008} (\bibinfo {year} {2020})},\ \Eprint
  {http://arxiv.org/abs/2003.07893} {arXiv:2003.07893 [hep-th]} \BibitemShut
  {NoStop}%
\bibitem [{\citenamefont {Bigazzi}\ \emph {et~al.}(2022)\citenamefont
  {Bigazzi}, \citenamefont {Cotrone},\ and\ \citenamefont
  {Olzi}}]{Bigazzi:2022luo}%
  \BibitemOpen
  \bibfield  {author} {\bibinfo {author} {\bibfnamefont {F.}~\bibnamefont
  {Bigazzi}}, \bibinfo {author} {\bibfnamefont {A.~L.}\ \bibnamefont
  {Cotrone}}, \ and\ \bibinfo {author} {\bibfnamefont {A.}~\bibnamefont
  {Olzi}},\ }\href@noop {} {\  (\bibinfo {year} {2022})},\ \Eprint
  {http://arxiv.org/abs/2211.05147} {arXiv:2211.05147 [hep-th]} \BibitemShut
  {NoStop}%
\bibitem [{\citenamefont {Liu}\ \emph {et~al.}(2019)\citenamefont {Liu},
  \citenamefont {Ma},\ and\ \citenamefont {Rho}}]{Liu:2018wgv}%
  \BibitemOpen
  \bibfield  {author} {\bibinfo {author} {\bibfnamefont {X.-H.}\ \bibnamefont
  {Liu}}, \bibinfo {author} {\bibfnamefont {Y.-L.}\ \bibnamefont {Ma}}, \ and\
  \bibinfo {author} {\bibfnamefont {M.}~\bibnamefont {Rho}},\ }\href {\doibase
  10.1103/PhysRevC.99.055808} {\bibfield  {journal} {\bibinfo  {journal} {Phys.
  Rev. C}\ }\textbf {\bibinfo {volume} {99}},\ \bibinfo {pages} {055808}
  (\bibinfo {year} {2019})},\ \Eprint {http://arxiv.org/abs/1811.10012}
  {arXiv:1811.10012 [nucl-th]} \BibitemShut {NoStop}%
\bibitem [{\citenamefont {Ma}\ \emph {et~al.}(2019{\natexlab{b}})\citenamefont
  {Ma}, \citenamefont {Nowak}, \citenamefont {Rho},\ and\ \citenamefont
  {Zahed}}]{Ma:2019xtx}%
  \BibitemOpen
  \bibfield  {author} {\bibinfo {author} {\bibfnamefont {Y.-L.}\ \bibnamefont
  {Ma}}, \bibinfo {author} {\bibfnamefont {M.~A.}\ \bibnamefont {Nowak}},
  \bibinfo {author} {\bibfnamefont {M.}~\bibnamefont {Rho}}, \ and\ \bibinfo
  {author} {\bibfnamefont {I.}~\bibnamefont {Zahed}},\ }\href {\doibase
  10.1103/PhysRevLett.123.172301} {\bibfield  {journal} {\bibinfo  {journal}
  {Phys. Rev. Lett.}\ }\textbf {\bibinfo {volume} {123}},\ \bibinfo {pages}
  {172301} (\bibinfo {year} {2019}{\natexlab{b}})},\ \Eprint
  {http://arxiv.org/abs/1907.00958} {arXiv:1907.00958 [hep-th]} \BibitemShut
  {NoStop}%
\bibitem [{\citenamefont {Callan}\ and\ \citenamefont
  {Harvey}(1985)}]{Callan:1984sa}%
  \BibitemOpen
  \bibfield  {author} {\bibinfo {author} {\bibfnamefont {C.~G.}\ \bibnamefont
  {Callan}, \bibfnamefont {Jr.}}\ and\ \bibinfo {author} {\bibfnamefont
  {J.~A.}\ \bibnamefont {Harvey}},\ }\href {\doibase
  10.1016/0550-3213(85)90489-4} {\bibfield  {journal} {\bibinfo  {journal}
  {Nucl. Phys. B}\ }\textbf {\bibinfo {volume} {250}},\ \bibinfo {pages} {427}
  (\bibinfo {year} {1985})}\BibitemShut {NoStop}%
\bibitem [{\citenamefont {Park}\ \emph {et~al.}(2019)\citenamefont {Park},
  \citenamefont {Paeng},\ and\ \citenamefont {Vento}}]{Park:2019bmi}%
  \BibitemOpen
  \bibfield  {author} {\bibinfo {author} {\bibfnamefont {B.-Y.}\ \bibnamefont
  {Park}}, \bibinfo {author} {\bibfnamefont {W.-G.}\ \bibnamefont {Paeng}}, \
  and\ \bibinfo {author} {\bibfnamefont {V.}~\bibnamefont {Vento}},\ }\href
  {\doibase 10.1016/j.nuclphysa.2019.06.010} {\bibfield  {journal} {\bibinfo
  {journal} {Nucl. Phys. A}\ }\textbf {\bibinfo {volume} {989}},\ \bibinfo
  {pages} {231} (\bibinfo {year} {2019})},\ \Eprint
  {http://arxiv.org/abs/1904.04483} {arXiv:1904.04483 [hep-ph]} \BibitemShut
  {NoStop}%
\bibitem [{\citenamefont {Canfora}(2018)}]{Canfora:2018rdz}%
  \BibitemOpen
  \bibfield  {author} {\bibinfo {author} {\bibfnamefont {F.}~\bibnamefont
  {Canfora}},\ }\href {\doibase 10.1140/epjc/s10052-018-6404-x} {\bibfield
  {journal} {\bibinfo  {journal} {Eur. Phys. J. C}\ }\textbf {\bibinfo {volume}
  {78}},\ \bibinfo {pages} {929} (\bibinfo {year} {2018})},\ \Eprint
  {http://arxiv.org/abs/1807.02090} {arXiv:1807.02090 [hep-th]} \BibitemShut
  {NoStop}%
\bibitem [{\citenamefont {Rho}(2022{\natexlab{b}})}]{Rho:2022meo}%
  \BibitemOpen
  \bibfield  {author} {\bibinfo {author} {\bibfnamefont {M.}~\bibnamefont
  {Rho}},\ }\href@noop {} {\  (\bibinfo {year} {2022}{\natexlab{b}})},\ \Eprint
  {http://arxiv.org/abs/2211.14890} {arXiv:2211.14890 [nucl-th]} \BibitemShut
  {NoStop}%
\bibitem [{\citenamefont {Sulejmanpasic}\ \emph {et~al.}(2017)\citenamefont
  {Sulejmanpasic}, \citenamefont {Shao}, \citenamefont {Sandvik},\ and\
  \citenamefont {Unsal}}]{Sulejmanpasic:2016uwq}%
  \BibitemOpen
  \bibfield  {author} {\bibinfo {author} {\bibfnamefont {T.}~\bibnamefont
  {Sulejmanpasic}}, \bibinfo {author} {\bibfnamefont {H.}~\bibnamefont {Shao}},
  \bibinfo {author} {\bibfnamefont {A.}~\bibnamefont {Sandvik}}, \ and\
  \bibinfo {author} {\bibfnamefont {M.}~\bibnamefont {Unsal}},\ }\href
  {\doibase 10.1103/PhysRevLett.119.091601} {\bibfield  {journal} {\bibinfo
  {journal} {Phys. Rev. Lett.}\ }\textbf {\bibinfo {volume} {119}},\ \bibinfo
  {pages} {091601} (\bibinfo {year} {2017})},\ \Eprint
  {http://arxiv.org/abs/1608.09011} {arXiv:1608.09011 [hep-th]} \BibitemShut
  {NoStop}%
\bibitem [{\citenamefont {Senthil}\ \emph {et~al.}(2004)\citenamefont
  {Senthil}, \citenamefont {Vishwanath}, \citenamefont {Balents}, \citenamefont
  {Sachdev},\ and\ \citenamefont {Fisher}}]{Senthil:2003eed}%
  \BibitemOpen
  \bibfield  {author} {\bibinfo {author} {\bibfnamefont {T.}~\bibnamefont
  {Senthil}}, \bibinfo {author} {\bibfnamefont {A.}~\bibnamefont {Vishwanath}},
  \bibinfo {author} {\bibfnamefont {L.}~\bibnamefont {Balents}}, \bibinfo
  {author} {\bibfnamefont {S.}~\bibnamefont {Sachdev}}, \ and\ \bibinfo
  {author} {\bibfnamefont {M.~P.~A.}\ \bibnamefont {Fisher}},\ }\href {\doibase
  10.1126/science.1091806} {\bibfield  {journal} {\bibinfo  {journal}
  {Science}\ }\textbf {\bibinfo {volume} {303}},\ \bibinfo {pages} {1490}
  (\bibinfo {year} {2004})},\ \Eprint {http://arxiv.org/abs/cond-mat/0311326}
  {arXiv:cond-mat/0311326} \BibitemShut {NoStop}%
\bibitem [{\citenamefont {Li}\ and\ \citenamefont {Ma}(2017)}]{Li:2017hqe}%
  \BibitemOpen
  \bibfield  {author} {\bibinfo {author} {\bibfnamefont {Y.-L.}\ \bibnamefont
  {Li}}\ and\ \bibinfo {author} {\bibfnamefont {Y.-L.}\ \bibnamefont {Ma}},\
  }\href@noop {} {\  (\bibinfo {year} {2017})},\ \Eprint
  {http://arxiv.org/abs/1710.02839} {arXiv:1710.02839 [hep-ph]} \BibitemShut
  {NoStop}%
\bibitem [{\citenamefont {Ma}\ and\ \citenamefont {Rho}(2018)}]{Ma:2016nki}%
  \BibitemOpen
  \bibfield  {author} {\bibinfo {author} {\bibfnamefont {Y.-L.}\ \bibnamefont
  {Ma}}\ and\ \bibinfo {author} {\bibfnamefont {M.}~\bibnamefont {Rho}},\
  }\href {\doibase 10.1103/PhysRevD.97.094017} {\bibfield  {journal} {\bibinfo
  {journal} {Phys. Rev. D}\ }\textbf {\bibinfo {volume} {97}},\ \bibinfo
  {pages} {094017} (\bibinfo {year} {2018})},\ \Eprint
  {http://arxiv.org/abs/1612.04079} {arXiv:1612.04079 [nucl-th]} \BibitemShut
  {NoStop}%
\bibitem [{\citenamefont {Wilkinson}(1973)}]{Wilkinson:1973zz}%
  \BibitemOpen
  \bibfield  {author} {\bibinfo {author} {\bibfnamefont {D.~H.}\ \bibnamefont
  {Wilkinson}},\ }\href {\doibase 10.1103/PhysRevC.7.930} {\bibfield  {journal}
  {\bibinfo  {journal} {Phys. Rev. C}\ }\textbf {\bibinfo {volume} {7}},\
  \bibinfo {pages} {930} (\bibinfo {year} {1973})}\BibitemShut {NoStop}%
\bibitem [{\citenamefont {Suhonen}(2017)}]{Suhonen:2017krv}%
  \BibitemOpen
  \bibfield  {author} {\bibinfo {author} {\bibfnamefont {J.~T.}\ \bibnamefont
  {Suhonen}},\ }\href {\doibase 10.3389/fphy.2017.00055} {\bibfield  {journal}
  {\bibinfo  {journal} {Front. in Phys.}\ }\textbf {\bibinfo {volume} {5}},\
  \bibinfo {pages} {55} (\bibinfo {year} {2017})},\ \Eprint
  {http://arxiv.org/abs/1712.01565} {arXiv:1712.01565 [nucl-th]} \BibitemShut
  {NoStop}%
\bibitem [{\citenamefont {Engel}\ and\ \citenamefont
  {Men\'endez}(2017)}]{Engel:2016xgb}%
  \BibitemOpen
  \bibfield  {author} {\bibinfo {author} {\bibfnamefont {J.}~\bibnamefont
  {Engel}}\ and\ \bibinfo {author} {\bibfnamefont {J.}~\bibnamefont
  {Men\'endez}},\ }\href {\doibase 10.1088/1361-6633/aa5bc5} {\bibfield
  {journal} {\bibinfo  {journal} {Rept. Prog. Phys.}\ }\textbf {\bibinfo
  {volume} {80}},\ \bibinfo {pages} {046301} (\bibinfo {year} {2017})},\
  \Eprint {http://arxiv.org/abs/1610.06548} {arXiv:1610.06548 [nucl-th]}
  \BibitemShut {NoStop}%
\bibitem [{\citenamefont {Rho}(2022{\natexlab{c}})}]{Rho:2022vju}%
  \BibitemOpen
  \bibfield  {author} {\bibinfo {author} {\bibfnamefont {M.}~\bibnamefont
  {Rho}},\ }\href@noop {} {\  (\bibinfo {year} {2022}{\natexlab{c}})},\ \Eprint
  {http://arxiv.org/abs/2212.05558} {arXiv:2212.05558 [nucl-th]} \BibitemShut
  {NoStop}%
\bibitem [{\citenamefont {Friman}\ and\ \citenamefont
  {Rho}(1996)}]{Friman:1996qc}%
  \BibitemOpen
  \bibfield  {author} {\bibinfo {author} {\bibfnamefont {B.}~\bibnamefont
  {Friman}}\ and\ \bibinfo {author} {\bibfnamefont {M.}~\bibnamefont {Rho}},\
  }\href {\doibase 10.1016/0375-9474(96)00215-1} {\bibfield  {journal}
  {\bibinfo  {journal} {Nucl. Phys. A}\ }\textbf {\bibinfo {volume} {606}},\
  \bibinfo {pages} {303} (\bibinfo {year} {1996})},\ \Eprint
  {http://arxiv.org/abs/nucl-th/9602025} {arXiv:nucl-th/9602025} \BibitemShut
  {NoStop}%
\bibitem [{\citenamefont {Kienle}\ and\ \citenamefont
  {Yamazaki}(2004)}]{Kienle:2004hq}%
  \BibitemOpen
  \bibfield  {author} {\bibinfo {author} {\bibfnamefont {P.}~\bibnamefont
  {Kienle}}\ and\ \bibinfo {author} {\bibfnamefont {T.}~\bibnamefont
  {Yamazaki}},\ }\href {\doibase 10.1016/j.ppnp.2003.09.001} {\bibfield
  {journal} {\bibinfo  {journal} {Prog. Part. Nucl. Phys.}\ }\textbf {\bibinfo
  {volume} {52}},\ \bibinfo {pages} {85} (\bibinfo {year} {2004})}\BibitemShut
  {NoStop}%
\bibitem [{\citenamefont {Lubos}\ \emph {et~al.}(2019)\citenamefont {Lubos}
  \emph {et~al.}}]{Lubos:2019nik}%
  \BibitemOpen
  \bibfield  {author} {\bibinfo {author} {\bibfnamefont {D.}~\bibnamefont
  {Lubos}} \emph {et~al.},\ }\href {\doibase 10.1103/PhysRevLett.122.222502}
  {\bibfield  {journal} {\bibinfo  {journal} {Phys. Rev. Lett.}\ }\textbf
  {\bibinfo {volume} {122}},\ \bibinfo {pages} {222502} (\bibinfo {year}
  {2019})}\BibitemShut {NoStop}%
\bibitem [{\citenamefont {van Kolck}(2019)}]{vanKolck:2019qea}%
  \BibitemOpen
  \bibfield  {author} {\bibinfo {author} {\bibfnamefont {U.}~\bibnamefont {van
  Kolck}},\ }\href {\doibase 10.1393/ncc/i2019-19052-7} {\bibfield  {journal}
  {\bibinfo  {journal} {Nuovo Cim. C}\ }\textbf {\bibinfo {volume} {42}},\
  \bibinfo {pages} {52} (\bibinfo {year} {2019})}\BibitemShut {NoStop}%
\bibitem [{\citenamefont {Tews}\ \emph {et~al.}(2017)\citenamefont {Tews},
  \citenamefont {Lattimer}, \citenamefont {Ohnishi},\ and\ \citenamefont
  {Kolomeitsev}}]{Tews:2016jhi}%
  \BibitemOpen
  \bibfield  {author} {\bibinfo {author} {\bibfnamefont {I.}~\bibnamefont
  {Tews}}, \bibinfo {author} {\bibfnamefont {J.~M.}\ \bibnamefont {Lattimer}},
  \bibinfo {author} {\bibfnamefont {A.}~\bibnamefont {Ohnishi}}, \ and\
  \bibinfo {author} {\bibfnamefont {E.~E.}\ \bibnamefont {Kolomeitsev}},\
  }\href {\doibase 10.3847/1538-4357/aa8db9} {\bibfield  {journal} {\bibinfo
  {journal} {Astrophys. J.}\ }\textbf {\bibinfo {volume} {848}},\ \bibinfo
  {pages} {105} (\bibinfo {year} {2017})},\ \Eprint
  {http://arxiv.org/abs/1611.07133} {arXiv:1611.07133 [nucl-th]} \BibitemShut
  {NoStop}%
\bibitem [{\citenamefont {Annala}\ \emph {et~al.}(2020)\citenamefont {Annala},
  \citenamefont {Gorda}, \citenamefont {Kurkela}, \citenamefont {N\"attil\"a},\
  and\ \citenamefont {Vuorinen}}]{Annala:2019puf}%
  \BibitemOpen
  \bibfield  {author} {\bibinfo {author} {\bibfnamefont {E.}~\bibnamefont
  {Annala}}, \bibinfo {author} {\bibfnamefont {T.}~\bibnamefont {Gorda}},
  \bibinfo {author} {\bibfnamefont {A.}~\bibnamefont {Kurkela}}, \bibinfo
  {author} {\bibfnamefont {J.}~\bibnamefont {N\"attil\"a}}, \ and\ \bibinfo
  {author} {\bibfnamefont {A.}~\bibnamefont {Vuorinen}},\ }\href {\doibase
  10.1038/s41567-020-0914-9} {\bibfield  {journal} {\bibinfo  {journal} {Nature
  Phys.}\ }\textbf {\bibinfo {volume} {16}},\ \bibinfo {pages} {907} (\bibinfo
  {year} {2020})},\ \Eprint {http://arxiv.org/abs/1903.09121} {arXiv:1903.09121
  [astro-ph.HE]} \BibitemShut {NoStop}%
\bibitem [{\citenamefont {Hippert}\ \emph {et~al.}(2021)\citenamefont
  {Hippert}, \citenamefont {Fraga},\ and\ \citenamefont
  {Noronha}}]{Hippert:2021gfs}%
  \BibitemOpen
  \bibfield  {author} {\bibinfo {author} {\bibfnamefont {M.}~\bibnamefont
  {Hippert}}, \bibinfo {author} {\bibfnamefont {E.~S.}\ \bibnamefont {Fraga}},
  \ and\ \bibinfo {author} {\bibfnamefont {J.}~\bibnamefont {Noronha}},\ }\href
  {\doibase 10.1103/PhysRevD.104.034011} {\bibfield  {journal} {\bibinfo
  {journal} {Phys. Rev. D}\ }\textbf {\bibinfo {volume} {104}},\ \bibinfo
  {pages} {034011} (\bibinfo {year} {2021})},\ \Eprint
  {http://arxiv.org/abs/2105.04535} {arXiv:2105.04535 [nucl-th]} \BibitemShut
  {NoStop}%
\bibitem [{\citenamefont {Tsang}\ \emph {et~al.}(2018)\citenamefont {Tsang},
  \citenamefont {Tsang}, \citenamefont {Danielewicz}, \citenamefont {Lynch},\
  and\ \citenamefont {Fattoyev}}]{Tsang:2018kqj}%
  \BibitemOpen
  \bibfield  {author} {\bibinfo {author} {\bibfnamefont {C.~Y.}\ \bibnamefont
  {Tsang}}, \bibinfo {author} {\bibfnamefont {M.~B.}\ \bibnamefont {Tsang}},
  \bibinfo {author} {\bibfnamefont {P.}~\bibnamefont {Danielewicz}}, \bibinfo
  {author} {\bibfnamefont {W.~G.}\ \bibnamefont {Lynch}}, \ and\ \bibinfo
  {author} {\bibfnamefont {F.~J.}\ \bibnamefont {Fattoyev}},\ }\href@noop {} {\
   (\bibinfo {year} {2018})},\ \Eprint {http://arxiv.org/abs/1807.06571}
  {arXiv:1807.06571 [nucl-ex]} \BibitemShut {NoStop}%
\bibitem [{\citenamefont {Demorest}\ \emph {et~al.}(2010)\citenamefont
  {Demorest}, \citenamefont {Pennucci}, \citenamefont {Ransom}, \citenamefont
  {Roberts},\ and\ \citenamefont {Hessels}}]{Demorest:2010bx}%
  \BibitemOpen
  \bibfield  {author} {\bibinfo {author} {\bibfnamefont {P.}~\bibnamefont
  {Demorest}}, \bibinfo {author} {\bibfnamefont {T.}~\bibnamefont {Pennucci}},
  \bibinfo {author} {\bibfnamefont {S.}~\bibnamefont {Ransom}}, \bibinfo
  {author} {\bibfnamefont {M.}~\bibnamefont {Roberts}}, \ and\ \bibinfo
  {author} {\bibfnamefont {J.}~\bibnamefont {Hessels}},\ }\href {\doibase
  10.1038/nature09466} {\bibfield  {journal} {\bibinfo  {journal} {Nature}\
  }\textbf {\bibinfo {volume} {467}},\ \bibinfo {pages} {1081} (\bibinfo {year}
  {2010})},\ \Eprint {http://arxiv.org/abs/1010.5788} {arXiv:1010.5788
  [astro-ph.HE]} \BibitemShut {NoStop}%
\bibitem [{\citenamefont {Antoniadis}\ \emph {et~al.}(2013)\citenamefont
  {Antoniadis} \emph {et~al.}}]{Antoniadis:2013pzd}%
  \BibitemOpen
  \bibfield  {author} {\bibinfo {author} {\bibfnamefont {J.}~\bibnamefont
  {Antoniadis}} \emph {et~al.},\ }\href {\doibase 10.1126/science.1233232}
  {\bibfield  {journal} {\bibinfo  {journal} {Science}\ }\textbf {\bibinfo
  {volume} {340}},\ \bibinfo {pages} {6131} (\bibinfo {year} {2013})},\ \Eprint
  {http://arxiv.org/abs/1304.6875} {arXiv:1304.6875 [astro-ph.HE]} \BibitemShut
  {NoStop}%
\bibitem [{\citenamefont {Cromartie}\ \emph {et~al.}(2019)\citenamefont
  {Cromartie} \emph {et~al.}}]{NANOGrav:2019jur}%
  \BibitemOpen
  \bibfield  {author} {\bibinfo {author} {\bibfnamefont {H.~T.}\ \bibnamefont
  {Cromartie}} \emph {et~al.} (\bibinfo {collaboration} {NANOGrav}),\ }\href
  {\doibase 10.1038/s41550-019-0880-2} {\bibfield  {journal} {\bibinfo
  {journal} {Nature Astron.}\ }\textbf {\bibinfo {volume} {4}},\ \bibinfo
  {pages} {72} (\bibinfo {year} {2019})},\ \Eprint
  {http://arxiv.org/abs/1904.06759} {arXiv:1904.06759 [astro-ph.HE]}
  \BibitemShut {NoStop}%
\bibitem [{\citenamefont {Raaijmakers}\ \emph {et~al.}(2019)\citenamefont
  {Raaijmakers} \emph {et~al.}}]{Raaijmakers:2019qny}%
  \BibitemOpen
  \bibfield  {author} {\bibinfo {author} {\bibfnamefont {G.}~\bibnamefont
  {Raaijmakers}} \emph {et~al.},\ }\href {\doibase 10.3847/2041-8213/ab451a}
  {\bibfield  {journal} {\bibinfo  {journal} {Astrophys. J. Lett.}\ }\textbf
  {\bibinfo {volume} {887}},\ \bibinfo {pages} {L22} (\bibinfo {year}
  {2019})},\ \Eprint {http://arxiv.org/abs/1912.05703} {arXiv:1912.05703
  [astro-ph.HE]} \BibitemShut {NoStop}%
\bibitem [{\citenamefont {Miller}\ \emph {et~al.}(2019)\citenamefont {Miller}
  \emph {et~al.}}]{Miller:2019cac}%
  \BibitemOpen
  \bibfield  {author} {\bibinfo {author} {\bibfnamefont {M.~C.}\ \bibnamefont
  {Miller}} \emph {et~al.},\ }\href {\doibase 10.3847/2041-8213/ab50c5}
  {\bibfield  {journal} {\bibinfo  {journal} {Astrophys. J. Lett.}\ }\textbf
  {\bibinfo {volume} {887}},\ \bibinfo {pages} {L24} (\bibinfo {year}
  {2019})},\ \Eprint {http://arxiv.org/abs/1912.05705} {arXiv:1912.05705
  [astro-ph.HE]} \BibitemShut {NoStop}%
\bibitem [{\citenamefont {Fattoyev}\ \emph {et~al.}(2018)\citenamefont
  {Fattoyev}, \citenamefont {Piekarewicz},\ and\ \citenamefont
  {Horowitz}}]{Fattoyev:2017jql}%
  \BibitemOpen
  \bibfield  {author} {\bibinfo {author} {\bibfnamefont {F.~J.}\ \bibnamefont
  {Fattoyev}}, \bibinfo {author} {\bibfnamefont {J.}~\bibnamefont
  {Piekarewicz}}, \ and\ \bibinfo {author} {\bibfnamefont {C.~J.}\ \bibnamefont
  {Horowitz}},\ }\href {\doibase 10.1103/PhysRevLett.120.172702} {\bibfield
  {journal} {\bibinfo  {journal} {Phys. Rev. Lett.}\ }\textbf {\bibinfo
  {volume} {120}},\ \bibinfo {pages} {172702} (\bibinfo {year} {2018})},\
  \Eprint {http://arxiv.org/abs/1711.06615} {arXiv:1711.06615 [nucl-th]}
  \BibitemShut {NoStop}%
\bibitem [{\citenamefont {Weise}(2019)}]{Weise:2019mou}%
  \BibitemOpen
  \bibfield  {author} {\bibinfo {author} {\bibfnamefont {W.}~\bibnamefont
  {Weise}},\ }\href {\doibase 10.7566/JPSCP.26.011002} {\bibfield  {journal}
  {\bibinfo  {journal} {JPS Conf. Proc.}\ }\textbf {\bibinfo {volume} {26}},\
  \bibinfo {pages} {011002} (\bibinfo {year} {2019})},\ \Eprint
  {http://arxiv.org/abs/1905.03955} {arXiv:1905.03955 [nucl-th]} \BibitemShut
  {NoStop}%
\bibitem [{\citenamefont {Takeda}\ \emph {et~al.}(2018)\citenamefont {Takeda},
  \citenamefont {Kim},\ and\ \citenamefont {Harada}}]{Takeda:2017mrm}%
  \BibitemOpen
  \bibfield  {author} {\bibinfo {author} {\bibfnamefont {Y.}~\bibnamefont
  {Takeda}}, \bibinfo {author} {\bibfnamefont {Y.}~\bibnamefont {Kim}}, \ and\
  \bibinfo {author} {\bibfnamefont {M.}~\bibnamefont {Harada}},\ }\href
  {\doibase 10.1103/PhysRevC.97.065202} {\bibfield  {journal} {\bibinfo
  {journal} {Phys. Rev. C}\ }\textbf {\bibinfo {volume} {97}},\ \bibinfo
  {pages} {065202} (\bibinfo {year} {2018})},\ \Eprint
  {http://arxiv.org/abs/1704.04357} {arXiv:1704.04357 [nucl-th]} \BibitemShut
  {NoStop}%
\bibitem [{\citenamefont {Marczenko}\ \emph
  {et~al.}(2022{\natexlab{b}})\citenamefont {Marczenko}, \citenamefont
  {Redlich},\ and\ \citenamefont {Sasaki}}]{Marczenko:2021uaj}%
  \BibitemOpen
  \bibfield  {author} {\bibinfo {author} {\bibfnamefont {M.}~\bibnamefont
  {Marczenko}}, \bibinfo {author} {\bibfnamefont {K.}~\bibnamefont {Redlich}},
  \ and\ \bibinfo {author} {\bibfnamefont {C.}~\bibnamefont {Sasaki}},\ }\href
  {\doibase 10.3847/2041-8213/ac4b61} {\bibfield  {journal} {\bibinfo
  {journal} {Astrophys. J. Lett.}\ }\textbf {\bibinfo {volume} {925}},\
  \bibinfo {pages} {L23} (\bibinfo {year} {2022}{\natexlab{b}})},\ \Eprint
  {http://arxiv.org/abs/2110.11056} {arXiv:2110.11056 [nucl-th]} \BibitemShut
  {NoStop}%
\bibitem [{\citenamefont {Yang}\ \emph {et~al.}(2021)\citenamefont {Yang},
  \citenamefont {Ma},\ and\ \citenamefont {Wu}}]{Yang:2020ucv}%
  \BibitemOpen
  \bibfield  {author} {\bibinfo {author} {\bibfnamefont {W.-C.}\ \bibnamefont
  {Yang}}, \bibinfo {author} {\bibfnamefont {Y.-L.}\ \bibnamefont {Ma}}, \ and\
  \bibinfo {author} {\bibfnamefont {Y.-L.}\ \bibnamefont {Wu}},\ }\href
  {\doibase 10.1007/s11433-020-1662-5} {\bibfield  {journal} {\bibinfo
  {journal} {Sci. China Phys. Mech. Astron.}\ }\textbf {\bibinfo {volume}
  {64}},\ \bibinfo {pages} {252011} (\bibinfo {year} {2021})},\ \Eprint
  {http://arxiv.org/abs/2011.03665} {arXiv:2011.03665 [nucl-th]} \BibitemShut
  {NoStop}%
\end{thebibliography}%

\end{document}